\documentclass[aps,prd,showpacs,nofootinbib,floats,floatfix,preprintnumbers,groupedaddress,twocolumn]{revtex4-1}
\usepackage{graphicx,epsfig}
\usepackage{dcolumn}
\usepackage{bm}
\usepackage{latexsym}
\usepackage[table]{xcolor}
\usepackage{booktabs}
\usepackage{color}
\usepackage{tabularx}
\usepackage{ulem}
\usepackage{hyperref}
\usepackage{float}
\usepackage{tabularx}
\usepackage{color}
\usepackage{comment}
\usepackage{physics}
\usepackage{subfigure}
\usepackage{tikz}
\usepackage{hyperref}
\usepackage{colortbl}
\usepackage{listings}
\usepackage{amsmath,amsfonts,amssymb}
\usepackage{fancyhdr}
\usepackage{orcidlink}
\usepackage{hyperref}
\usepackage{natbib}
\usepackage{multirow}
\hypersetup{
	colorlinks   = true, 
	urlcolor     = blue, 
	linkcolor    = blue, 
	citecolor    = red    
}

\hypersetup{
	colorlinks   = true, 
	urlcolor     = blue, 
	linkcolor    = blue, 
	citecolor    = red 
}

\def\a{\alpha}

\definecolor{lightblue}{RGB}{173,216,230}
\definecolor{lightgreen}{RGB}{144,238,144}
\definecolor{magenta}{RGB}{255,0,255}
\definecolor{myolive}{RGB}{128,128,0}
\definecolor{masteredyellow}{RGB}{255,255,102}
\definecolor{mymaroon}{RGB}{128,0,0}

\begin{document}

	\title{Near-horizon chaos beyond Einstein gravity}
	
        \author{Surajit Das$^{1,2,3}$\orcidlink{0000-0003-2994-6951}}
        \email{surajitdas@mail.ustc.edu.cn,\\
        surajit.cbpbu20@gmail.com}

	\author{Surojit Dalui$^{4}$\orcidlink{0000-0003-1003-8451}}
        \email{surojitdalui@shu.edu.cn,\\
        surojitdalui003@gmail.com}

        \author{Rickmoy Samanta$^{3}$\orcidlink{0000-0002-4735-7774}}
        \email{rickmoy.samanta@hyderabad.bits-pilani.ac.in}

        \affiliation{$^{1}$Department of Astronomy, School of Physical Sciences, \textcolor{blue}{University of Science and Technology of China}, Hefei, Anhui 230026, China\\
\\
        $^{2}$CAS Key Laboratory for Researches in Galaxies and Cosmology, School of Astronomy and Space Science, \textcolor{blue}{University of Science and Technology of China}, Hefei, Anhui 230026, China\\
\\
        $^{3}$Department of Physics, \textcolor{blue}{Birla Institute of Technology and Science - Pilani}, Hyderabad Campus, Telangana 500078, India\\
\\
        $^{4}$Department of Physics, \textcolor{blue}{Shanghai University}, 99 Shangda Road, Baoshan District, Shanghai 200444, China}


\begin{abstract}
    We investigate chaos in the dynamics of massless particles near the horizon of static spherically symmetric black holes in two well-motivated models of $f(R)$ gravity. In both these models, we probe chaos in the particle trajectories (under suitable harmonic confinement) in the vicinity of the black hole horizons, for a set of initial conditions. The particle trajectories, associated Poincar$\acute{e}$ sections, and Lyapunov exponents clearly illustrate the role played by the black hole horizon in the growth of chaos. We find that with increasing energy, the particle trajectories explore regions closer to the black hole horizon, with reduced overlap between two initially close trajectories. We demonstrate how this energy range is controlled by the parameters of the modified gravity theory under consideration. The growth of chaos in such a classical setting is known to respect a surface gravity bound arising from universal aspects of particle dynamics close to the black hole horizon [K. Hashimoto and N. Tanahashi, \textcolor{blue}{Phys. Rev. D {\bf{95}}, 024007 (2017)}], analogous to the quantum Maldacena, Shenker, and Stanford bound [J. Maldacena {\it{et al.,}}\textcolor{blue}{J. High Energy Phys. 08 (2016) 106}]. Interestingly, both models studied in our work respect the bound, in contrast to some of the other models of $f(R)$ gravity in the existing literature. The work serves as a motivation to use chaos as an additional tool to probe Einstein gravity in the strong gravity regime in the vicinity of black hole horizons. 
\end{abstract}
	

	\maketitle

	\section{Introduction}\label{s1}
    The recent detection of gravitational waves by LIGO \cite{abbott1,abbott2,abbott3,abbott4} and black hole images captured by EHT \cite{telescope1,telescope2} have transformed black holes from a purely theoretical concept into real physical systems within our Universe. The study of near-horizon physics, encompassing both classical and quantum aspects, holds significant importance. Within the classical framework, extensive research has explored the intriguing influence of the black hole horizon on integrable systems, leading to their transition into chaotic states.  Investigations on the influence of black hole horizons on inducing chaos in particle motion have a rich history in prior works \cite{suzuki,kenta,i1,i2,i3,i4,i5,i6,i7,i8,i9,i10,Ran,DeFalco:2020yys,DeFalco:2021uak,dalui,yu,Wenfu,Andrea1,Addazi,hashimoto,hashi}, including chaotic string motion in Schwarzschild geometry in $AdS_5$ \cite{Leopoldo}. These studies encompass various scenarios, including spinning \cite{i7,i8,i9}, or magnetized \cite{i10} black hole systems, and involve test particles of varying mass, charge \cite{Ran}, or spin \cite{i4,i7,i8}. A recent analysis \cite{hashimoto} investigates the role of a Schwarzschild black hole on inducing chaos in particle trajectories close to the horizon.  Over the recent years, we are beginning to explore a variety of tools to test Einstein gravity in the strong field regime close to black hole event horizons \cite{telescope1,telescope2}. This motivates one to study the impact of black hole horizons on the onset and development of chaos in particle trajectories in modified theories of gravity. Such chaos probes may serve as an additional tool to probe Einstein gravity in the vicinity of the horizon. The focus of the current work is to systematically analyze the chaotic trajectories of particles, the associated Poincar$\Acute{e}$ sections and Lyapunov exponents as a quantification of chaos in two well motivated models of $f(R)$ gravity. We further compare our results with similar results obtained for the Schwarzschild black hole arising in Einstein 
    gravity.
    
    In the initial part of the paper, we construct static spherically symmetric (SSS) black hole geometries in two specific models (model-I and model-II) of $f(R)$ gravity. The first model considered by Saffari {\it{et al.}}, \cite{Saffa} is a black hole solution that successfully reproduces the flat rotation curve of spiral galaxies and also incorporates the late-time cosmic acceleration of the Universe. The second model utilizes both charged and neutral black hole solutions that asymptotically approach flat spacetime \cite{main1,main2}. Our focus is on understanding the behavior of massless particles close to the event horizon of the aforementioned black hole solutions in $f(R)$ gravity. The resulting nonlinear dynamical equations are constructed in these models, and the particle trajectories, associated Poincar$\Acute{e}$ sections and Lyapunov exponents are analyzed using standard numerical techniques. Our calculations reveal that the outgoing radial trajectories of a particle near the event horizon exhibit exponential growth over time. This observation suggests the possibility of chaotic behavior in the dynamics of particles, particularly when the particle is initially part of an integrable system, say under harmonic confinement. In both models, we probe chaos in the confined particle trajectories in the vicinity of the black hole horizon for a set of 200 initial conditions. The results clearly demonstrate the role played by the black hole horizon on the onset of chaos and the transition to fully chaotic behavior within a specific range of energies. This range is controlled by the parameters of the modified gravity theory, namely $\beta$ and $a$ in model-I and model-II, respectively, to be introduced in later sections. One of the novel outcomes of our study is presented in Fig.\ref{T1}, which clearly exhibits the impact of the black hole horizons on the onset and spread of chaos in modified gravity models considered in this work. In this figure, we track two initially close trajectories near the event horizon of model-I and model-II. We find that with increasing energy, the particle trajectories explore regions closer to the black hole horizon, with reduced overlap between the two trajectories, a signature of chaos which we later confirm by an explicit numerical computation of the Lyapunov exponents.  We further compare these observations with similar investigations in Schwarzschild geometry in Einstein gravity. Interestingly, for both models explored in this work, we find that the Lyapunov exponents obey the surface gravity bound arising from universal nature of particle dynamics close to black hole horizon, as shown in the works of Hashimoto and Tanahashi \cite{hashimoto}, in contrast to some of the other models of $f(R)$  and $f(T)$ gravity in the existing literature  \cite{Andrea1,Addazi} that violate such a bound. Let us note that the surface gravity bound arising in this  classical context is identical to the quantum bound on chaos by MSS (Maldacena, Shenker and Stanford) arising via the AdS/ CFT correspondence \cite{maldacena}. Moreover, our numerical investigations demonstrate that   for model-I, the total Lyapunov exponent increases with increasing energy and attains the largest value at the highest possible energies, for all values of $\beta$ allowed by current observations \cite{Saffa,Anderson}. Secondly, in the context of model-II, we find that the Lyapunov exponent decreases with an increase in modified gravity parameter ``a" for both the charged and neutral black hole solutions, consistent with increasing distance from the horizon.
    
    Remarkably, our findings suggest that particles emitted from the horizon, once liberated from its gravitational grasp, exhibit chaotic motion. Moreover, an early or delayed onset of chaos in particle dynamics as one approaches the horizon or deviations from the Lyapunov exponents predicted by Einstein gravity may act as an indirect probe of departure from Einstein gravity in the strong field regime.
	
    The paper is organized as follows: In Secs. \ref{s2a} and \ref{s2b}, we construct the black hole backgrounds in modified gravity model-I and model-II, respectively. In Sec. \ref{s2b-a}, we study the solution for the charged black hole and in Sec. \ref{s2b-b}, the neutral black hole solution has been derived. Within Sec. \ref{sec3}, we construct the dynamical equations of motion. Next, we numerically track two initially close trajectories of the massless probe particle near the black hole horizon in the modified gravity backgrounds, needed for the construction of Fig.\ref{T1}. In Sec. \ref{s4b}, we construct the various Poincar$\Acute{e}$ sections in  model-I and model-II, followed by a computation of the associated Lyapunov exponents in Sec. \ref{s4c}. Finally, we summarize our results in modified gravity model-I and model-II and compare them with similar results for the Schwarzschild black hole in Einstein gravity in Sec. \ref{sec5}.

	\section{Background geometries in $f(R)$ gravity}\label{sec2}
    For the sake of completeness, let us revisit the static spherically symmetric (SSS) black hole solutions in two separate models of $f(R)$ gravity. In the later sections, we will study the chaotic dynamics of massless particles moving in the background geometries constructed in this section. The detailed construction of the SSS solution in the first model of $f(R)$ gravity (model I) is carried out in Sec. \ref{s2a}, following Refs. \cite{Saffa,Saheb}. The second model, hereafter referred to as model II is presented in detail in Sec. \ref{s2b}, where we construct both charged and neutral black hole solutions following \cite{main1,main2,main3}.

     The generic action for $f(R)$ gravity is given by 
	\begin{equation}
		\mathcal{A}=\frac{1}{2\kappa}\int d^4x\sqrt{-g}~f(R)+\mathcal{A}_{matter},\label{1}
	\end{equation}
    where $g$ is the determinant of  metric $g_{\mu\nu}$ in a 4D spacetime and $\mathcal{A}_{matter}$ refers to the matter part of the action. Taking the variation of the generic $f(R)$ action Eq.\eqref{1} with respect to the metric, we get the standard field equations of $f(R)$ gravity (setting $\kappa=8\pi G=1$) as 
        \begin{equation}
            f_{R}(R) ~R_{\mu\nu}-\frac{1}{2}f(R)~g_{\mu\nu}-\left(\nabla_{\mu}\nabla_{\nu}-g_{\mu\nu}\Box\right)f_{R}(R)=T_{\mu\nu},\label{2}
        \end{equation}
	where $f_{R}(R)\equiv\frac{df(R)}{dR}$, $\Box\equiv \nabla_{\mu}\nabla^{\mu}$, and $T_{\mu\nu}$ is the energy-momentum tensor. Taking the trace of the field equation Eq.\eqref{2} gives
        \begin{equation}
            f(R)=\frac{1}{2}\left(3~\Box f_{R}+R~f_{R}- T\right)\label{3}.
        \end{equation}
    We can rewrite the field equation Eq.\eqref{2} in terms of $f_{R}$ as
        \begin{eqnarray}
        R_{\mu\nu}-\frac{1}{4}R~ g_{\mu\nu}=\frac{1}{f_{R}}\left(T_{\mu\nu}-\frac{1}{4}Tg_{\mu\nu}\right)\nonumber\\
        +\frac{1}{f_{R}}\left(\nabla_{\mu}\nabla_{\nu}f_{R}-\frac{1}{4}g_{\mu\nu}\Box f_{R}\right)\label{4}.
        \end{eqnarray}

        \subsection{Model I}\label{s2a} 
    Let us now consider the first model (model I) of $f(R)$ gravity by setting the matter contribution $T_{\mu\nu}=0$, whose action $\mathcal{A}_{I}$ is given by \cite{Saffa,Saheb}
        \begin{equation}
            \mathcal{A}_{I}=\frac{1}{2}\int d^4x\sqrt{-g}\left[R+6{\beta}^2~\ln(\frac{R}{R_c})\right],\label{5}
        \end{equation}
    where $R_c$ is an integration constant and $\beta$ is a real constant. The vacuum field equation for Eq.\eqref{5} can be written as
        \begin{eqnarray}
            R_{\mu\nu}-\frac{1}{2} g_{\mu\nu} R=\left(\nabla_{\mu}\nabla_{\nu}-g_{\mu\nu}\Box\right)\frac{6{\beta}^2}{R}-\frac{6{\beta}^2}{R}R_{\mu\nu}\nonumber\\
            +3{\beta}^2g_{\mu\nu}\ln{\frac{R}{R_c}}.\label{6}
        \end{eqnarray}
    Considering the metric of a static spherically symmetric 4D spacetime of the form,
        \begin{equation}
		dS^2=-B(r)dt^2+\frac{dr^2}{B(r)}+r^2d\Omega^2,\label{7}
	\end{equation}
    $d\Omega^2=d\theta^2+\sin^2{\theta}d{\phi}^2$ is the metric element of a unit $2D$ sphere. The metric solution of $B(r)$ up to  first order in $\beta$ (setting $G=1,~c=1$) is given by \cite{Saffa,Saheb}
        \begin{equation}
            B(r)=1-\frac{2M}{r}+\beta r,\label{8}
        \end{equation}
    where $M$ is a typical mass of the black hole.

    \subsubsection{Details of the derivation of the black hole solution in model I}\label{s2a-1}
\begin{figure}[H]
		\begin{center}
	    \includegraphics[width=1.0\linewidth]{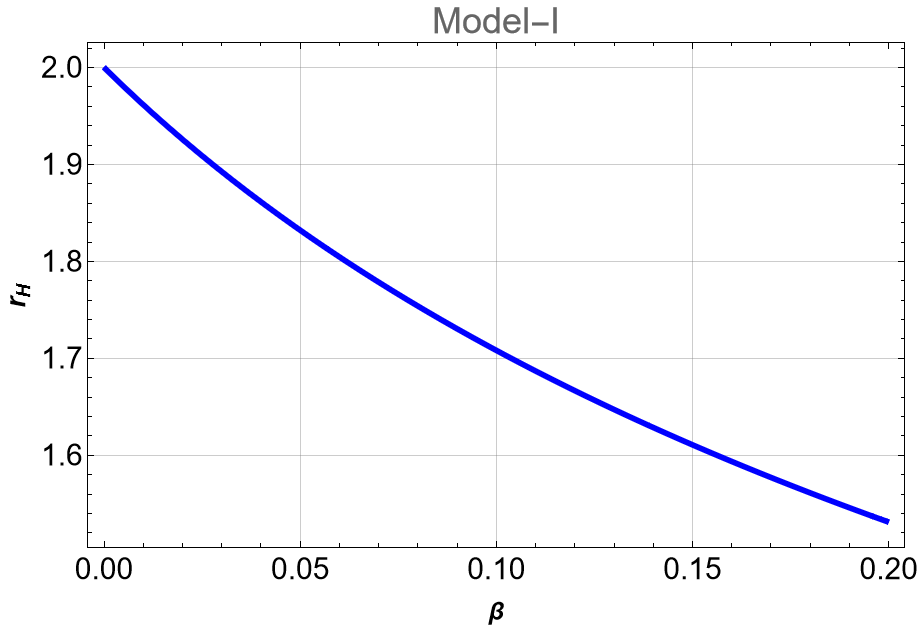}
		\end{center}
		\caption{Plot of horizon radius $r_{H}$ versus free parameter $\beta$ for $M=1$ in model I.}
		\label{f1}
	\end{figure}

    \noindent
    Let us now pause here and present a few technical details on how to arrive at the above solution for model I following the works of Refs. \cite{Saffa,Saheb,delacruz,mult,o1,o2,o3,o4,o5,o6}. Plugging the SSS metric ansatz  Eq.\eqref{7} into the field equations Eq.\eqref{4} (setting $T_{\mu\nu}=0$) , we have
        \begin{eqnarray}
            &&\frac{d^2B}{dr^2}-\frac{f^{\prime}_{R}}{f_{R}}\left(\frac{2}{r}B-\frac{dB}{dr}\right)+\frac{2}{r^2}\left(1-B\right)=0,\label{9}\\
            &&\text{and,}\nonumber\\
            && r\frac{d^2f_{R}}{dr^2}=0.\label{10}    
        \end{eqnarray}
    We can solve for $B(r)$ from equations Eq.\eqref{9} and Eq.\eqref{10} for a given model of $f(R)$ and its derivative with respect to $R$, denoted by $f_{R}(R)$. To arrive at model I, we consider the form of $f_{R}(R)$ as \cite{Saffa,Saheb}
        \begin{equation}
            f_{R}(R)=\Big(1+\frac{r}{d}\Big)^{-\psi},\label{11}
        \end{equation}
    where the dimensionless parameter $\psi\ll1$ and $d$ is a characteristic length scale of the order of galactic size. Restricting to length scales relevant for the solar system, i.e., $r\ll d$, one can expand the action in terms of the small parameters $\frac{r}{d}$ and $\frac{\psi r}{d}$. Thus, Eq.\eqref{11} can be written as, 
        \begin{equation}
            f_{R}(R)=1-\frac{\psi r}{d}.\label{12}
        \end{equation}
    Plugging this into the differential equation Eq.\eqref{9}, we obtain the solution of $B(r)$, i.e., Eq.\eqref{8} to first order in $\beta \equiv \frac{\psi}{d}$ in the metric. The $\beta \rightarrow 0$ limit of Eq.\eqref{8} is the Schwarzschild black hole solution as expected. In a typical spiral galaxy, the dimensionless parameter $\psi\approx 10^{-6}$ provides a flat rotation curve for stars and $d\sim10~kpc$ \cite{Saffa,Anderson,Saheb}.  
    Using the metric elements Eq.\eqref{8}, the corresponding  Ricci scalar is given by (up to the first order in $\beta$)
        \begin{equation}
            R(r)=-\frac{6\beta}{r}\label{13}.
        \end{equation}
   Finally substituting $r$ in terms of $R(r)$ in Eq.\eqref{12} and integrating, we obtain the $f(R)$ gravity action model I  as introduced in Eq.\eqref{5}, i.e., $f(R)=R+6{\beta}^2~\ln(\frac{R}{R_c})$.
    The horizon radius, employing $B(r)=0$ gives us
        \begin{equation}
            r_{H}=\frac{-1+\sqrt{(1+8M\beta)}}{2\beta}.\label{14}
        \end{equation}
    Expanding the horizon radius $r_H$ [Eq.\eqref{14}] in series, one can get
        \begin{equation}
            r_H=2M-4M^2\beta+\mathcal{O}(\beta^2).\label{14.a}
        \end{equation}
    For the small value of $\beta$ ($\beta\ll 1$), the horizon radius $r_{H}$ reduces to the Schwarzschild radius $2M$ at the leading order. The plot of horizon radius vs $\beta$ is presented in Fig.\ref{f1}. 

    As an additional comment, let us mention that the particular form of $f(R)$ considered in model I is useful in two regimes. The first regime is relevant at solar system length scales when $R\gg\Lambda$ ($\Lambda$ being the cosmological constant) and $\frac{R}{6\beta^2}\gg\frac{2}{\psi}$.  On the other hand, when $R\simeq 6\beta^2\simeq\Lambda$ and $\psi\ll 1$, the corresponding action reduces to $f(R)=R+\Lambda$, which resembles the late-time accelerating expansion of the Universe at the cosmological length scales \cite{Saffa}. However, in the present work, we will focus on the chaotic dynamics of massless particles using Eq.\eqref{8}, which is valid at solar system length scales.

	\subsection{Model II}\label{s2b}
    For model II, we consider the following $f(R)$ action  \cite{main1,main2,main3}:
        \begin{equation}
            \mathcal{A}_{II}=\frac{1}{2\kappa}\int d^4x\sqrt{-g}\left[R-2a\sqrt{R}\right]+\mathcal{A}_{matter},\label{15}
        \end{equation}
        where $a>0$ is a dimensional parameter. Following Eq.\eqref{2}, the field equation for the action Eq.\eqref{15} is given by
        \begin{eqnarray}
            R_{\mu\nu}-\frac{1}{2}Rg_{\mu\nu}=\frac{a}{\sqrt{R}}\left(R_{\mu\nu}-Rg_{\mu\nu}\right)+\kappa T_{\mu\nu}\nonumber\\
            +\left(\nabla_{\mu}\nabla_{\nu}-g_{\mu\nu}\Box\right)\frac{a}{\sqrt{R}}.\label{16}
        \end{eqnarray}
    We choose a static spherically symmetric metric ansatz have the form,
        \begin{equation}
		dS^2=-P(r)~dt^2+\frac{dr^2}{P(r)}+r^2d\Omega^2.\label{17}
	\end{equation}
    \hrulefill

        \subsubsection{Charged black hole solution}\label{s2b-a}
\begin{figure}[H]
		\begin{center}
			\includegraphics[width=1.0\linewidth]{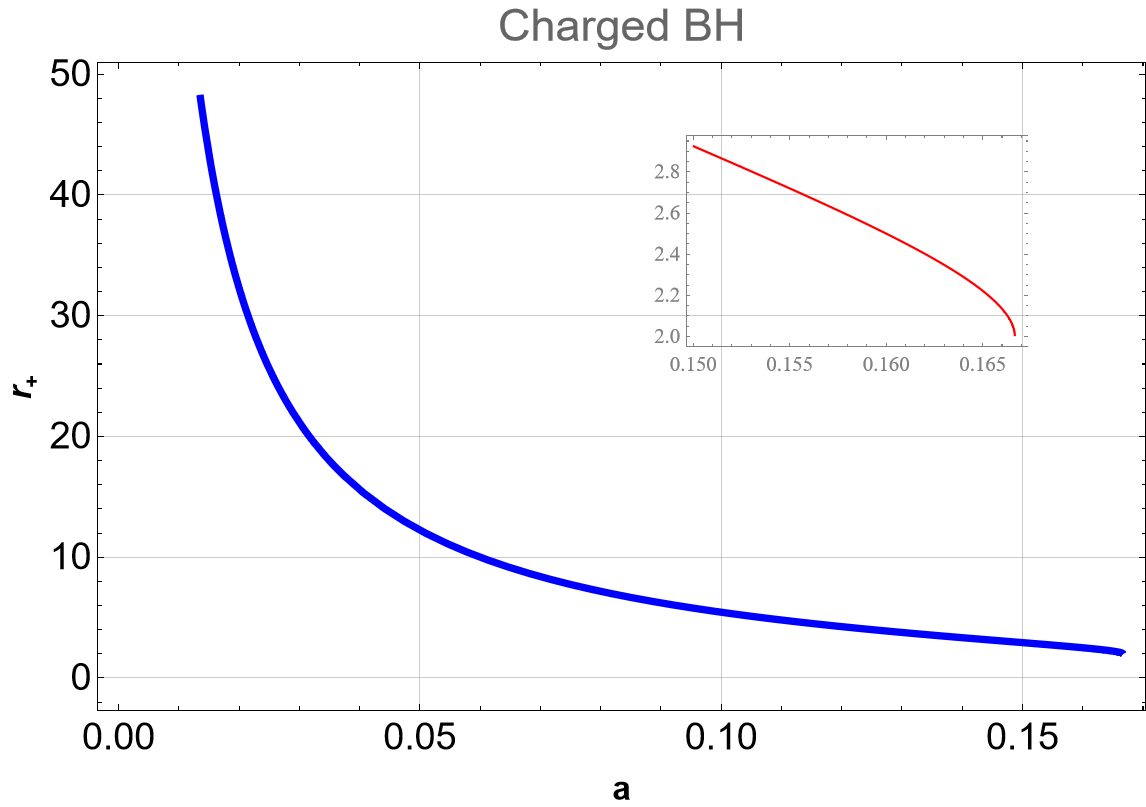}
		\end{center}
		\caption{Plot of the outer horizon radius $r_+$ with the modified gravity parameter $a$. The inset shows a zoomed version of the variation of the outer horizon radius for $a>0.15$.}
		\label{f2}
	\end{figure}

    To get the exact charged black hole solution, the matter action $\mathcal{A}_{matter}$ of Eq.\eqref{15} is chosen to be 
	\begin{eqnarray}
	   \mathcal{A}_{EM}=-\frac{1}{2}\int d^4x\sqrt{-g}F_{\mu\nu}F^{\mu\nu}\label{18},
	\end{eqnarray}
	where $F_{\mu\nu}$ is the electromagnetic field tensor, defined as $F_{\mu\nu}=\partial_{\nu}A_{\mu}-\partial_{\mu}A_{\nu}$ with $A_{\mu}$ is the one-form gauge potential. By varying the corresponding action with respect to the metric tensor $g_{\mu\nu}$, we obtain the field equations (with $\kappa=8\pi G=1$) as
	\begin{eqnarray}
	       I_{\mu\nu}&&\equiv R_{\mu\nu}-\frac{1}{2}R~g_{\mu\nu}-2\Big(g_{\sigma\epsilon}F^{\sigma}_{\nu}F^{\epsilon}_{\mu}-\frac{1}{4}g_{\mu\nu}F_{\sigma\epsilon}F^{\sigma\epsilon}\Big)\nonumber\\
        &&-\left(\nabla_{\mu}\nabla_{\nu}-g_{\mu\nu}\Box\right)\frac{a}{\sqrt{R}}-\frac{a}{\sqrt{R}}\left(R_{\mu\nu}-Rg_{\mu\nu}\right)=0,\nonumber\\
       \label{19}\\
       &&~~~~~~\text{with,}\nonumber\\
	&&\frac{1}{\sqrt{-g}}\partial_{\nu}(\sqrt{-g}F^{\mu\nu})=0.\label{20}
	\end{eqnarray}
    Now, the exact charged black hole solution for model II (using a timelike gauge field) is given by \cite{main2,main3}
        \begin{eqnarray}
            &&P(r)=\Big(\frac{1}{2}-\frac{1}{3ar}+\frac{1}{3ar^2}\Big),\label{21}\\
            &&\text{and,}\nonumber\\
            &&A=\frac{1}{\sqrt{3a}r}dt,\label{21a}
        \end{eqnarray}
    where $A$ is the gauge field. We will use the above charged black hole background Eq.\eqref{21} while studying the chaotic dynamics in later sections. In summary, the exact charged black hole solution for Eq.\eqref{17} is given by
        \begin{eqnarray}
           &&dS^2=-\Big(\frac{1}{2}-\frac{1}{3ar}+\frac{1}{3ar^2}\Big)dt^2\nonumber\\
    &&~~~~~~~~~~~~~+{\Big(\frac{1}{2}-\frac{1}{3ar}+\frac{1}{3ar^2}\Big)}^{-1}dr^2+r^2 d\Omega^2.\nonumber\\
 \label{22} 
        \end{eqnarray}
    The above metric solution is asymptotically flat. The horizon of the charged black hole solution is found by solving the equation [Eq.\eqref{21}] $P(r)=0$, 
	\begin{eqnarray}
		r_{\pm}=\frac{1\pm\sqrt{1-6a}}{3a},\label{23}
	\end{eqnarray}
 where $\pm$ denotes the outer/inner horizon, respectively. The allowed values of $a$ lie in the range $0<a\leq \frac{1}{6}$. The plot of the outer horizon radius $r_{+}$ versus $a$ is shown in Fig.\ref{f2}. In this paper, we will focus on the chaotic dynamics in the region $r>r_{+}$.

        \subsubsection{Neutral black hole solution}\label{s2b-b}
    One can also find a neutral black hole solution in model II, i.e., Eq.\eqref{15}, by taking the matter action $\mathcal{A}_{matter}=0$ (with $\kappa=8\pi G=1$). The relevant field equations are
        \begin{eqnarray}
             R_{\mu\nu}-\frac{1}{2}Rg_{\mu\nu}=\frac{a}{\sqrt{R}}\left(R_{\mu\nu}-Rg_{\mu\nu}\right)\nonumber\\
            +\left(\nabla_{\mu}\nabla_{\nu}-g_{\mu\nu}\Box\right)\frac{a}{\sqrt{R}}\label{24}
        \end{eqnarray}
    The exact solution for the neutral black hole adopting the SSS ansatz Eq.\eqref{17} is given by,
	\begin{eqnarray}
		P(r)=\frac{1}{2}-\frac{1}{3ar}.\label{25}
	\end{eqnarray}
    In summary, the neutral black hole solution for Eq.\eqref{17} is given by
        \begin{eqnarray}
           &&dS^2=-\Big(\frac{1}{2}-\frac{1}{3a}\Big)dt^2+{\Big(\frac{1}{2}-\frac{1}{3ar}\Big)}^{-1}dr^2+r^2 d\Omega^2.\nonumber\\
            \label{25.1} 
        \end{eqnarray}
    
\begin{figure}[H]
		\begin{center}
			\includegraphics[width=1.0\linewidth]{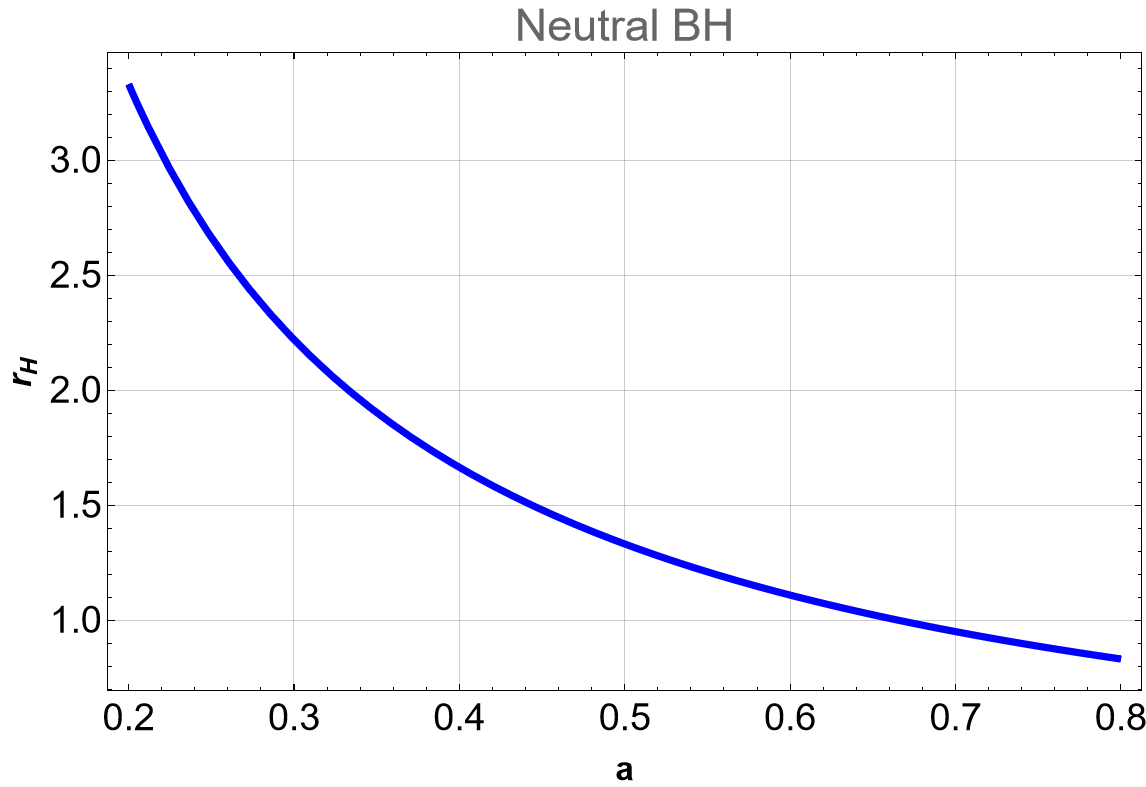}
		\end{center}
		\caption{Plot of the horizon radius $r_H$ with the modified gravity parameter $a$ for the neutral black hole.}
		\label{f3}
	\end{figure}

    By solving $P(r)\equiv\Big(\frac{1}{2}-\frac{1}{3ar}\Big)=0$, we get the neutral black hole horizon radius $r_H=\frac{2}{3a}$. Hence, we need $a>0$. Moreover, the black hole horizon radius is thus inversely proportional to a. The plot of horizon radius $r_H$ vs $a$ is presented in Fig.\ref{f3}.

    \hrulefill

        \subsubsection{Details of the derivation of charged and neutral black hole solutions in model II}\label{s2b-1}
        \noindent
    Let us provide some technical details of the derivation of the above black hole solutions in model II, closely following \cite{main1} and Nashed {\it{et al.}}, \cite{main2,main3}, along with the references therein. One starts with a Lagrangian derivation of equations of motion for an SSS metric. A general SSS metric in $4D$ spacetime can be written as
        \begin{equation}
            dS^2=-P(r)e^{2\a(r)}dt^2+\frac{dr^2}{P(r)}+r^2d\Omega^2.\label{26}
        \end{equation}
    Here, $\a(r),~P(r)$ are unknown functions of $r$, which we wish to determine. In terms of Lagrange multipliers $\lambda$ and in absence of matter, the action Eq.\eqref{1} can be written as follows \cite{main1,safko}:
	\begin{eqnarray}
		&&\mathcal{A}\equiv\frac{1}{2}\int dt\int e^{\a(r)}r^2 dr \Bigg[f(R)-\lambda\Bigg(R+\Big(P''(r)\nonumber\\
  &&+~3P'(r)\a'(r)+2P(r){\a^\prime}^2(r)+2P(r)\a''(r)+4\frac{P'(r)}{r}\nonumber\\
  &&+~4\frac{P(r)\a'(r)}{r}+2\frac{P(r)}{r^2}-\frac{2}{r^2}\Big)\Bigg)\Bigg],\label{27}
	\end{eqnarray}
    where we have used the value of the curvature scalar using Eq.\eqref{26}, given by
	\begin{eqnarray}
	R=-3P'(r)\a'(r)-2P(r){\a^\prime}^2(r)-P''(r)+\frac{2}{r^2}\nonumber\\
        -~2P(r)\a''(r)-4\frac{P'(r)}{r}-4\frac{P(r)\a'(r)}{r}-2\frac{P(r)}{r^2}.\nonumber\\
        \label{28}
	\end{eqnarray}
    The prime denotes the derivative with respect to radial coordinate $r$. Taking the variation of the given action Eq.\eqref{27} with respect to $R$, one gets
	\begin{eqnarray}
		\frac{df(R)}{dR}=\lambda\label{29}.
	\end{eqnarray}
    Finally, by making an integration by parts of Eq.\eqref{27} and using Eq.\eqref{29}, one can obtain the Lagrangian of the system in this context, which takes the following form:
    \begin{widetext}
	\begin{eqnarray}
		\mathcal{L}=e^{\a(r)}\Bigg[r^2\Big(f(R)-Rf_R(R)\Big)+2f_R(R)\Big(1-P(r)-rP'(r)\Big)
		+f_{RR}~(R)r^2R'(r)\Big(P'(r)+2P(r)\a'(r)\Big)\Bigg],\label{30}
	\end{eqnarray}
    \end{widetext}
    where $f_R(R),~f_{RR}(R)$ denote the first and second derivatives of $f$ with respect to Ricci scalar $R$, respectively. We can get three equations of motion for the Lagrangian by varying with respect to three dynamical variables $\a(r),R(r)$, and $P(r)$, respectively.\\
    Now, for the first and third equations of motion, we get the following:
	\begin{eqnarray}
		&&\frac{d^2R}{dr^2}+\Big[\frac{2}{r}+\frac{P'(r)}{2P(r)}\Big]\frac{dR}{dr}+\frac{f_{RRR}}{f_{RR}}{\Big(\frac{dR}{dr}\Big)}^2+\frac{Rf_R-f(R)}{2P(r)f_{RR}}\nonumber\\
		&&~~~~~~~~~-\frac{f_R}{r^2 P(r)f_{RR}}\Big[1-P(r)-rP'(r)\Big]=0~\label{31}\\
		&&\text{and,}\nonumber
        \end{eqnarray}
        \begin{eqnarray}
		&&{f_{RR}}\frac{d^2 R}{dr^2}-\a'(r)\Big(\frac{2f_R}{r}+f_{RR}{\frac{dR}{dr}}\Big)\nonumber\\
		&&~~~~~~~~~~~~~~~~~~~~~~~~~~+f_{RRR}{\Big(\frac{dR}{dr}\Big)}^2=0.\label{32}
	\end{eqnarray}
    For constant $\alpha$, the Ricci scalar $R$ in Eq.\eqref{28} has to be of the form,
	\begin{equation}
		\frac{d^2P(r)}{dr^2}+\frac{4}{r}\frac{dP(r)}{dr}+2\frac{P(r)}{r^2}-\frac{2}{r^2}=-R,\label{34}
	\end{equation}
    and we solve Eq.\eqref{32} for a constant $\a$, leading to
	\begin{equation}
		f_R(R)=\gamma r+\delta,\label{33}
	\end{equation}
    where $\gamma$ and $\delta$ are the arbitrary integration constants.\\
    Following \cite{main1}, from Eq.\eqref{31}, we can now solve for the metric function $P(r)$  using Eq.\eqref{34}, Eq.\eqref{33}, and the identity  $\frac{df_R}{dr}=\gamma\equiv f_{RR}\frac{dR}{dr}$, 
	\begin{eqnarray}
		\Big(\gamma+\frac{\delta}{r}\Big)\frac{d^2P(r)}{dr^2}+\frac{\gamma}{r}\frac{dP(r)}{dr}-\frac{2\delta}{r^3}\Big(P(r)-1\Big)\nonumber\\
        -\frac{2\gamma}{r^2}\Big(2P(r)-1\Big)=0.\label{35}
	\end{eqnarray}
    Since $\delta$ is dimensionless, choosing $\delta=1$, the generic solution of Eq.\eqref{35} have the following form:
	\begin{eqnarray}
		P(r)&&=\frac{1}{2}\Big[2+r\big(2rC_1+3\gamma^2 r-2\gamma\big)+\gamma C_2\big(1-2\gamma r\big)\Big]\nonumber\\
		&&-\gamma^2 r^2\Big[\ln r-\ln(1+\gamma r)\Big]\big(1+\gamma C_2\big)-\frac{C_2}{3r}~,\label{36}
	\end{eqnarray}
    where $C_1$ and $C_2$ are two arbitrary integration constants. 
    Choosing $C_1=-\frac{3\gamma^2}{2}$ and $C_2=-\frac{1}{\gamma}$, we have the general solution of $P(r)$ from Eq.\eqref{36} as
        \begin{equation}
           P(r)=\frac{1}{2}+\frac{1}{3\gamma r}.\label{36.1} 
        \end{equation} 
    The horizon radius of $P(r)$ is thus given by $r_H=-\frac{2}{3\gamma}$. Positivity of $r_H$ implies $\gamma \equiv -a<0$ with $a>0$.
    
    Now to find the Ricci scalar as a function of $r$, applying Eq.\eqref{36} in Eq.\eqref{34} by using the same chosen constants $C_1$ and $C_2$ stated above, we find that the Ricci scalar satisfies the inverse square law, i.e., $R=\frac{1}{r^2}$. Therefore substituting the value of $r=\frac{1}{\sqrt{R}}$ in Eq.\eqref{33} finally leads us to the action of model II \cite{main1,main2,main3},
	\begin{equation}
		f(R)=R-2a\sqrt{R},\label{37}
	\end{equation}
    which is mentioned in Eq.\eqref{15}.

        \begin{itemize}
        \item[(i)]{Charged black hole solution:}
        \end{itemize}
    
    Following \cite{main2,main3}, we now sketch the steps needed to arrive at the charged black hole solution. Let us note that the conformal invariance of the Maxwell field requires the vanishing of the trace of the Maxwell field energy-momentum tensor $T$, as follows:
        \begin{equation}
            3\Big(a\sqrt{R}-\Box\frac{a}{\sqrt{R}}\Big)-R=0.\label{38}
        \end{equation}
    Therefore, by assuming the SSS background spacetime as given in Eq.\eqref{17} and using the field equations Eq.\eqref{19}, Eq.\eqref{20}, and Eq.\eqref{38}, by solving the system of equations $(I^t_t-I^r_r)$ and $I^{\theta}_{\theta}$ and also using the expression for the Ricci scalar Eq.\eqref{34}, we get the exact charged black hole solution, which is mentioned in Eq.\eqref{21}.

        \begin{itemize}
        \item[(ii)]{Neutral black hole solution:}
        \end{itemize}
    
    The neutral black hole solution is already derived in Eq.\eqref{36.1} with $\gamma =-a$. Therefore, we obtain Eq.\eqref{25}. Let us note that the additional term $\frac{1}{3ar^2}$ in the charged black hole case [Eq.\eqref{21}] vanishes when one switches off the gauge field, leading to the neutral black hole solution [Eq.\eqref{25}] as expected.

        \section{Dynamics of a massless particle in SSS black hole spacetime}\label{sec3}
    In this section, we summarize the dynamical equations of motion of a massless particle in a static spherically symmetric (SSS) black hole spacetime. For this purpose, we consider the usual SSS metric as
	\begin{equation}
		ds^2=-A(r)dt^2+\frac{dr^2}{A(r)}+r^2 d\Omega^2.\label{3.1}
	\end{equation}
    Due to the presence of horizon singularities, a better choice of coordinates, namely the Painlev$\Acute{e}$-Gullstrand coordinates, is often useful \cite{Pain},
	\begin{equation}
		dt\longrightarrow dt-\frac{\sqrt{1-A(r)}}{A(r)} dr.\label{3.2}
	\end{equation}
	Using Eq.\eqref{3.2}, the metric Eq.\eqref{3.1} takes the following form as
	\begin{equation}
		ds^2=-A(r)dt^2+2\sqrt{1-A(r)}dt dr+dr^2+r^2 d\Omega^2.\label{3.3}
	\end{equation}
    The metric admits a timelike Killing vector $\zeta^{\alpha}=(1,0,0,0)$, so that the conserved energy is given by $E=-\zeta^{\alpha}p_{\alpha}$, where $p_{\alpha}$ is the particle's four momentum. Next, to find the energy of the particle in terms of other components of four-momentum, we use the covariant form of the dispersion relation given by
	\begin{equation}
		g^{\alpha\beta}p_{\alpha}p_{\beta}=-m^2 c^2,\label{3.4}
	\end{equation}
	with $m$, being the mass of the particle.
	
    Now, using the dispersion relation Eq.\eqref{3.4} (with $c=1$) for the metric Eq.\eqref{3.3}, we can get the energy of a massless particle as follows:
	\begin{equation}
		E=-\sqrt{1-A(r)}p_r\pm\sqrt{p^2_{r}+\frac{p^2_{\theta}}{r^2}}.\label{3.5}
	\end{equation}
    Here, we have assumed that the particle is moving only along the radial $r$ and the angular $\theta$ directions; i.e., the motion of the particle is in the poloidal plane with $p_{\phi}=0$. The negative sign denotes the energy for the ingoing particle, while the positive sign is for the outgoing particle. In this paper, we study the dynamics of the outgoing particle only. Now a general question arises: \textit{What is the generic behavior of a radially moving particle in the vicinity of the event horizon?} This can be understood from the energy expression Eq.\eqref{3.5} with the choice  of $p_{\theta}=0$, which is given by
	\begin{equation}
		\dot{r}=\frac{\partial E}{\partial p_r}=1-\sqrt{1-A(r)}\simeq\kappa(r-r_H),\label{3.6}
	\end{equation}
    where the derivative is taken with respect to an affine parameter, and we have considered only the first-order expansion of $A(r)$ near the horizon as $A(r)\simeq 2\kappa(r-r_H)$. The term $\kappa=\frac{A'(r_H)}{2}$ is the surface gravity of the black hole. Now the solution of the radial geodesic Eq.\eqref{3.6} is given by

	\begin{equation}
		r=r_H+\Bar{c}~ r_H e^{\kappa\tau},\label{3.7}
	\end{equation}
    where $\Bar{c}$ is an integration constant and $\tau$ is an affine parameter. On the other hand, considering the leading order equation of the radial momentum in the near-horizon region,
	\begin{equation}
		\dot{p_r}\simeq-\kappa p_r,\label{3.8}
	\end{equation}
    which leads to the solution of $p_r$ as $p_r=p_{r_0}e^{-\kappa\tau}$ ($p_{r_0}$ is an arbitrary integration constant). Hence it is clear from the solutions of the equations Eq.\eqref{3.7} and Eq.\eqref{3.8} that depending on the sign of the affine parameter $\tau$, either $r$ or $p_r$ shows the exponential growth with an increase of $|\tau|$. In various scenarios, it has been demonstrated that particle trajectories can become unstable in the vicinity of black hole horizons \cite{hashimoto,hashi}. Studies have revealed that when examining a model involving either a static spherically symmetric black hole \cite{sd1,sd2} or a rotating Kerr black hole \cite{sd3}, an outgoing massless and uncharged particle can exhibit instability. This instability manifests through a Hamiltonian of the form $H\sim xp$ (as far as radial motion is concerned), indicating an unstable system. Furthermore, in the context of quantum BH, it has been observed that this instability correlates with the emergence of thermality within the system. This observation suggests a close connection between the instability and thermality of the black hole horizon. Therefore, under the influence of the horizon, the exponential growth of the radial motion can be interpreted as the presence of chaos in an integrable system \cite{hashimoto,hashi,dalui,bera}. 
 
    On the other hand, the value of the maximum Lyapunov exponent in this context is defined as \cite{Sandri,Strog}
	\begin{equation}
		\lambda_{L,max}=\lim_{\tau\to\infty}\frac{1}{\tau}\ln\Big(\frac{\delta r(\tau)}{\delta r(0)}\Big).\label{3.9}
	\end{equation} 
    The term $\delta r(\tau)$ in the numerator represents the separation between two initially close trajectories at time $\tau$, and $\delta r(0)$ in the denominator represents the separation at the initial time.  Since we are considering null geodesics, Eq.\eqref{3.5} is used to implicitly define the affine parameter $\tau$. In this situation, the Lyapunov coefficient $\lambda_{L}$ is bounded as \cite{hashimoto}
	\begin{equation}
		\lambda_{L}\leq \kappa,\label{3.10}
	\end{equation}
	where $\kappa$ represents the surface gravity of the black hole. A violation of this bound has been reported in some works, for a charged probe in Kerr-Newman-AdS black hole \cite{Gwak} and for a charged massive particle around a charged black hole, balanced by the Lorentz force \cite{Zhao}. On the other hand, this bound can also be violated in the case of Einstein's theory of gravity \cite{Lei,Kan}, in the case of $f(R)$ or $f(T)$ teleparallel theories of gravity \cite{Andrea1,Addazi}, or around a black hole that coexist with anisotropic matter fields \cite{Soy}.
	
    Let us now construct the dynamical equations of motion of a probe massless particle in the background geometries arising in the modified gravity models considered in the previous sections.  In this paper, our motivation is to study the effect of the black hole's event horizon in modified gravity on the particle's trajectories. Further, in order to ensure that the particle trajectories do not cross the horizon, we subject the probe particle to harmonic confinement along radial  $r$ and angular $\theta$ directions with the oscillator strength $K_r$ and $K_{\theta}$ respectively. By adjusting these strengths, we can confine the particle trajectories in any finite region. The inclusion of harmonic oscillators in the radial and angular directions is a choice  of the external potential, rather than a result of some underlying physics of the black hole background. Far from the horizon, the system is thus integrable, with periodic orbits and the appearance of unbroken tori. However, as one approaches the horizon, the near-horizon gravity makes the system highly nonlinear, leading to the breakdown of the regular KAM  tori and subsequent chaos. It is this onset and development of chaos that we wish to investigate in this work.
    
    We thus consider the situation where a probe massless particle is subjected to two harmonic potentials   $\frac{1}{2}K_r(r-r_c)^2$ and $\frac{1}{2}K_\theta(y-y_c)^2$ along $r$ and $\theta$ directions, respectively. The terms $K_r$ and $K_\theta$ represent the oscillator strengths along $r$ and $\theta$ directions, respectively (we also introduce a new variable $y=r_{H}\theta$ while writing the dynamical equations). Here, $r_c$ and $y_c$ are the equilibrium positions of these two harmonic potentials. Such a model was suggested in references \big(Ref. \cite{hashimoto} for massive particle and Refs. \cite{dalui,bera} for massless particle\big). It is worth noting that replacing the harmonic potential with a different potential could alter the dynamics of the particle's trajectory. However, the motion along the radial direction remains unaffected for massless particles, as discussed in the literature \cite{dalui}.
	
    Now the total energy of the probe particle under the influence of harmonic potentials for the metric Eq.\eqref{3.3} is given by
	\begin{eqnarray}
		&&E=-\sqrt{1-A(r)}p_r+\sqrt{p^2_{r}+\frac{p^2_{\theta}}{r^2}}+\frac{1}{2}K_r(r-r_c)^2\nonumber\\
		&&~~~~~~~~~~~~+\frac{1}{2}K_\theta~r^2_{H}(\theta-\theta_c)^2.\label{3.11}
	\end{eqnarray}
    Let us note that the radial momenta $p_{r}$ and cross radial momenta $p_{\theta}$, which are appearing in the above energy expression are  the usual canonical momenta that can be derived from the Lagrangian of a massless particle in the Painlev$\Acute{e}$-Gullstrand coordinate system.\footnote{The choice of Painlev$\Acute{e}$-Gullstrand coordinates is primarily motivated by the fact that they are non-singular at the horizon and hence leads to a better numerical implementation of particle dynamics close to the horizon \cite{pg1, pg2}. Moreover, the Painlev$\Acute{e}$-Gullstrand time coordinate corresponds to slices of spacetime that are regular at horizon. These slices avoid the problematic ``freezing" of time at the horizon, which occurs in Schwarzschild time due to the gravitational time dilation. In Appendix-A, we have also carried out the same analysis in Schwarzschild coordinates for Model-I, results remaining unchanged.}It is also important to note that the harmonic oscillators in the radial ($r$) and angular directions ($\theta$) are added by hand in order to confine the particle so that it does not fall into the horizon. Our approach closely follows the setup described in Sec. IV ``Numerical Analyses" and the associated Eq.(38) of Hashimoto {\it{et al}}. \cite{hashimoto}. Moreover, the harmonic oscillator being an integrable system ensures that the chaotic features observed in the model receives contributions from the black hole background.\\
    Correspondingly, the dynamical equations of motion have the following form:
	\begin{eqnarray}
		\dot{r}&=&\frac{\partial E}{\partial p_r}=-\sqrt{1-A(r)}+\frac{p_r}{\sqrt{p^2_r+\frac{p^2_{\theta}}{r^2}}},\label{3.12}
		\\
		\dot{p_r}&=&-\frac{\partial E}{\partial r}=-\dfrac{A'(r)}{2\sqrt{1-A(r)}}p_r+\dfrac{p^2_{\theta}/r^3}{\sqrt{p^2_r+\frac{p^2_{\theta}}{r^2}}}\nonumber\\
		&&-K_r(r-r_c),\label{3.13}
		\\
		\dot{\theta}&=&\frac{\partial E}{\partial p_{\theta}}=\dfrac{p_{\theta}/r^2}{\sqrt{p^2_r+\frac{p^2_{\theta}}{r^2}}},\label{3.14}
		\\
		\dot{p_{\theta}}&=&-\frac{\partial E}{\partial\theta}=-K_{\theta}~r^2_H(\theta-\theta_c).\label{3.15}
	\end{eqnarray}
    These are the main equations for the numerical studies to be performed in the later sections. Let us add that here, we have neglected the interaction between the harmonic potentials and the black hole spacetime, assuming such interactions to be weak.

        \section{Numerical analysis}\label{sec4}
    In this section, we will examine the role played by the black hole's event horizon in modified gravity theories towards the onset of chaos (i.e., the first appearance of broken tori in the associated Poincar$\Acute{e}$ section) for outgoing massless particles. For this purpose, we will consider the black hole solutions in the modified gravity model I and model II constructed in the previous sections and then analyze the phase space of a probe particle moving in these backgrounds. This essentially leads us to an investigation of the nonlinear dynamics dictated by Eqs.\eqref{3.12},\eqref{3.13},\eqref{3.14}, and Eq.\eqref{3.15} and the associated Poincar$\Acute{e}$ sections and Lyapunov exponents of the system.

        \subsection{Orbits of the probe particle}\label{s4a}
\begin{figure*}
    \centering
    \begin{tabular}{|p{0.22\linewidth}|c|c|c|}
        \hline
        Model-I for $\beta=10^{-2}.$ &
        \includegraphics[width=0.25\linewidth]{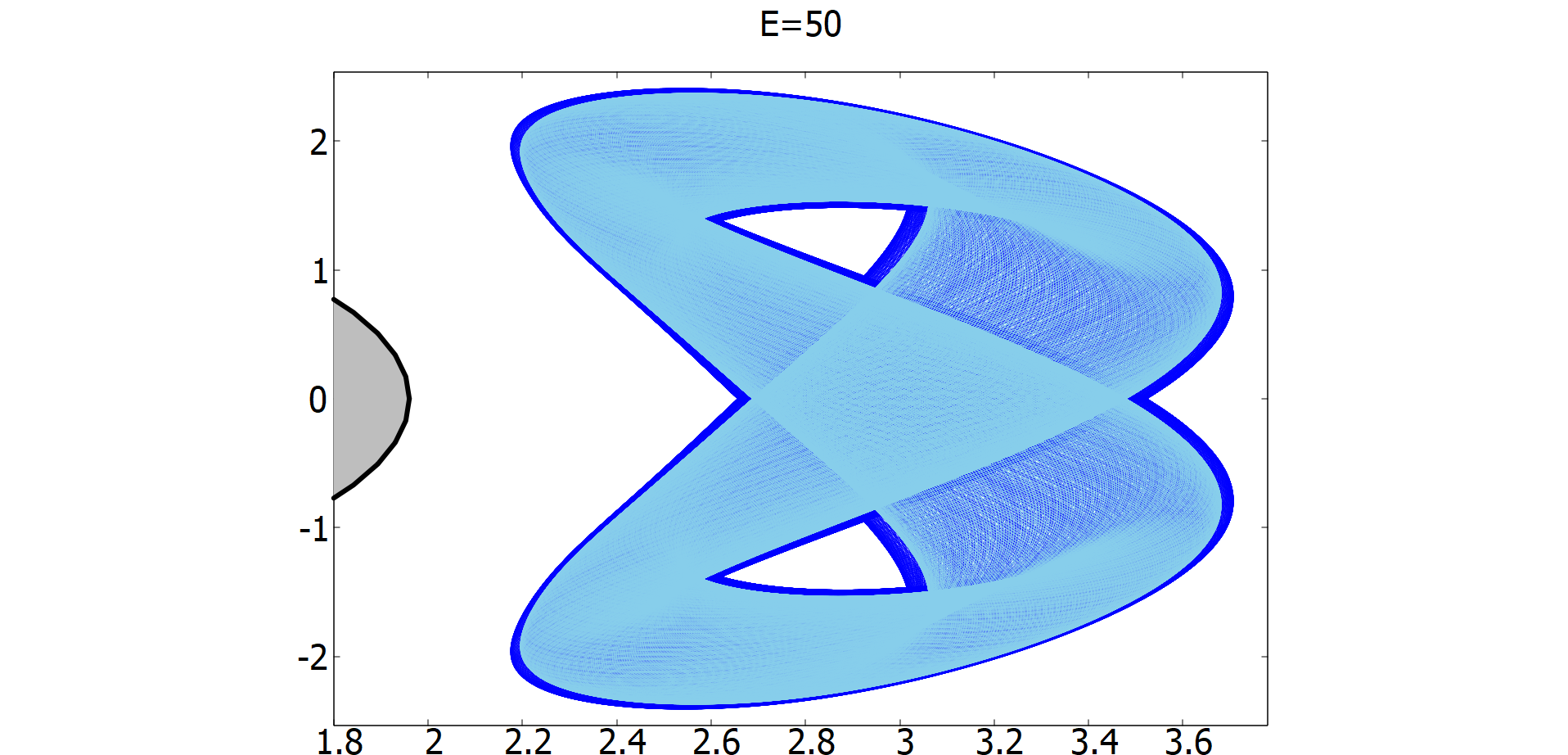} &
        \includegraphics[width=0.25\linewidth]{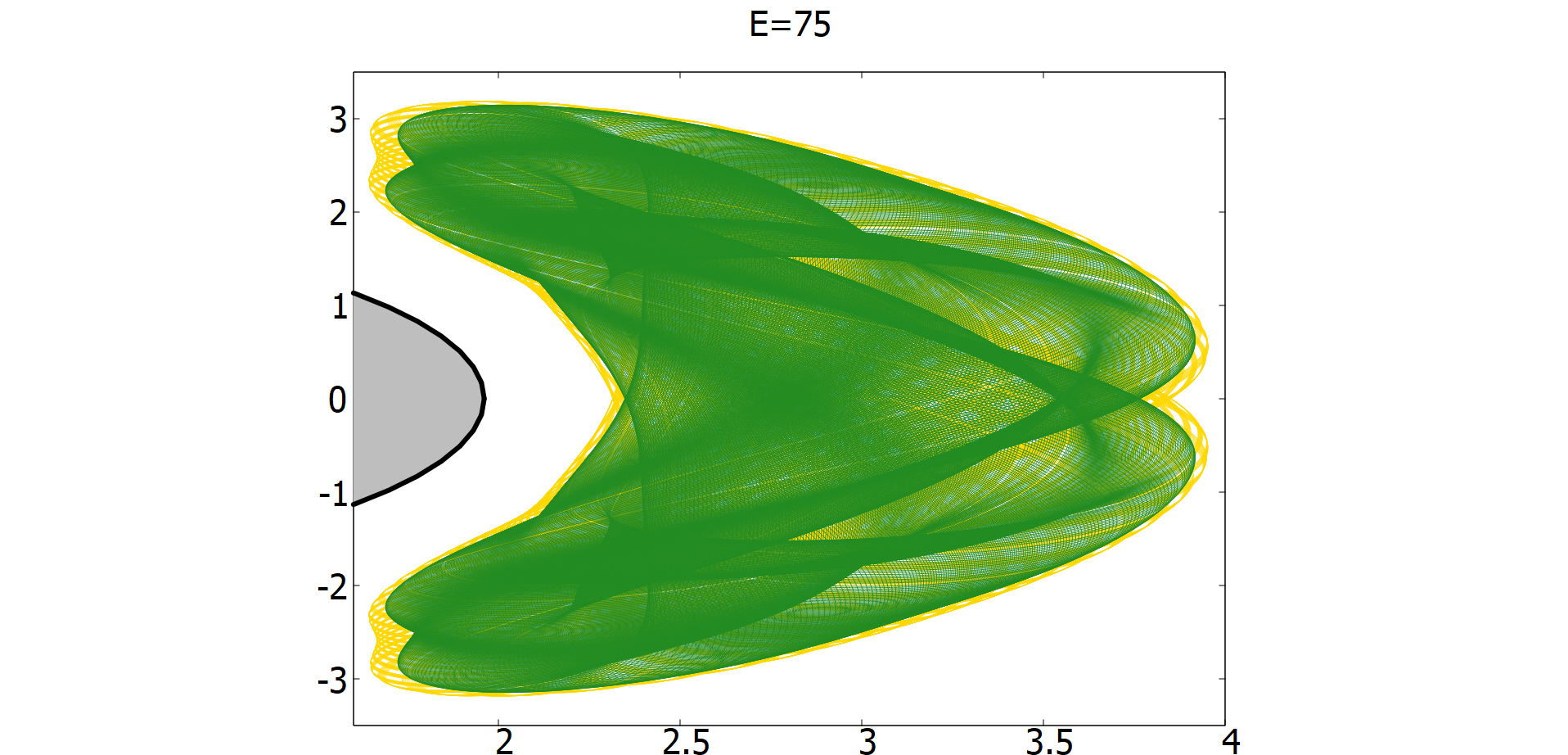} & \includegraphics[width=0.25\linewidth]{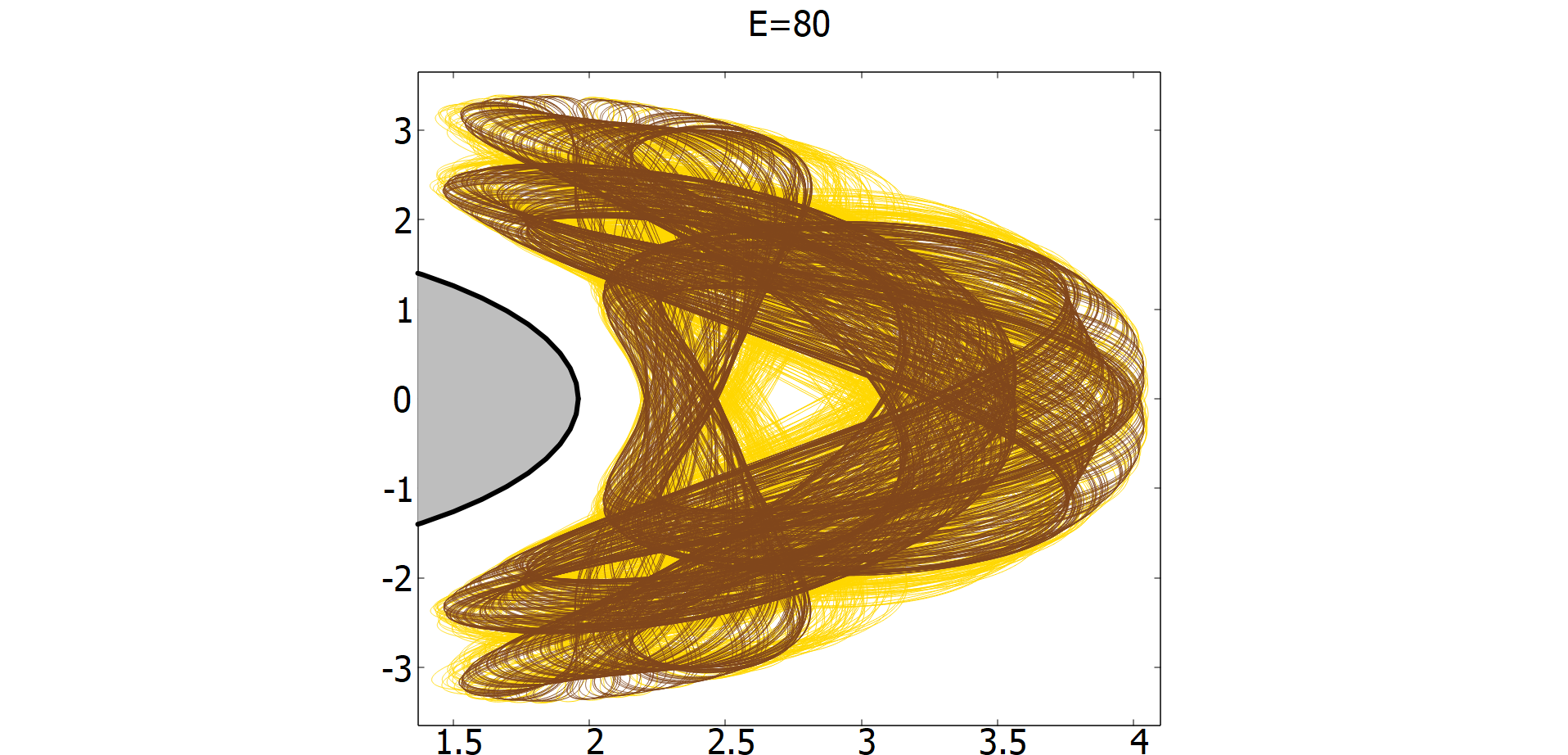} \\ 
        \hline
        Model-II Charged BH with fixed $a=0.166.$ & \includegraphics[width=0.25\linewidth]{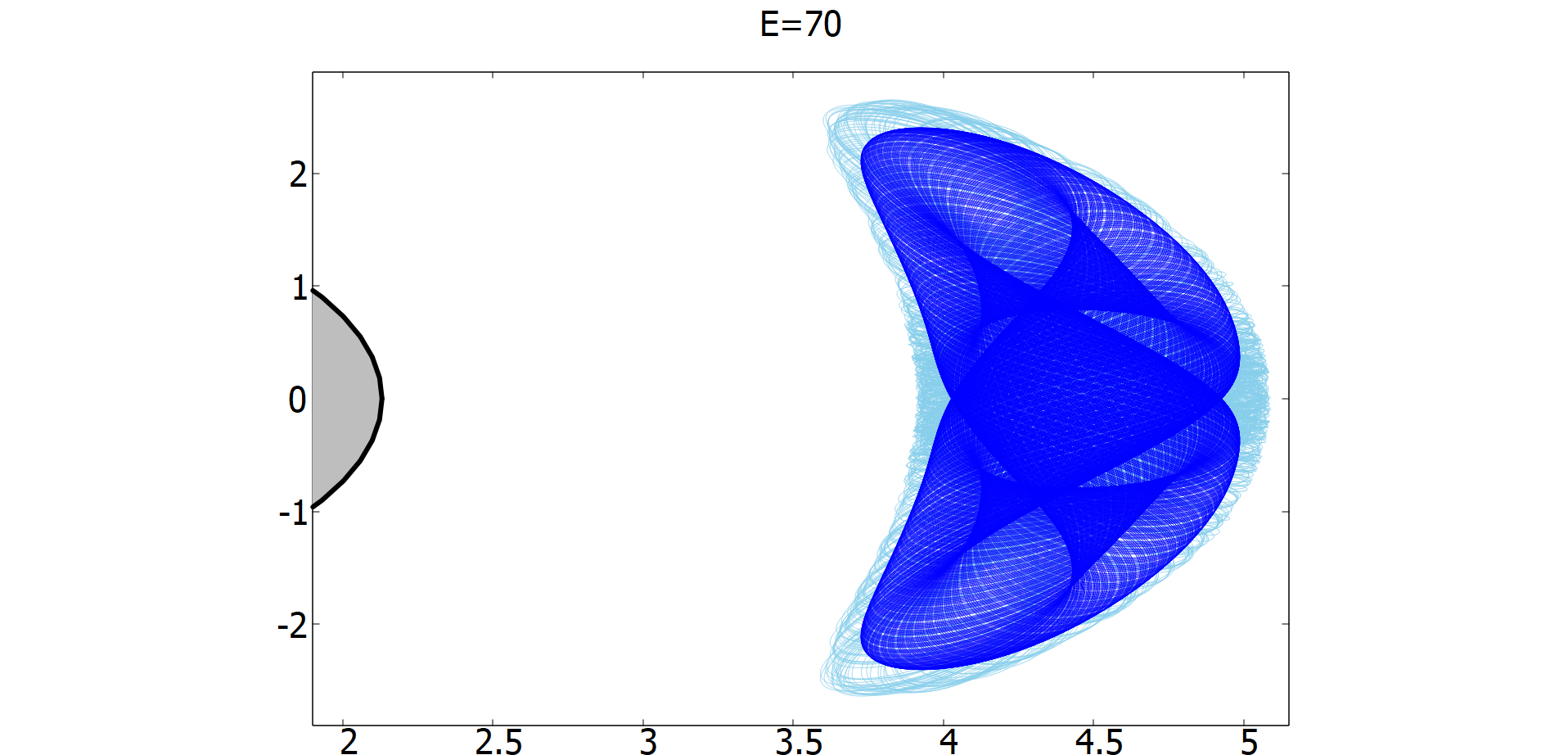} & \includegraphics[width=0.25\linewidth]{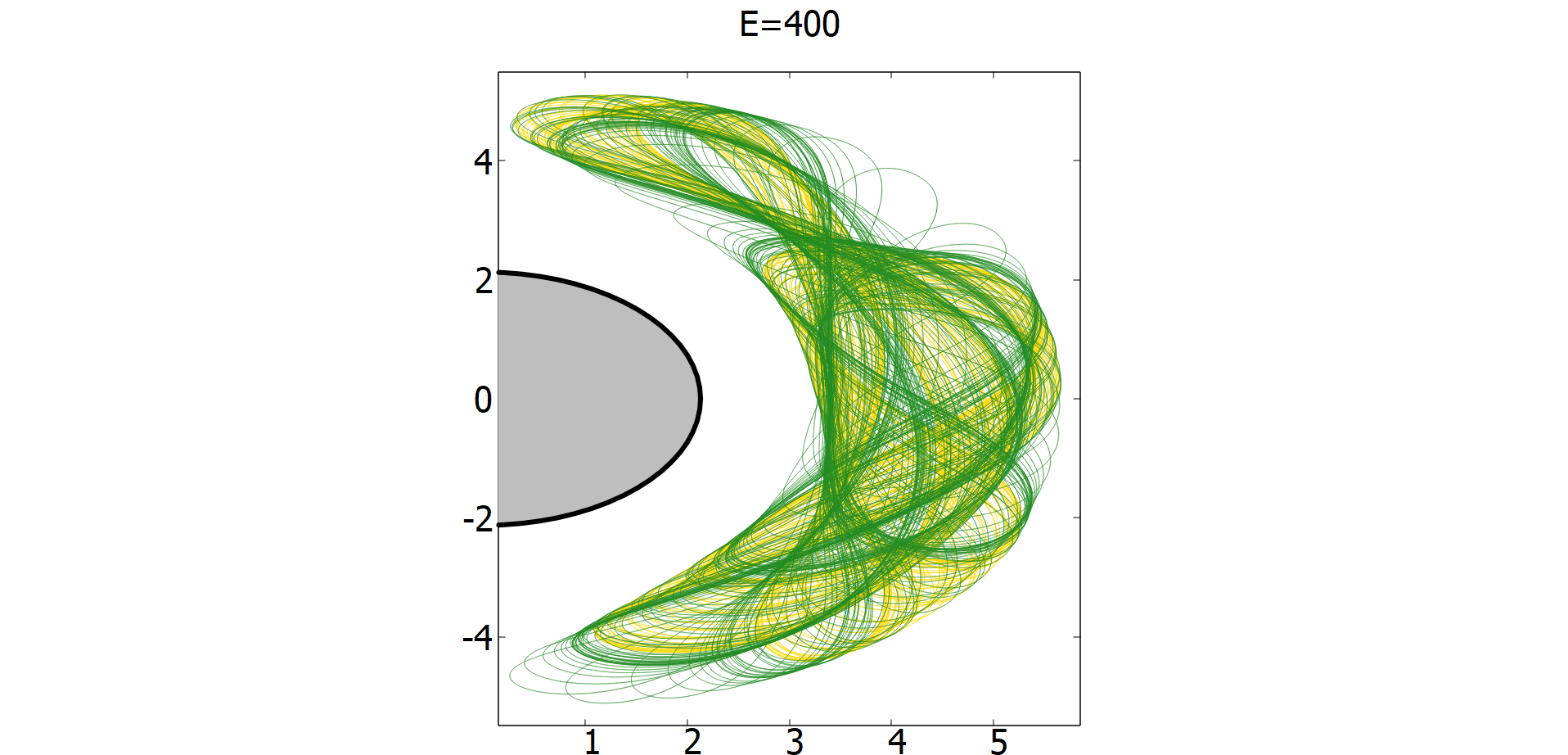} & \includegraphics[width=0.25\linewidth]{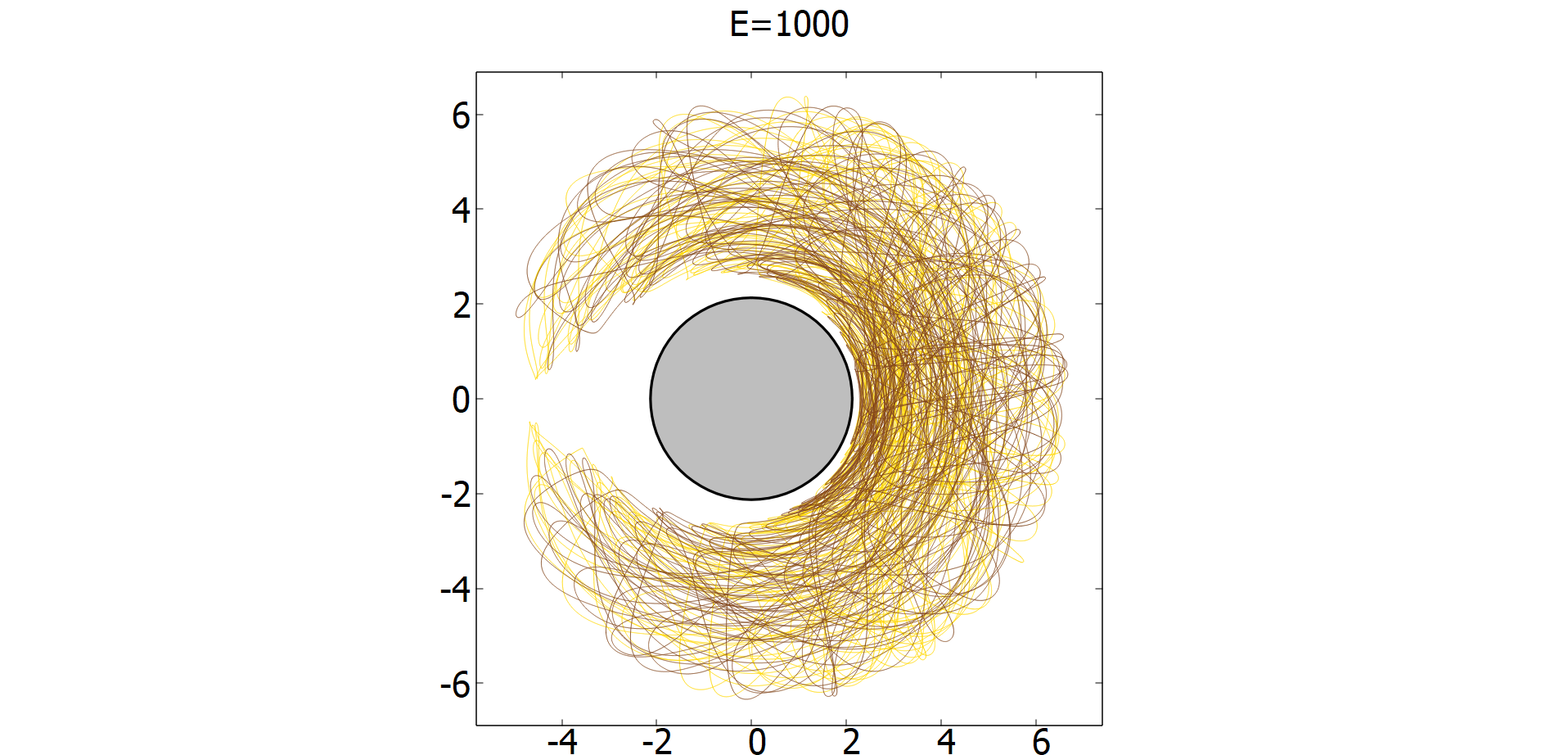} \\
        \hline
         Model-II Neutral BH with fixed $a=0.5.$ & \includegraphics[width=0.25\linewidth]{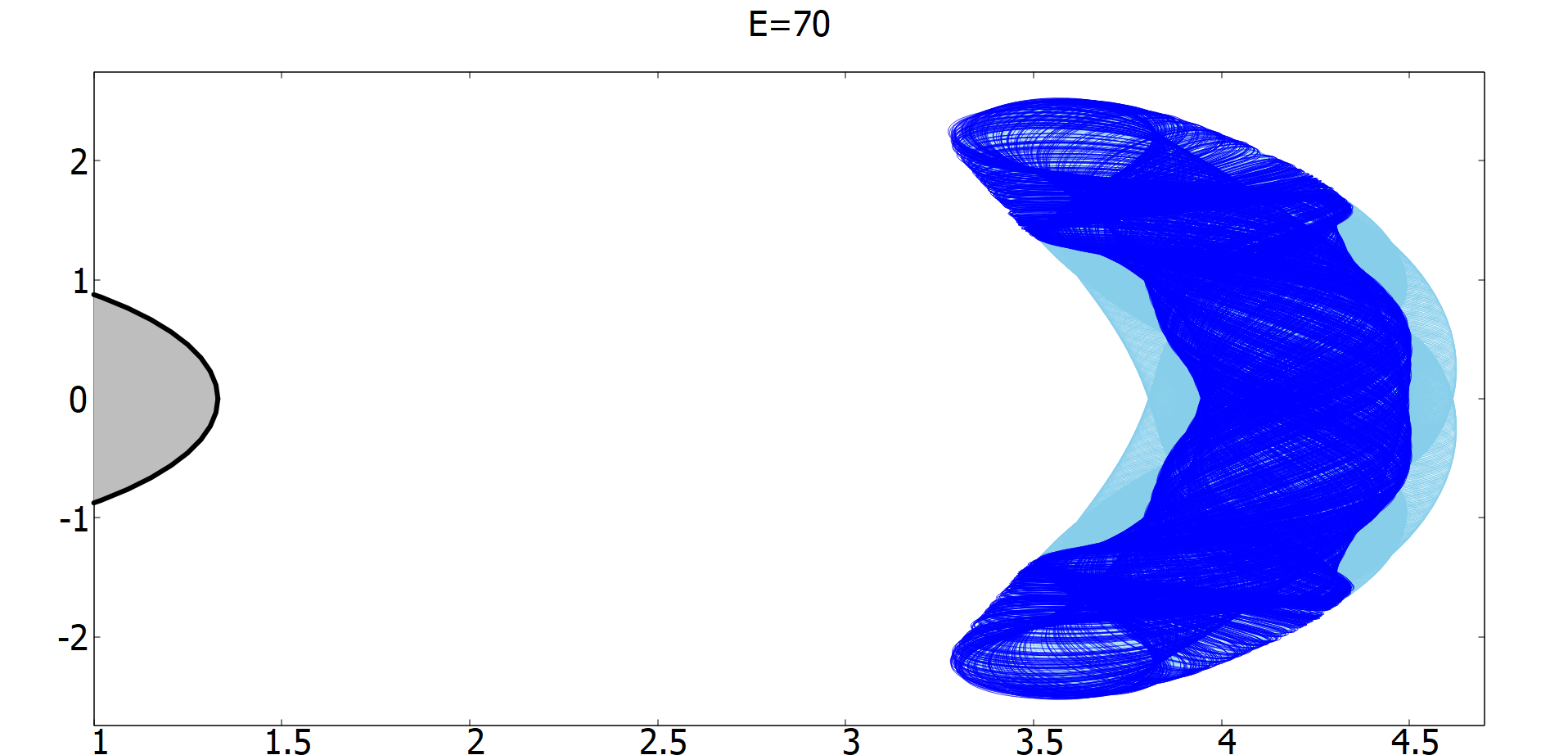} & \includegraphics[width=0.25\linewidth]{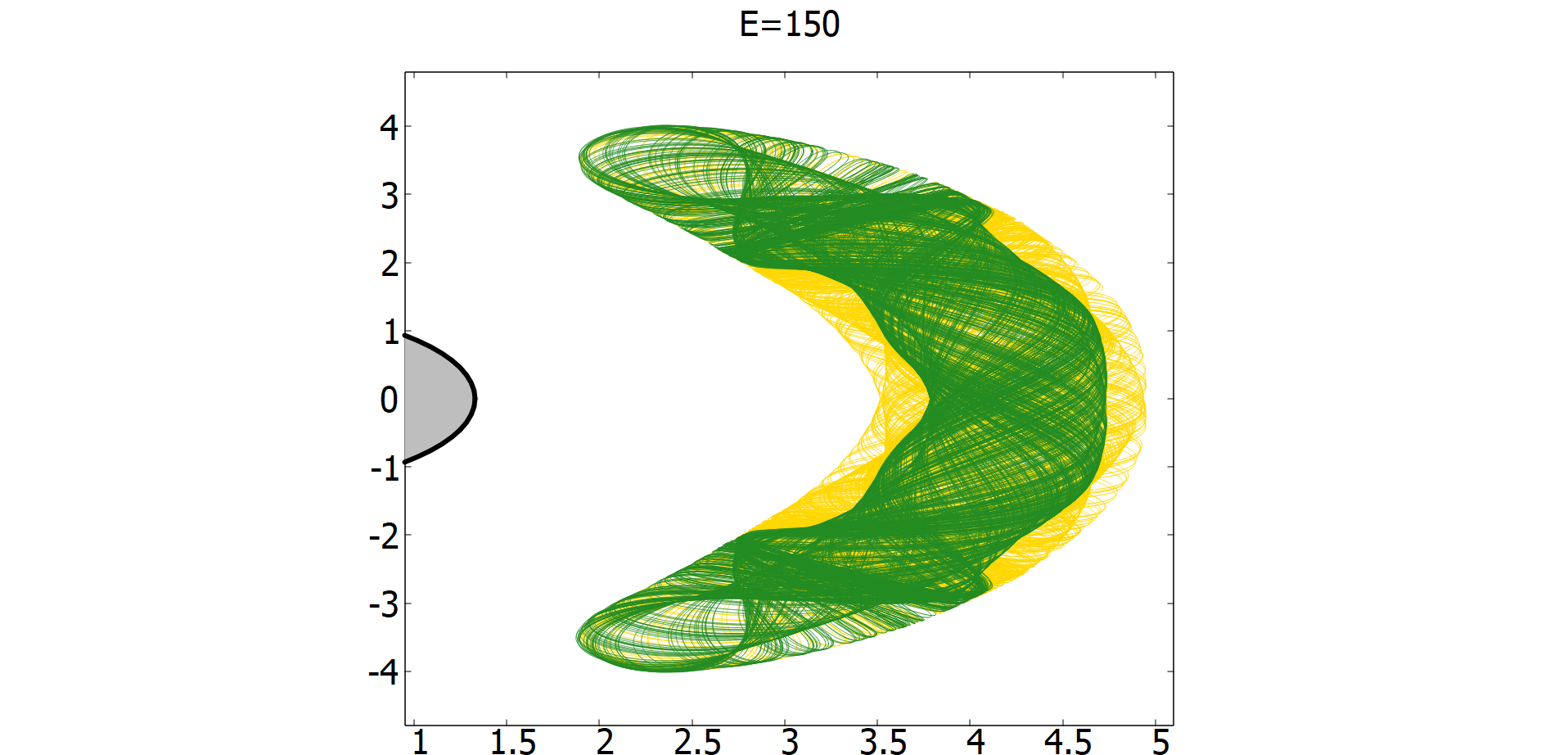} & \includegraphics[width=0.25\linewidth]{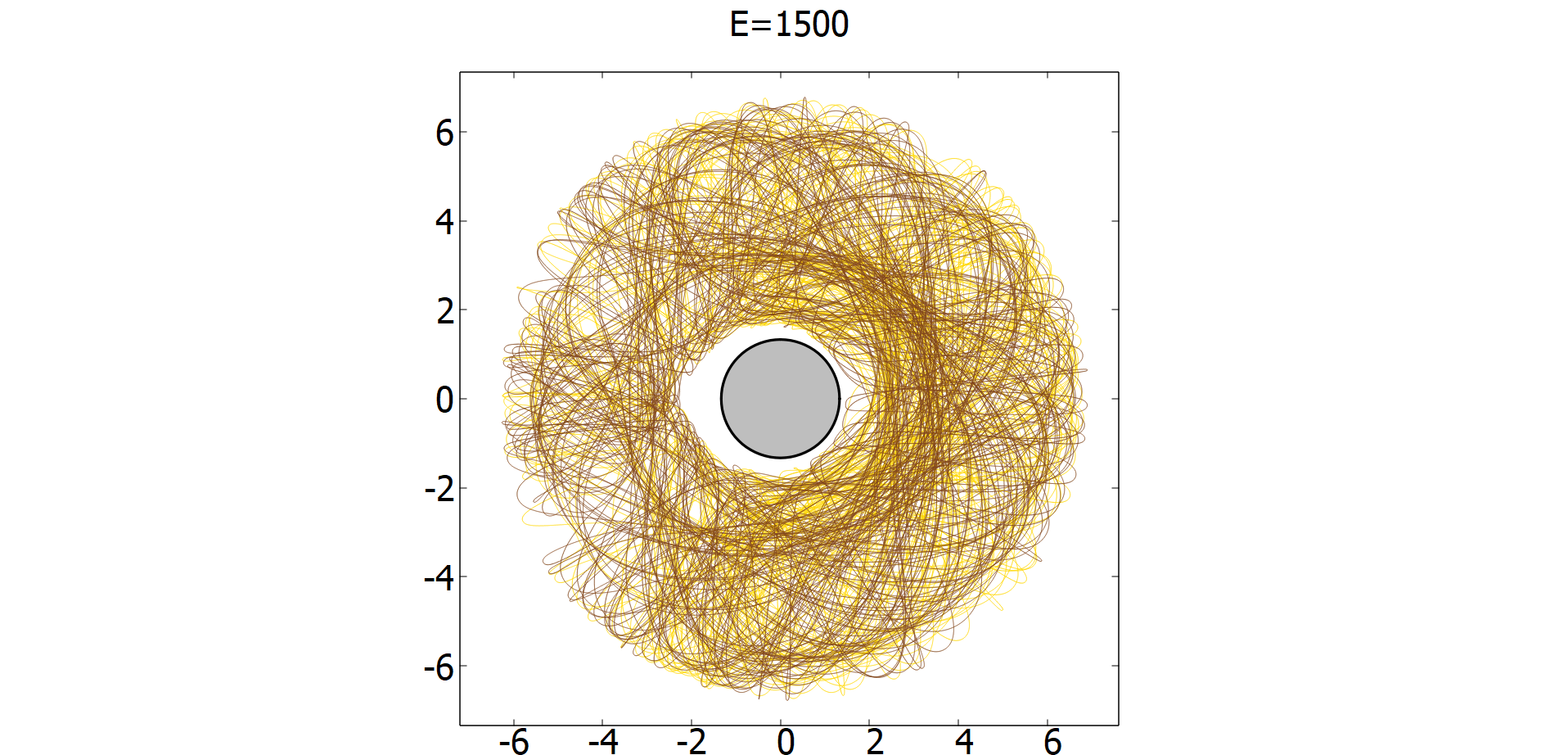} \\ \hline
    \end{tabular}
    \caption{Dynamics of a probe particle near the black hole's horizon for different models of $f(R)$ gravity considered in this work \big[in $(x-y)$ chart, see text for details\big]. In all these figures, we track two initially close trajectories within the allowed range of parameters (see text). The gray-shaded region represents the region inside the event horizon of the black hole. In each row, the leftmost figure represents periodic orbit for low values of energy, the middle row represents the onset of chaos (i.e., the first appearance of broken tori in the associated Poincar$\Acute{e}$ section) at moderate energies, and the rightmost figure represents the full development of chaos at high energies. With increase in energy from left to right, the particle comes closer to  the horizon with reduced overlap between nearby trajectories, indicating growth of chaos.}
    \label{T1}
    \end{figure*}

    Before analyzing the Poincar$\Acute{e}$ sections and Lyapunov exponents, let us provide an overview of typical particle's trajectories for a set of two initial conditions for the models discussed in the previous section.  This will help us to gain intuition for the construction and better visualization of the Poincar$\Acute{e}$ sections. This is presented in Fig.\ref{T1}. The top row shows the particle trajectories in model I for $\beta=10^{-2}$ for three different energies. We plot the trajectories in the  2D $(x-y)$ chart where $x= r \cos \theta, y= r \sin \theta$ as usual and $\theta \in(- \pi, \pi)$.  The second and third row presents similar figures for the charged and neutral black hole  backgrounds in model II, respectively.\\
    Two things are apparent from these plots. Firstly, we clearly observe the confining nature of the harmonic potentials introduced along the $r$ and $\theta$ directions. In the absence of the black hole, these harmonic potentials present an integrable system. Secondly, we observe that in the presence of the black hole horizon, the integrability of the system is broken, and there is an onset of chaos at increasing energies (so that the particle trajectories come in the vicinity of the horizon). Here, by ``onset" of chaos, we imply the first appearance of broken tori in the associated Poincar$\Acute{e}$ section. In all these figures, we show the horizon contour.

    \noindent

        \subsection{Analysis of the Poincar$\Acute{e}$ sections}\label{s4b}
    The Poincar$\Acute{e}$ map, an essential tool for studying nonlinear dynamics, is defined as the intersection of periodic/ aperiodic orbits with a subspace transverse to the trajectories residing in the full state space. The essential idea is to map the higher dimensional phase-space trajectories into the lower one using the Poincar$\Acute{e}$ map \cite{Strog}. For model I and model II, we choose the $\theta=0$ plane as the Poincar$\Acute{e}$ section. Within this space, we plot the points on the $(r-p_r)$ plane when the particle intersects the Poincar$\Acute{e}$ section, guided by the constraint of fixed energy $E$ and $p_{\theta}>0$. For the periodic case, the plotted points will lie on a torus in the phase space, while for the chaotic scenario, some of these tori will be broken. Such broken tori in different corners of the phase space is a well-recognized signature of chaos \cite{Strog,Stock}. 
    Now, we will numerically solve the dynamical equations of motion \Big[Eq.\eqref{3.12}, Eq.\eqref{3.13}, Eq.\eqref{3.14}, and Eq.\eqref{3.15}\Big] for the black hole backgrounds arising in modified gravity (model I and model II). These numerical solutions will aid in the construction of Poincar$\Acute{e}$ section. We've already shown analytically that the radial trajectory of the particle induces exponential growth in the presence of a horizon, which indicates a possibility of chaotic behavior (see Sec. \ref{sec3}). Here, we present a numerical confirmation of this expectation by examining the Poincar$\Acute{e}$ sections for the SSS black hole solutions, where we systematically analyze the effects of the modified gravity parameters $\beta$ and  $a$ in model I and model II, respectively. Special emphasis is placed on a careful comparison of chaos in such modified gravity backgrounds with similar investigations in Einstein gravity \cite{i1,i2,i4,i7,i8,hashimoto,dalui,rindler}.

        \subsubsection{Model I:}\label{s4b-1}
    For the construction of the Poincar$\Acute{e}$ section in model I, we have considered two values of the modified gravity parameter $\beta$ i.e., $\beta=10^{-5}$ and $10^{-2}$, relevant for typical spiral galaxies of solar system length scales, where  $\beta\approx 10^{-26}~m^{-1}$ \cite{Saffa,Saheb}. Therefore, the lower value  $\beta$, i.e., $\beta=10^{-5}$, corresponds to a background similar to the Schwarzschild black hole arising in Einstein gravity. Increasing $\beta$ to $10^{-2}$ allows us to probe the effects of modified gravity on the chaotic dynamics of particles close to the event horizon. Let us also add that as $\beta$ increases, the size of the black hole shrinks, a fact which will be of considerable importance in the analysis presented below.\\
    In model I, demanding $\sqrt{1-A(r)} > 0$ in Eq.\eqref{3.13} constrains $r$ to be in the range, $r_H<r<\sqrt{\frac{2}{\beta}}$, where the expression for $r_H$ is given by Eq.\eqref{14}. We will focus on the chaotic dynamics in this range of $r$. We solve  the dynamical equations of motion \Big[Eq.\eqref{3.12}, Eq.\eqref{3.13}, Eq.\eqref{3.14} and Eq.\eqref{3.15}\Big] using the fourth order Runge-Kutta method with fixed step size $h=0.01$. In this analysis, we have chosen $K_r=100$, $K_{\theta}=25$ and $\theta_c=0$. For $\beta=10^{-5}$, the upper bound on the radial coordinate is 447.21, whereas for $\beta=10^{-2}$, it is 14.14. We wish to constrain the motion of the particle within this specific range of $r$. We set the equilibrium position of the harmonic oscillator $r_c$ to $3.2$ (for both $\beta$ values), ensuring that the particle resides near the event horizon. The other free variables $r,~p_r$ and $\theta$ are initialized randomly within the range $3.0<r<3.5~,-0.5<p_r<0.5$ and $-0.05<\theta<0.05$, respectively. The value of $p_{\theta}$ is obtained from Eq.\eqref{3.11} for a fixed value of the conserved energy $E$ for outgoing massless particle. The different colors in the following figures denote trajectories of the particle solved for those randomly chosen initial conditions.

\begin{widetext}
        \begin{figure}[H]
	\centering
	\begin{center} 
	$\begin{array}{ccc}
	\subfigure[] 
        {\includegraphics[width=0.7\linewidth,height=0.5\linewidth]{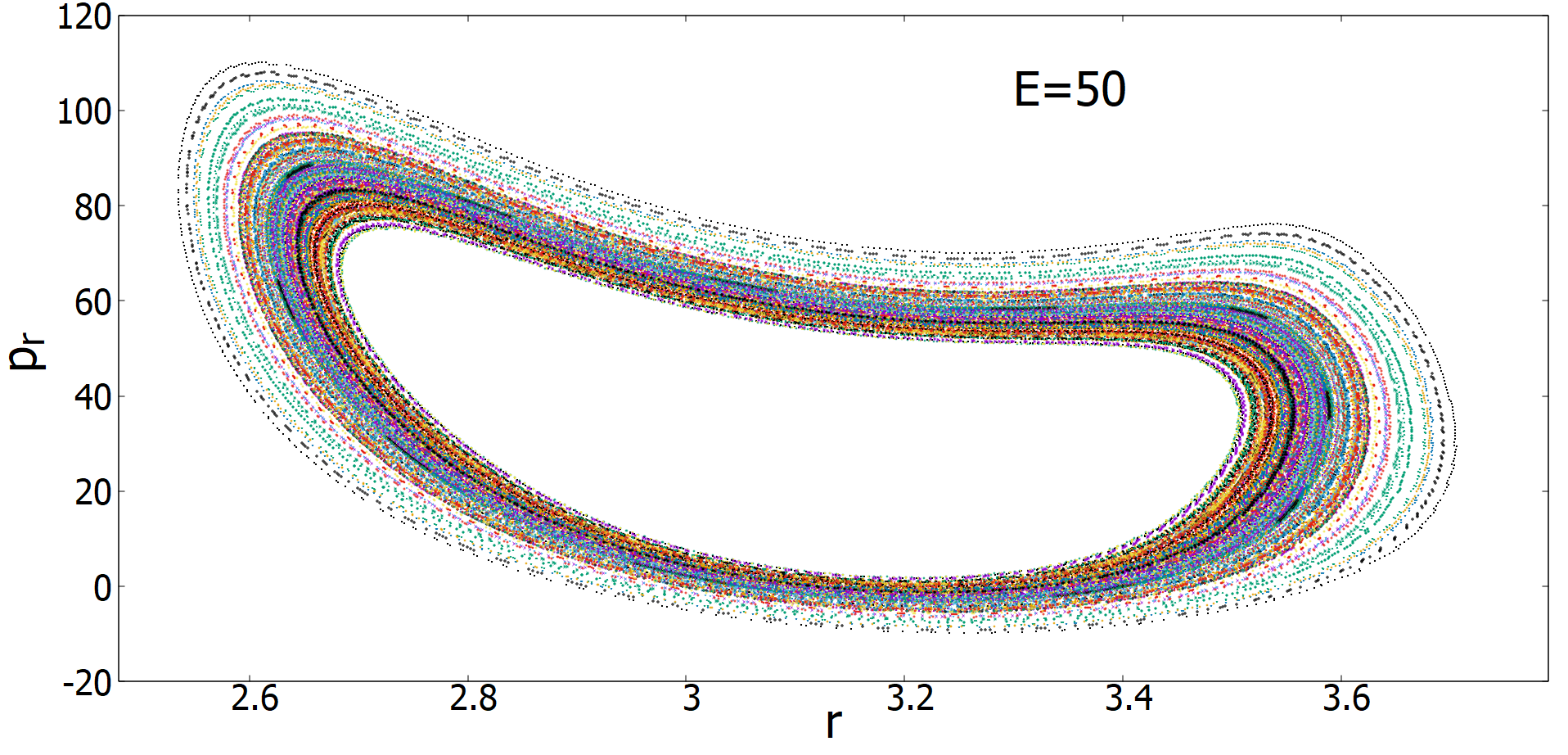}\label{4a}}
        \subfigure[]{\includegraphics[width=0.7\linewidth,height=0.5\linewidth]{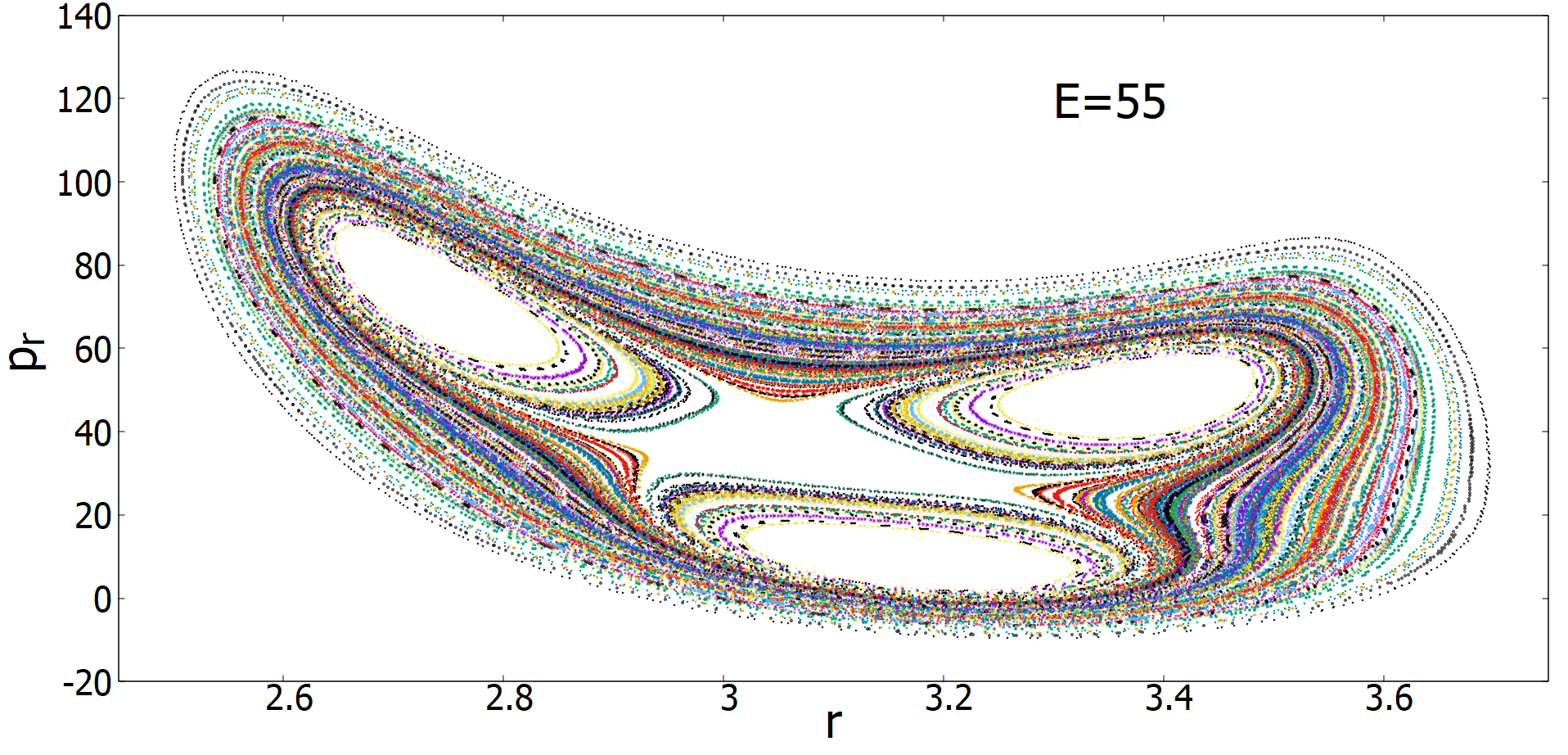}\label{4b}}
	\subfigure[] 
        {\includegraphics[width=0.7\linewidth,height=0.5\linewidth]{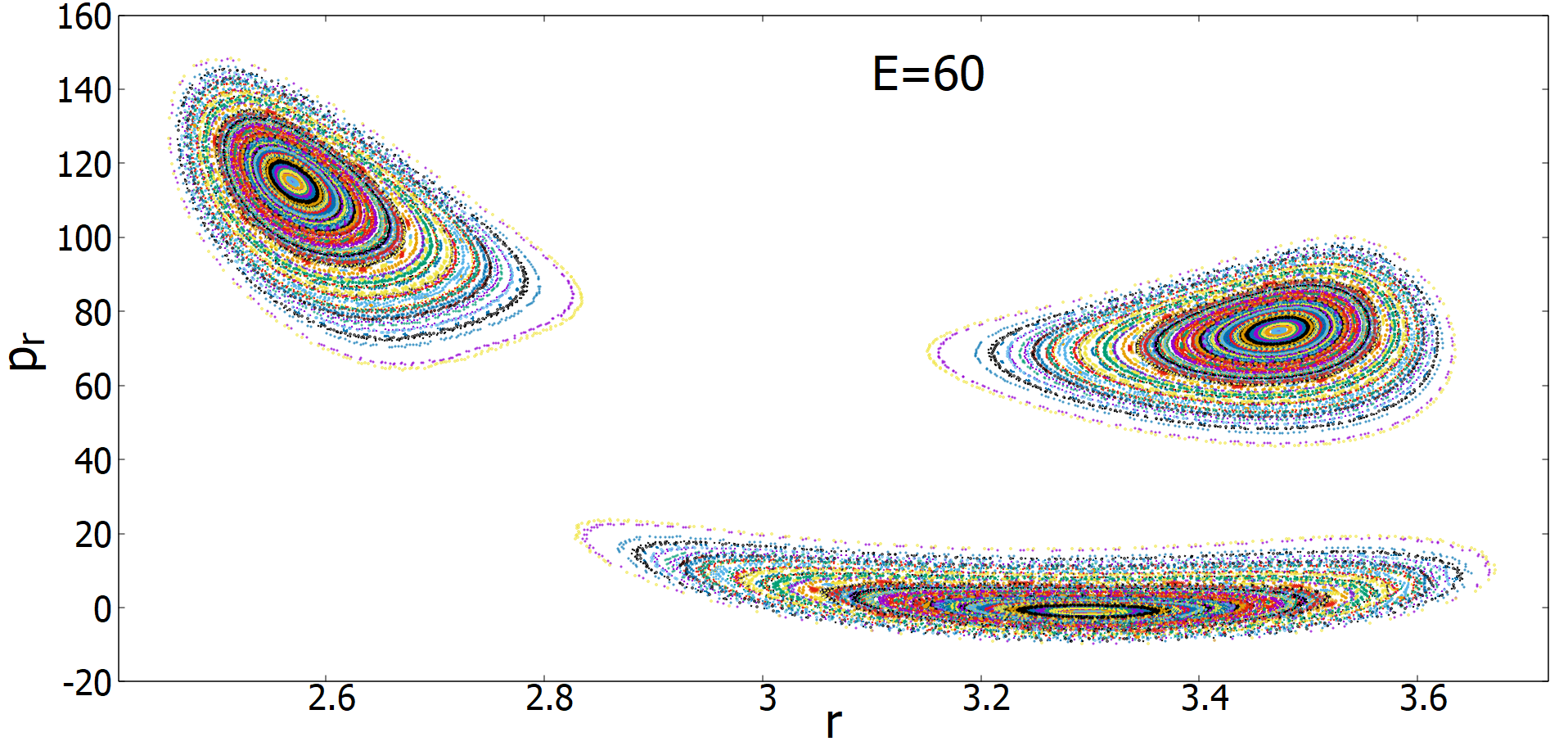}\label{4c}}\\
	\subfigure[] 
        {\includegraphics[width=0.7\linewidth,height=0.5\linewidth]{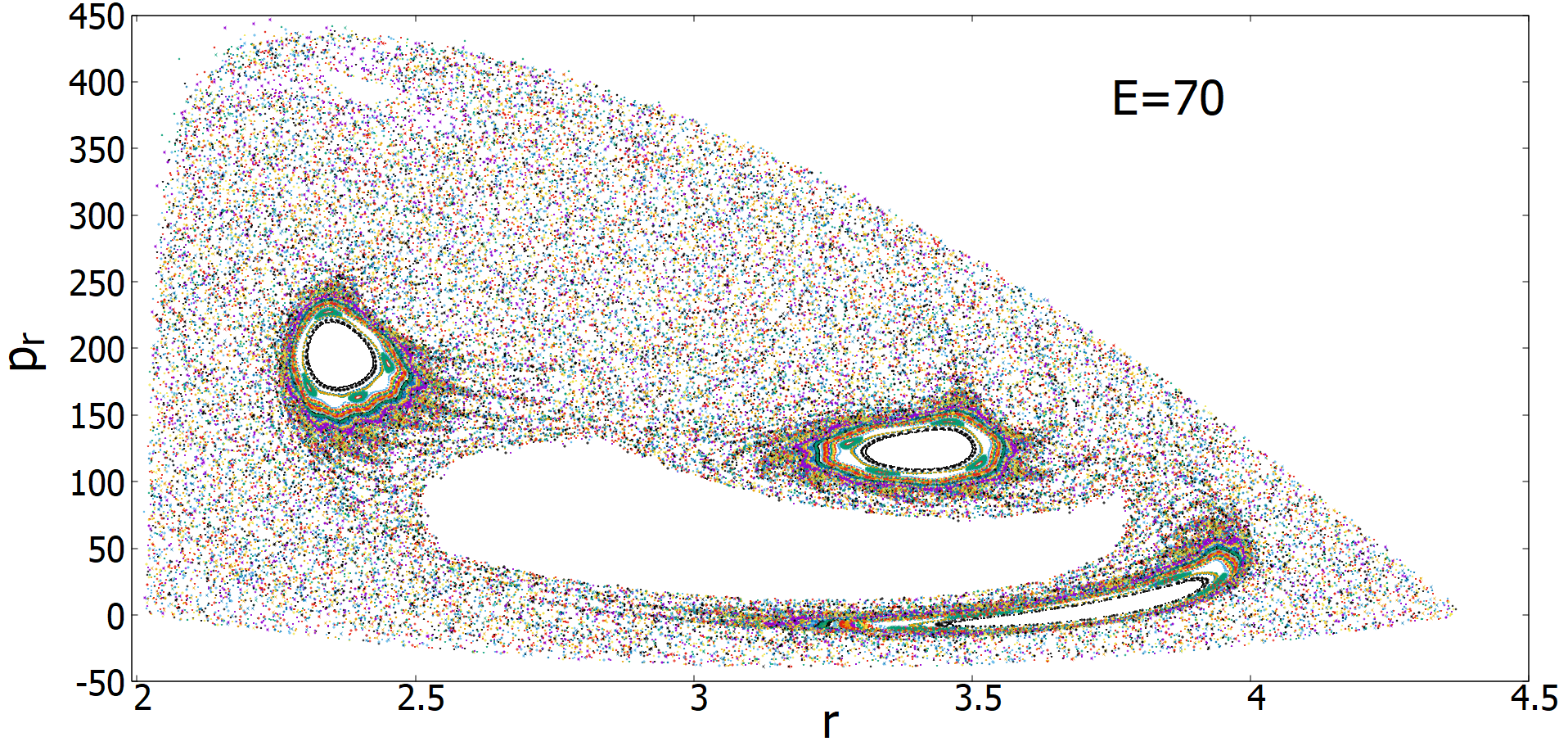}\label{4d}}
	\subfigure[] 
        {\includegraphics[width=0.7\linewidth,height=0.5\linewidth]{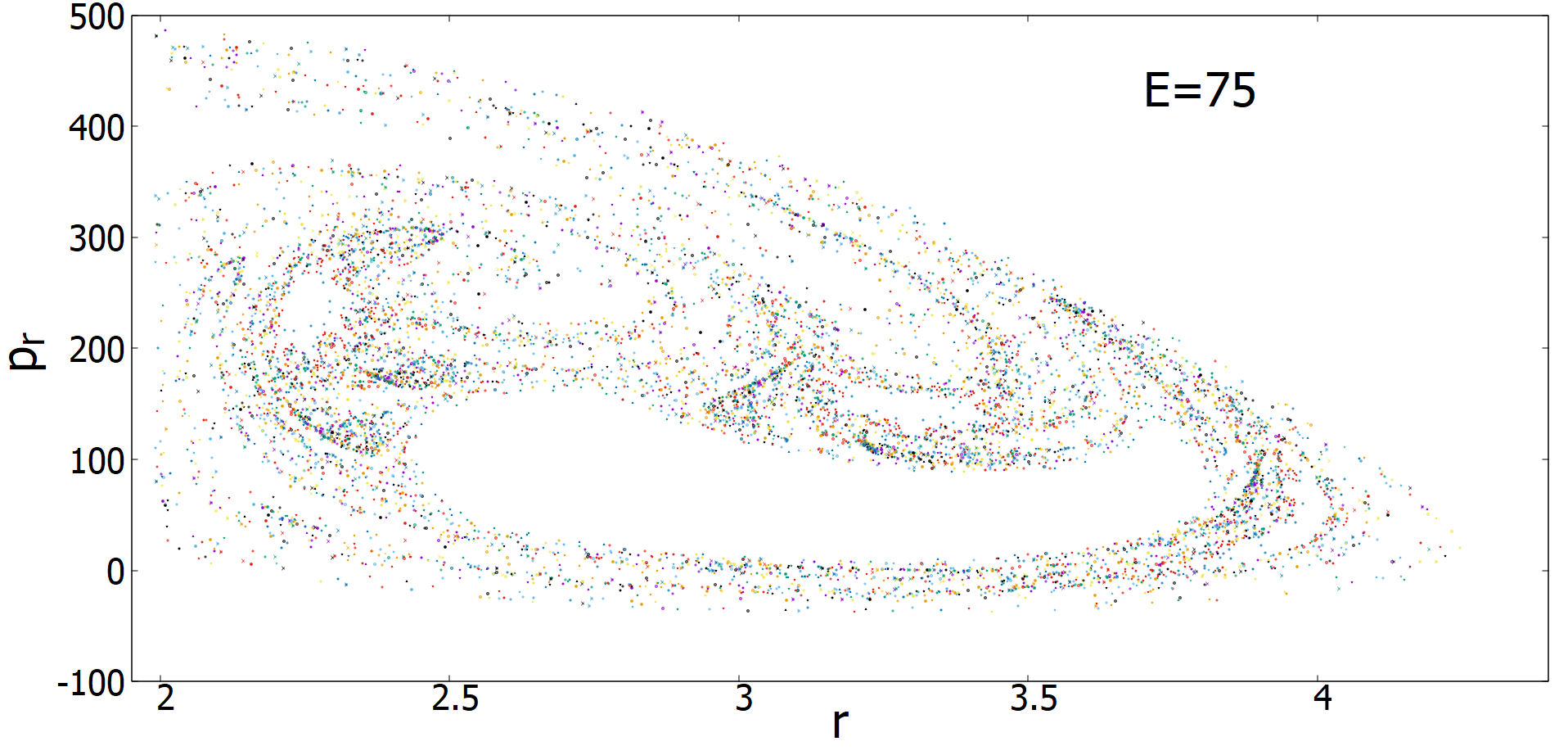}\label{4e}}
	\end{array}$
        \end{center}
        \begin{minipage}{\textwidth}
        \caption{The Poincar$\Acute{e}$ sections in the $(r-p_r)$ phase plane with $\theta=0$ and $p_{\theta}>0$ for different energies with $\beta=10^{-5}$.}\label{f4}
        \hrulefill
        \end{minipage}
        \begin{center}
        $\begin{array}{ccc}
	\subfigure[] 
        {\includegraphics[width=0.7\linewidth,height=0.5\linewidth]{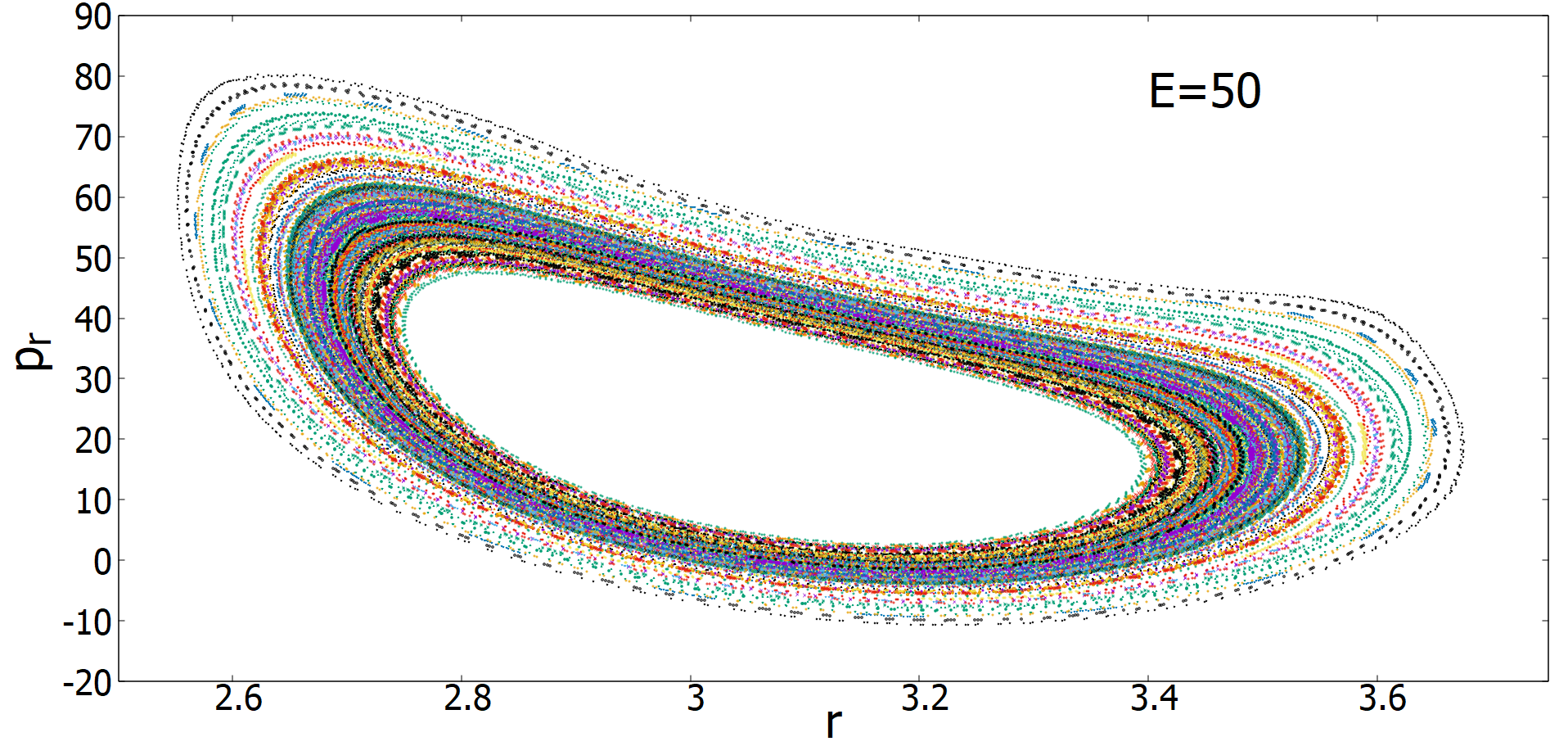}\label{5a}}
        \subfigure[]{\includegraphics[width=0.7\linewidth,height=0.5\linewidth]{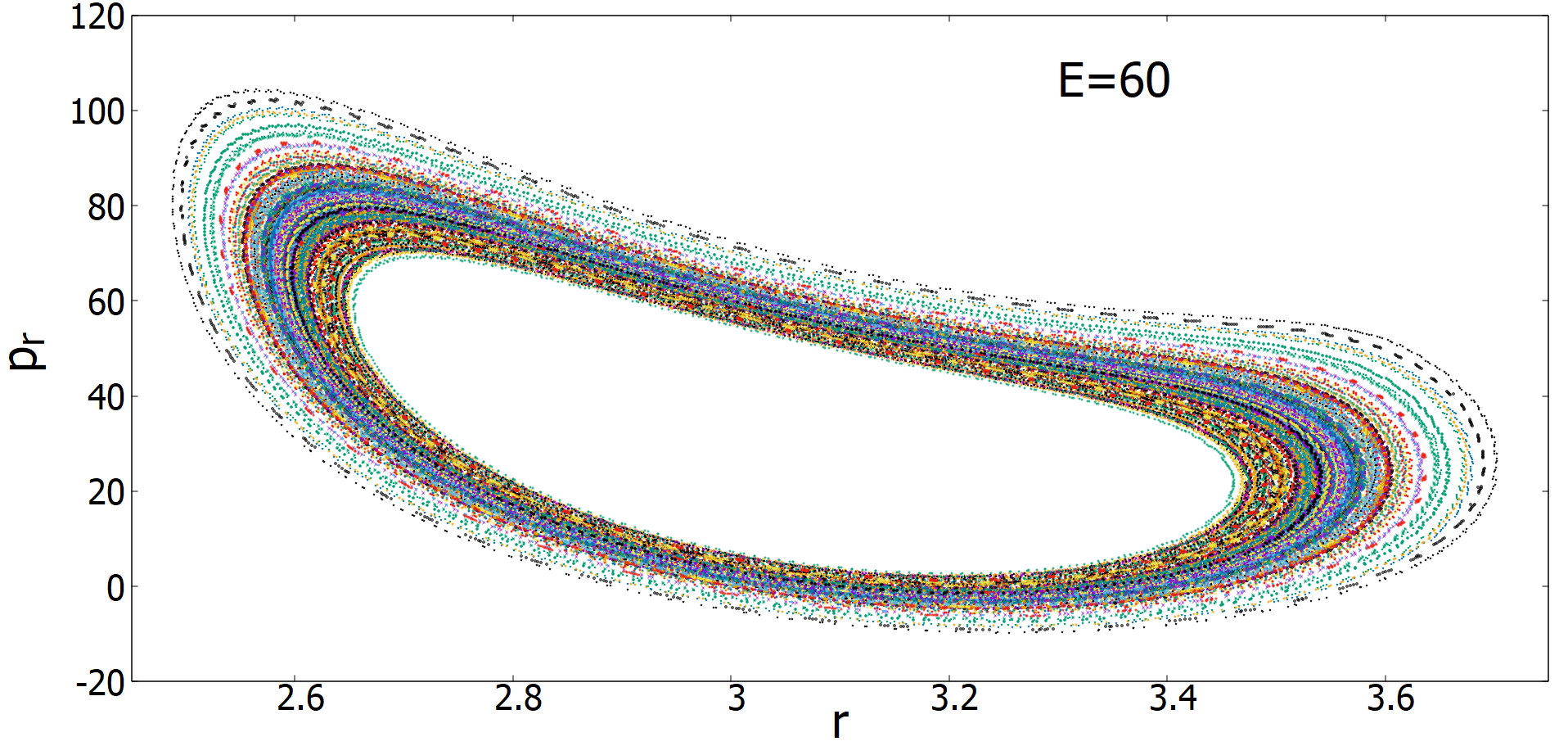}\label{5b}}
	\subfigure[] 
        {\includegraphics[width=0.7\linewidth,height=0.5\linewidth]{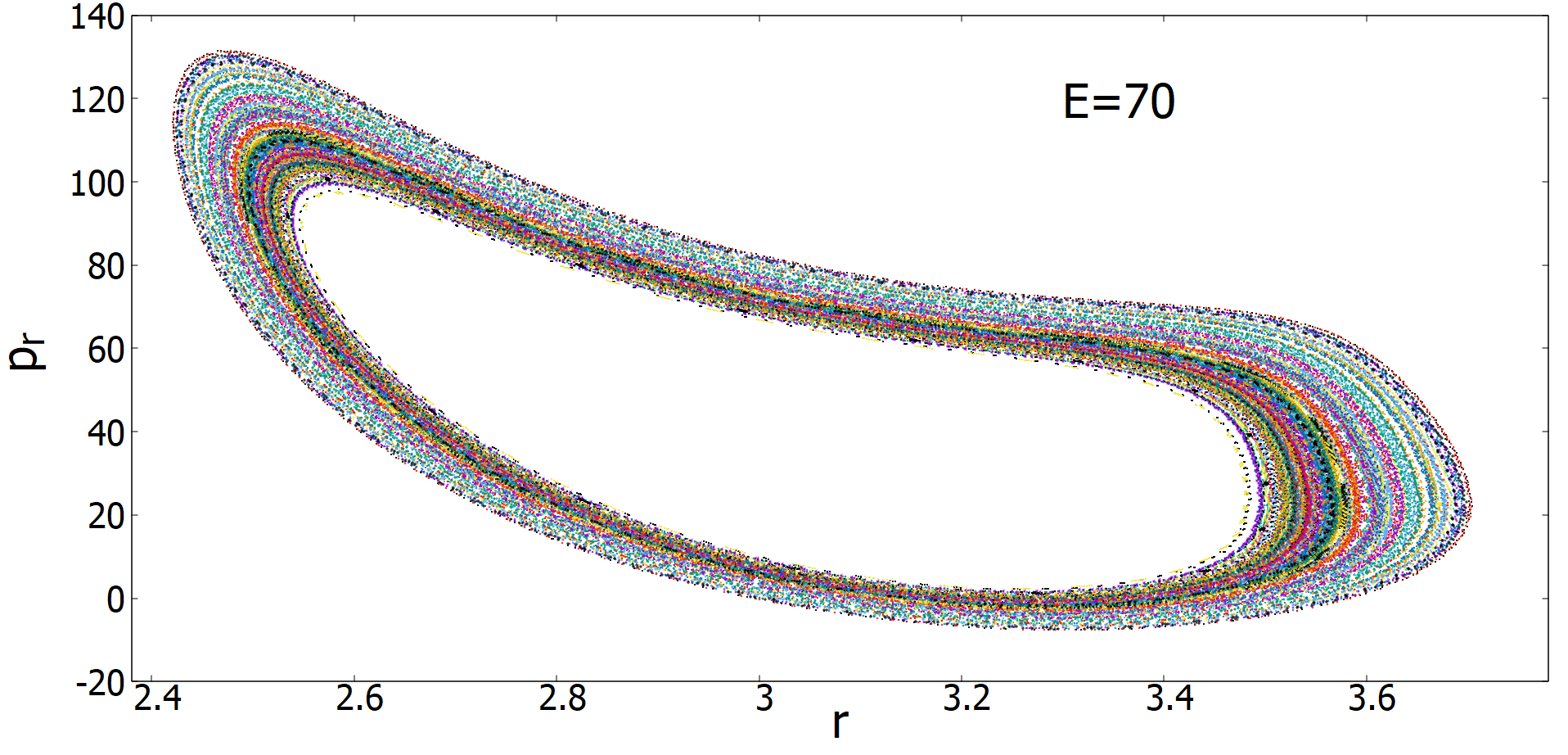}\label{5c}}\\
        \subfigure[]{\includegraphics[width=0.7\linewidth,height=0.5\linewidth]{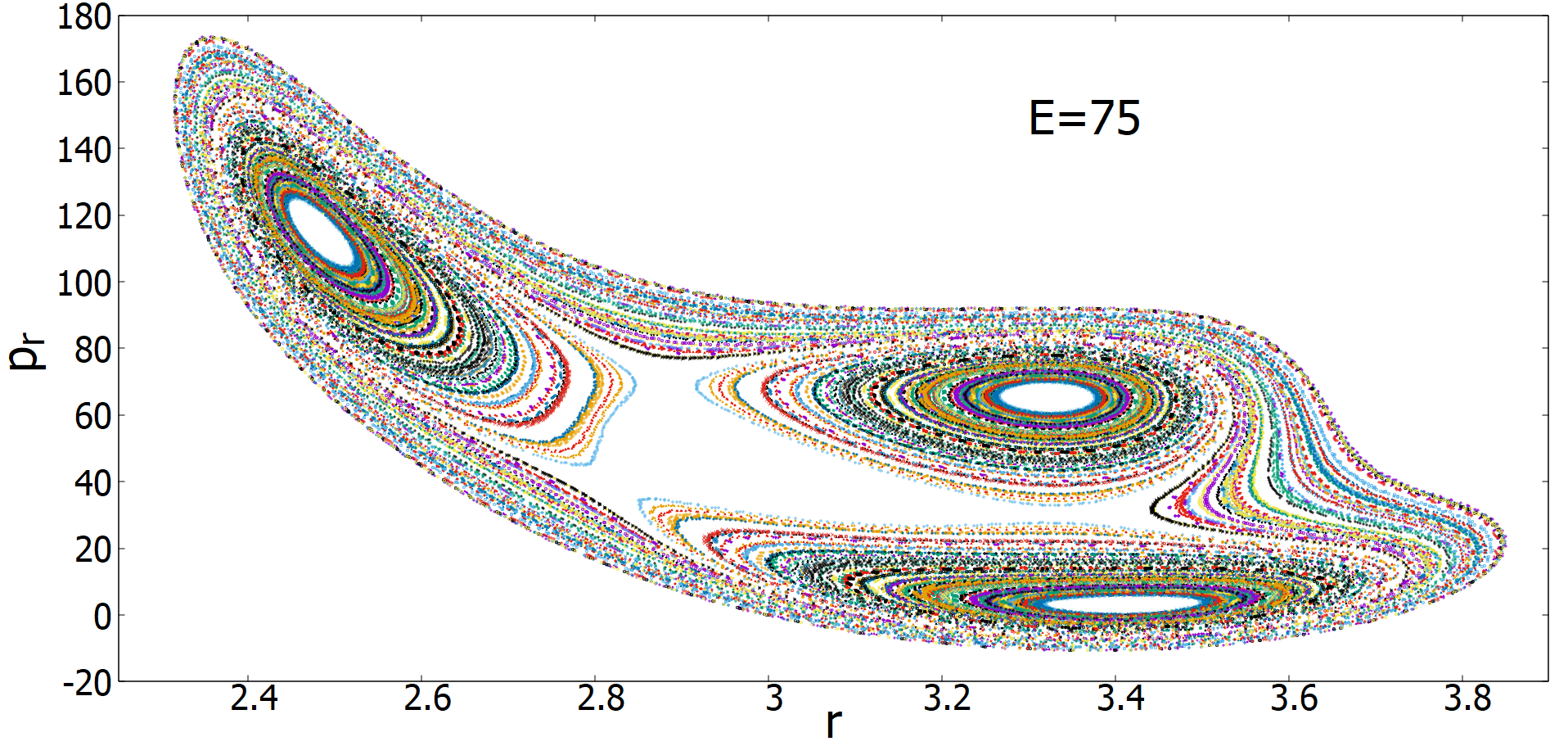}\label{5d}}
        \subfigure[]{\includegraphics[width=0.7\linewidth,height=0.5\linewidth]{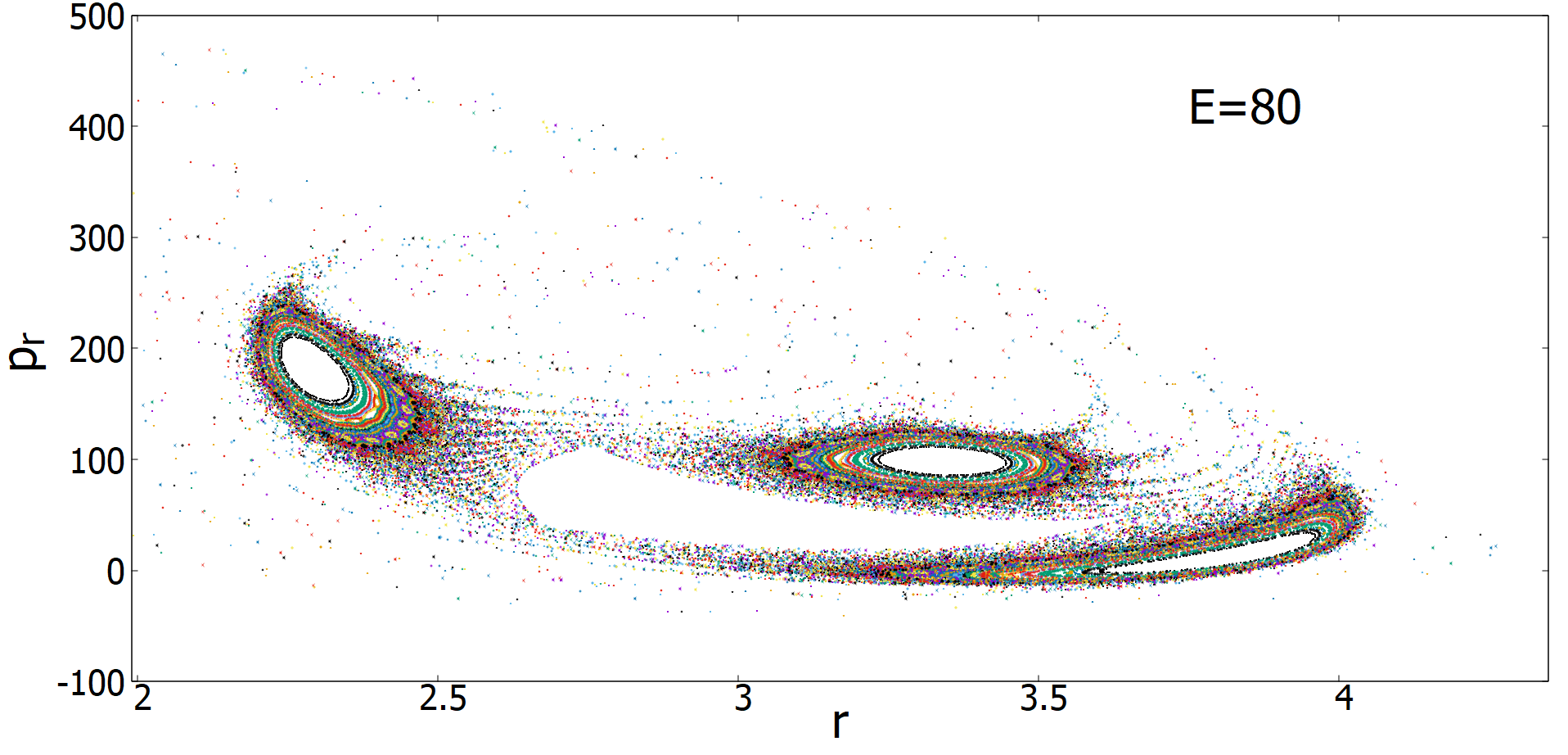}\label{5e}}
        \subfigure[]{\includegraphics[width=0.7\linewidth,height=0.5\linewidth]{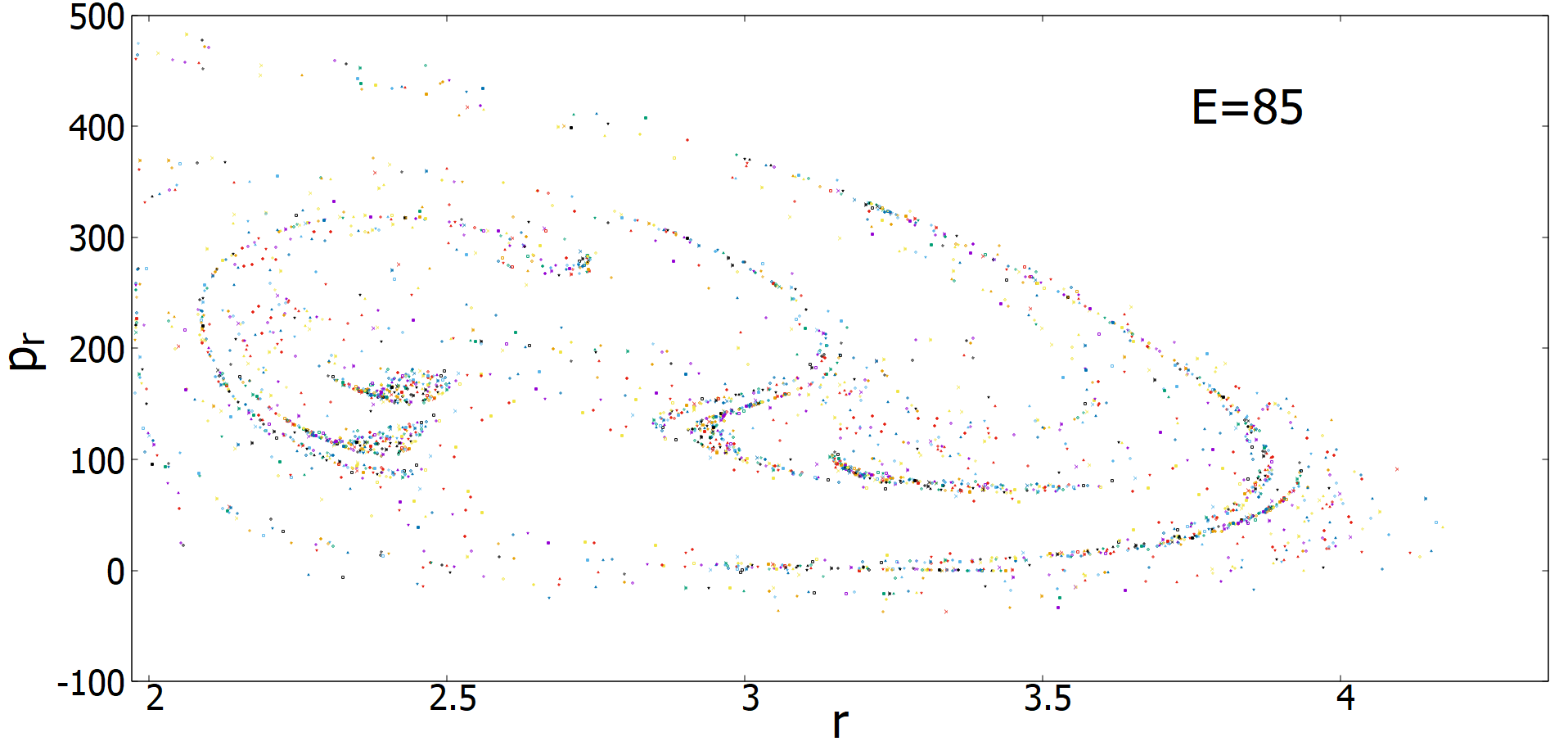}\label{5f}}
        \end{array}$
        \end{center}
        \begin{minipage}{\textwidth}
        \caption{The Poincar$\Acute{e}$ sections in the $(r-p_r)$ phase plane with $\theta=0$ and $p_{\theta}>0$ for different energies with $\beta=10^{-2}$.}\label{f5}
        \end{minipage}
        \end{figure}
    \end{widetext}

    In Fig.\ref{f4}, we illustrate the Poincar$\Acute{e}$ section of a particle's trajectory projected onto the $(r-p_r)$ phase plane, setting $\theta=0$  with $p_{\theta}>0$, considering five different energies: $E=50, 55, 60, 70$, and $75$ for $\beta=10^{-5}$. At the lowest energy, $E=50$, the Poincaré sections display regular Kolmogorov-Arnold-Moser (KAM) tori \cite{KAM}, indicating that the orbit is predominantly confined near the center of the harmonic potential, which is taken as $r_c=3.2$ and only a single frequency presents in the system. However, due to the conservation of the system's Hamiltonian, as the total energy increases, so does the momentum, causing the trajectory to approach the event horizon of the black hole. Consequently, at $E=55$, the Poincar$\Acute{e}$ section features distorted tori and for $E=60$, they separate completely as shown in Fig.\ref{4c}. As depicted in Fig.\ref{4d}, $E=70$ results in the breakdown of the regular tori, with scattered points appearing in the phase plane, indicative of chaotic behavior. Finally, at the highest energy ($E=75$), depicted in Fig.\ref{4e}, the KAM tori completely disintegrates, and the phase plane becomes filled with points distributed randomly, indicative of fully chaotic behavior.
    
    The manifestation of chaotic behavior in model I varies significantly for two distinct values of $\beta$, as evident from the Poincaré maps \Big(see the differences in Fig.\ref{f4} and Fig.\ref{f5}\Big). For the larger value of $\beta=10^{-2}$, the tori begin to disperse at relatively higher energy, as depicted in Fig.\ref{f5}. This can be understood as follows: far away from the horizon, the particle is essentially confined in a harmonic trap, which is an integrable system exhibiting periodic orbits. In contrast, if the particle is closer to the horizon, the nonlinear dynamics are dictated by the underlying curved spacetime background, which results in chaotic dynamics. Thus, the proximity of the particle with respect to the event horizon plays a crucial role in the onset of chaos, as observed in \cite{dalui} for Einstein gravity.\\
    Increasing energy of the particle results in higher momentum and a shift of the corresponding orbits towards the event horizon, ultimately disrupting the regular tori. Moreover, as we remarked in the last section, the larger value of the modified gravity parameter $\beta$ shrinks the size of the event horizon, effectively reducing the impact of the horizon on the particle's trajectory (since the particle is now confined at $r_c$, which is far from the location of the event horizon). Thus, we expect that for higher values of the modified gravity parameter $\beta$, we need to raise the energy (and equivalently, the radial momentum) so that the particle trajectory can come in close proximity to the horizon. Hence, the onset of chaos should be observed at higher values of energy for large values of the modified gravity parameter $\beta$.   On the other hand, in the limit where $\beta$ approaches $0$ (or equivalently, for the smaller value of $\beta$), all Poincaré maps exhibit behavior akin to that presented in \cite{dalui}. All these effects are clearly evident from Fig.\ref{f4}. Let us add that one cannot choose arbitrarily high values of energy in this setup since we restrict the investigation only to trajectories outside the event horizon, i.e. $r_H<r<\sqrt{\frac{2}{\beta}}$. Beyond this region, one encounters numerical instabilities as expected. We have also carried out the same analysis in Schwarzschild coordinates for the model I (see Figs. \ref{af1} and \ref{af2} for details).

        \subsubsection{Model II: Charged black hole background}\label{s4b-2}
    We now carry out a similar investigation for the charged SSS black hole solution in model II [Eq.\eqref{21}]. The dynamical equations of motion  \Big[Eq.\eqref{3.12}, Eq.\eqref{3.13}, Eq.\eqref{3.14}, and Eq.\eqref{3.15}\Big] are again numerically solved using the fourth order Runge-Kutta method with step size $h=0.01$. We then investigate the Poincar$\Acute{e}$ maps for varying energy $E$ and the modified gravity parameter $a$. In the whole scenario, we have considered $K_r=400$, $K_{\theta}=47$, $\theta_c=0$. 
    
    In this model, demanding  $\sqrt{1-A(r)}>0$ in Eq.\eqref{3.13} implies $r>\frac{-1+\sqrt{1+6a}}{3a}>r_{+}$. Thus, we explore the 
    
    \newpage

    \noindent
    particle dynamics in the region $r>r_{+}$. In order to make sure that the particle resides near the horizon and also to get better Poincar$\Acute{e}$ maps, we set $r_c=4.5$. The other free variables $r$, $p_r$, and $\theta$ are initialized with random numbers given in the range $4.0<r<4.9,~-0.5<p_r<0.5$, and $-0.05<\theta<0.05$. Moreover, just like in model I, $p_{\theta}$ is obtained from Eq.\eqref{3.11} for a fixed value of energy $E$. 
       
\begin{widetext}
        \begin{figure}[H]
	\centering
	\begin{center} 
	$\begin{array}{ccc}
	\subfigure[]              
        {\includegraphics[width=0.7\linewidth,height=0.5\linewidth]{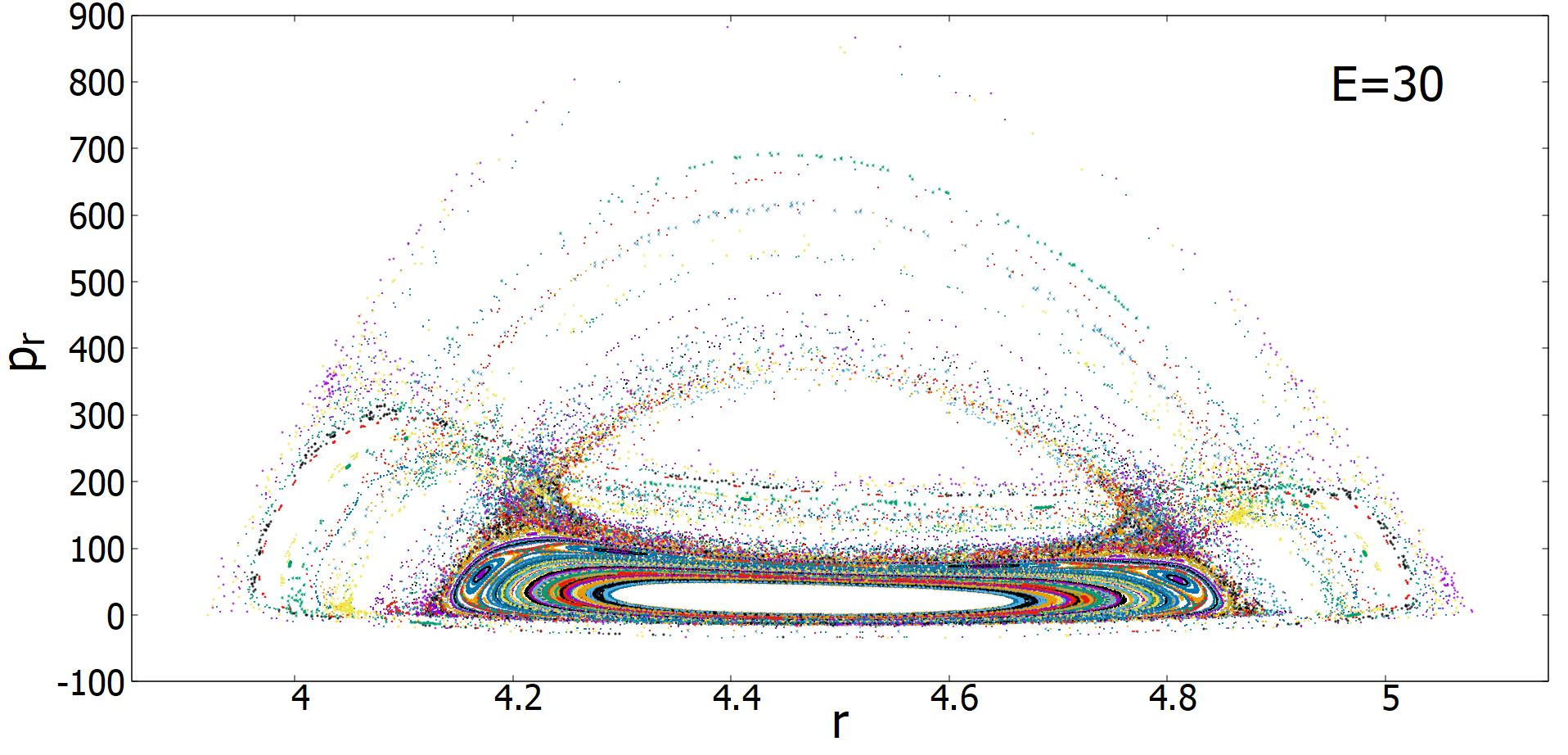}\label{6a}}
	\subfigure[]           
        {\includegraphics[width=0.7\linewidth,height=0.5\linewidth]{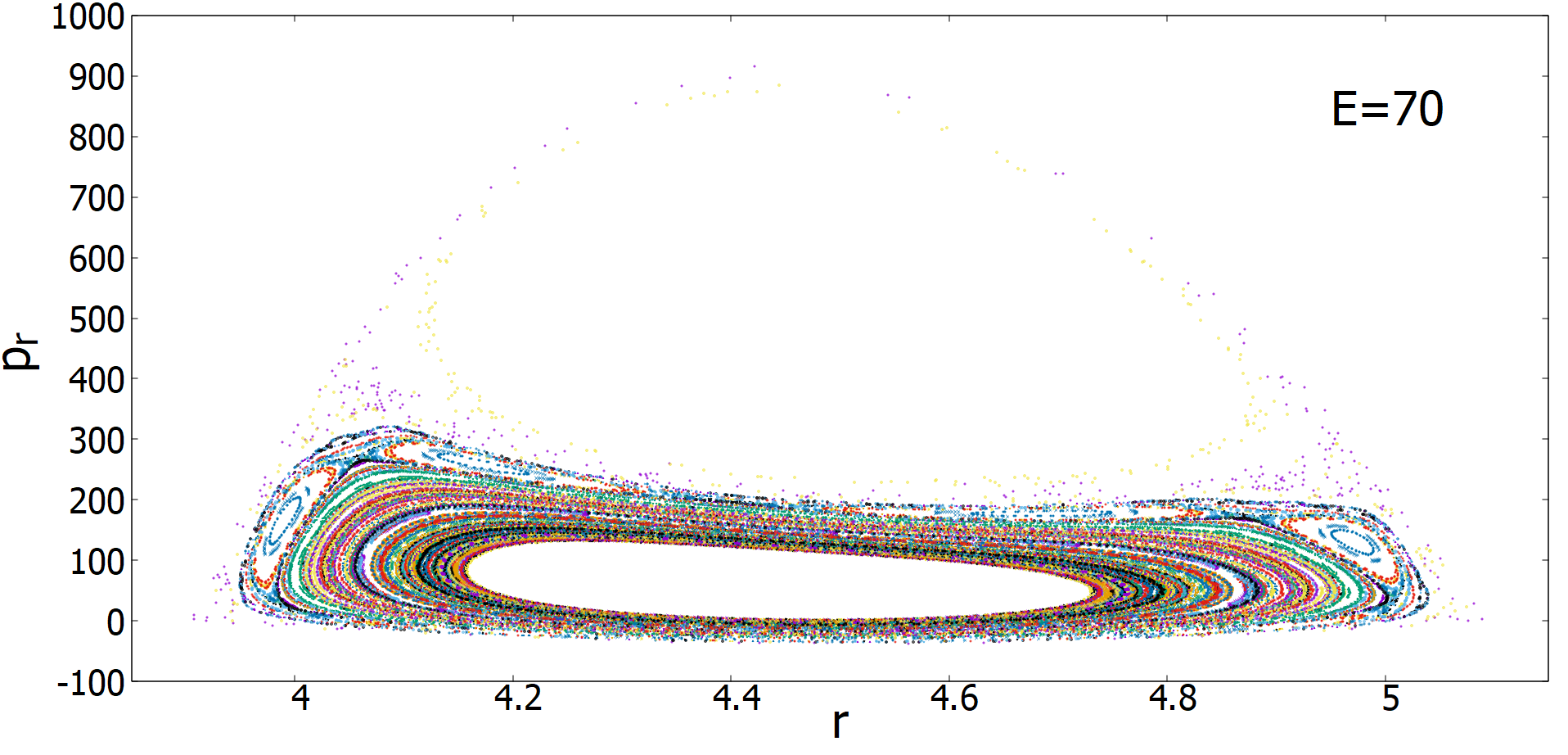}\label{6b}}
	\subfigure[]               
        {\includegraphics[width=0.7\linewidth,height=0.5\linewidth]{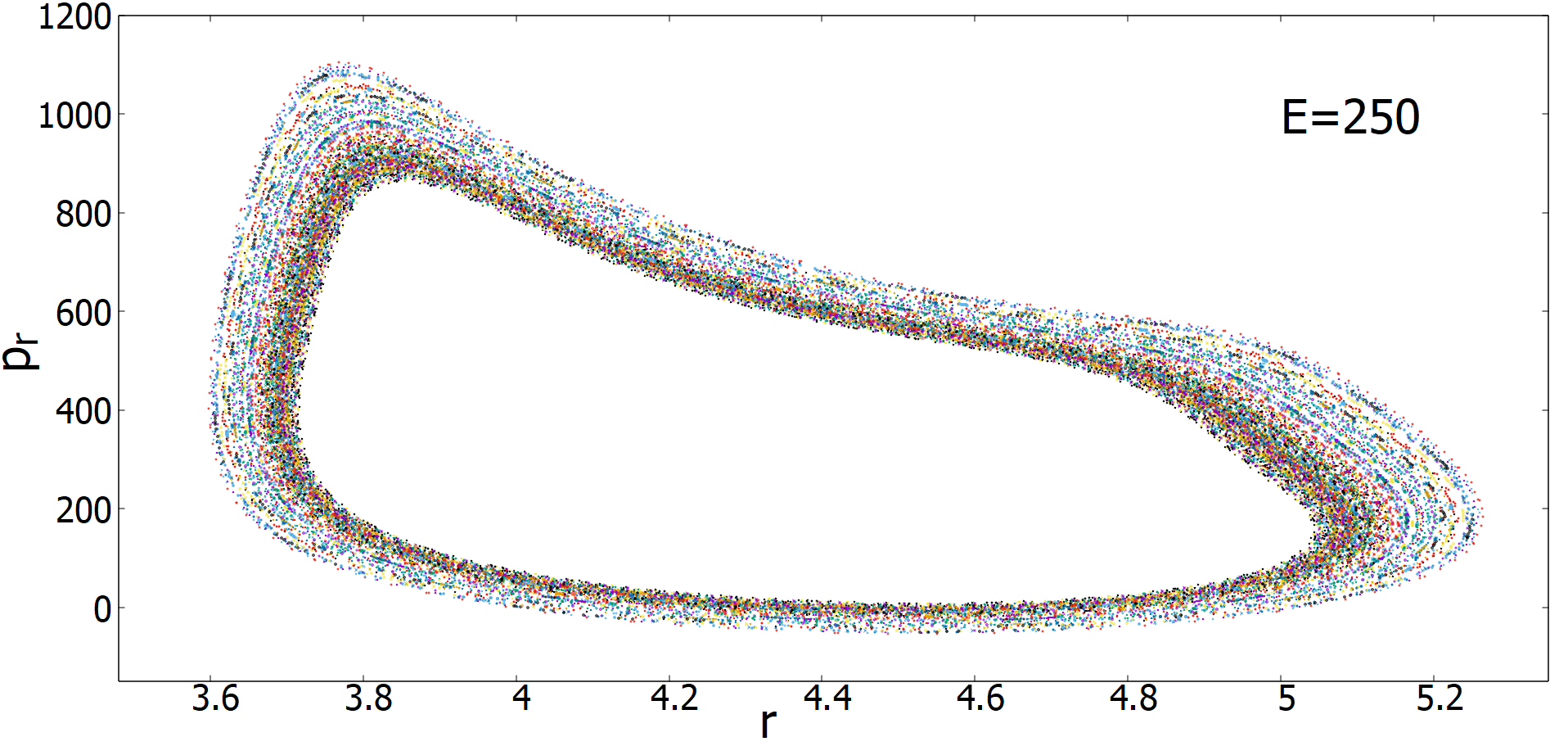}\label{6c}}\\
	\subfigure[] 
        {\includegraphics[width=0.7\linewidth,height=0.5\linewidth]{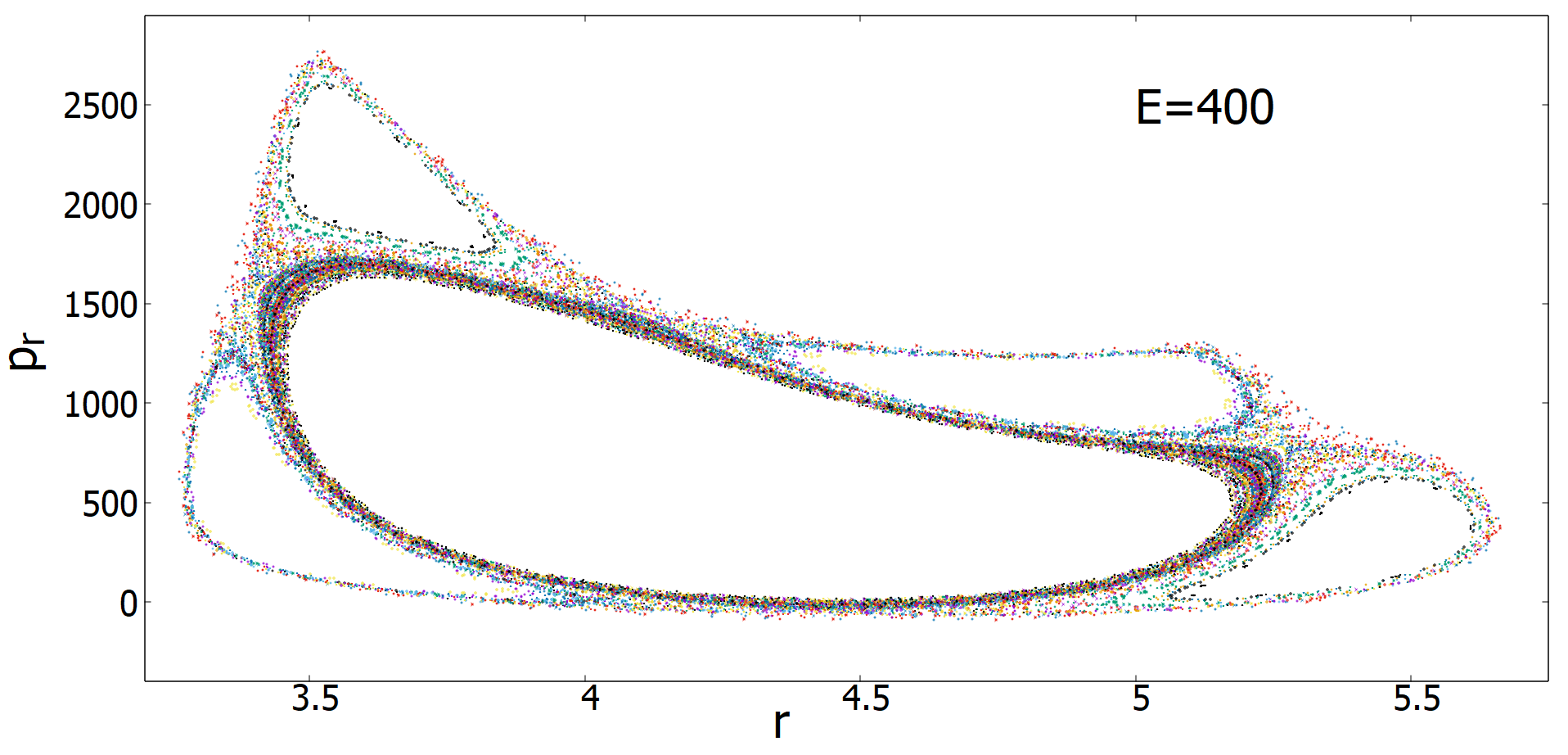}\label{6d}}
	\subfigure[] 
        {\includegraphics[width=0.7\linewidth,height=0.5\linewidth]{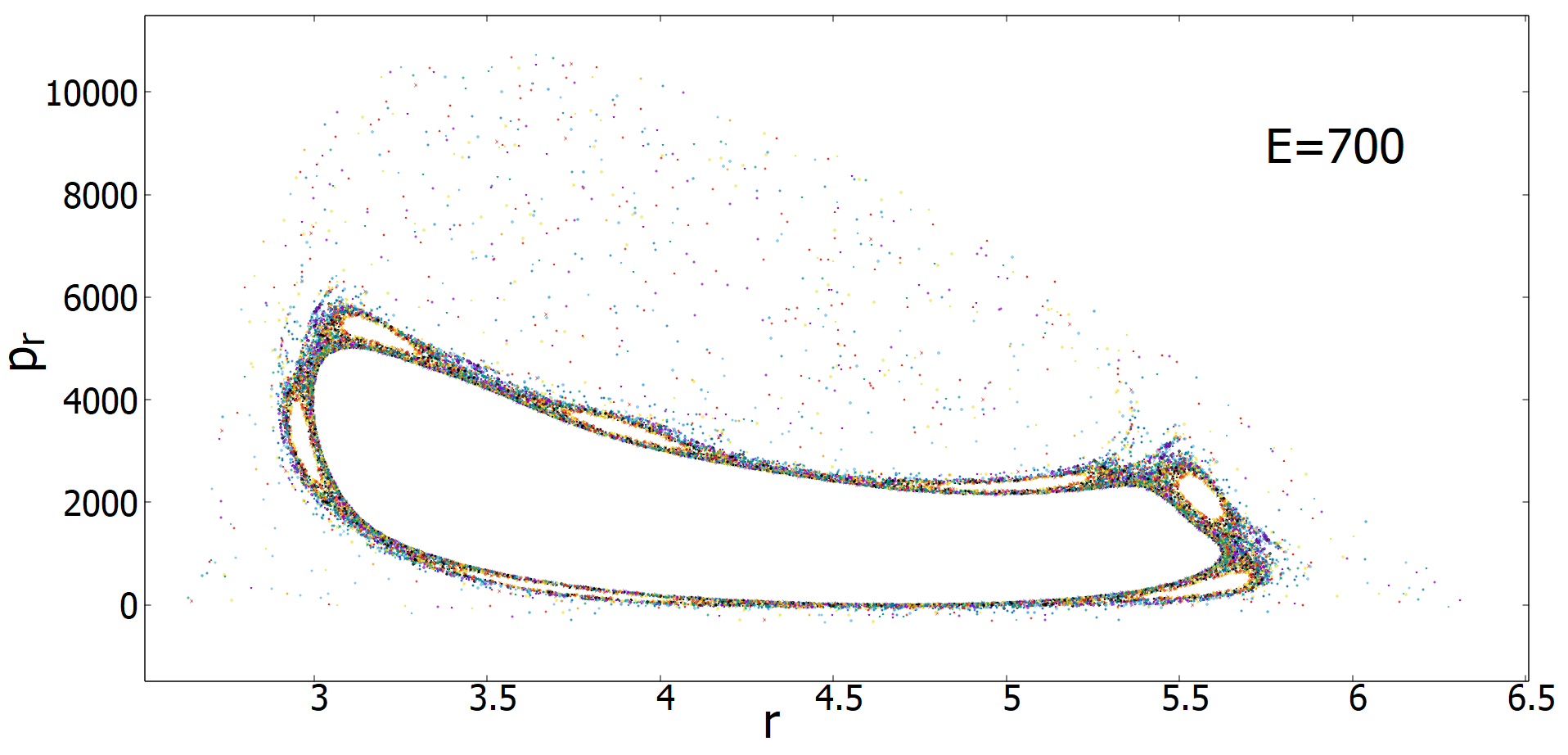}\label{6e}}
        \subfigure[]{\includegraphics[width=0.7\linewidth,height=0.5\linewidth]{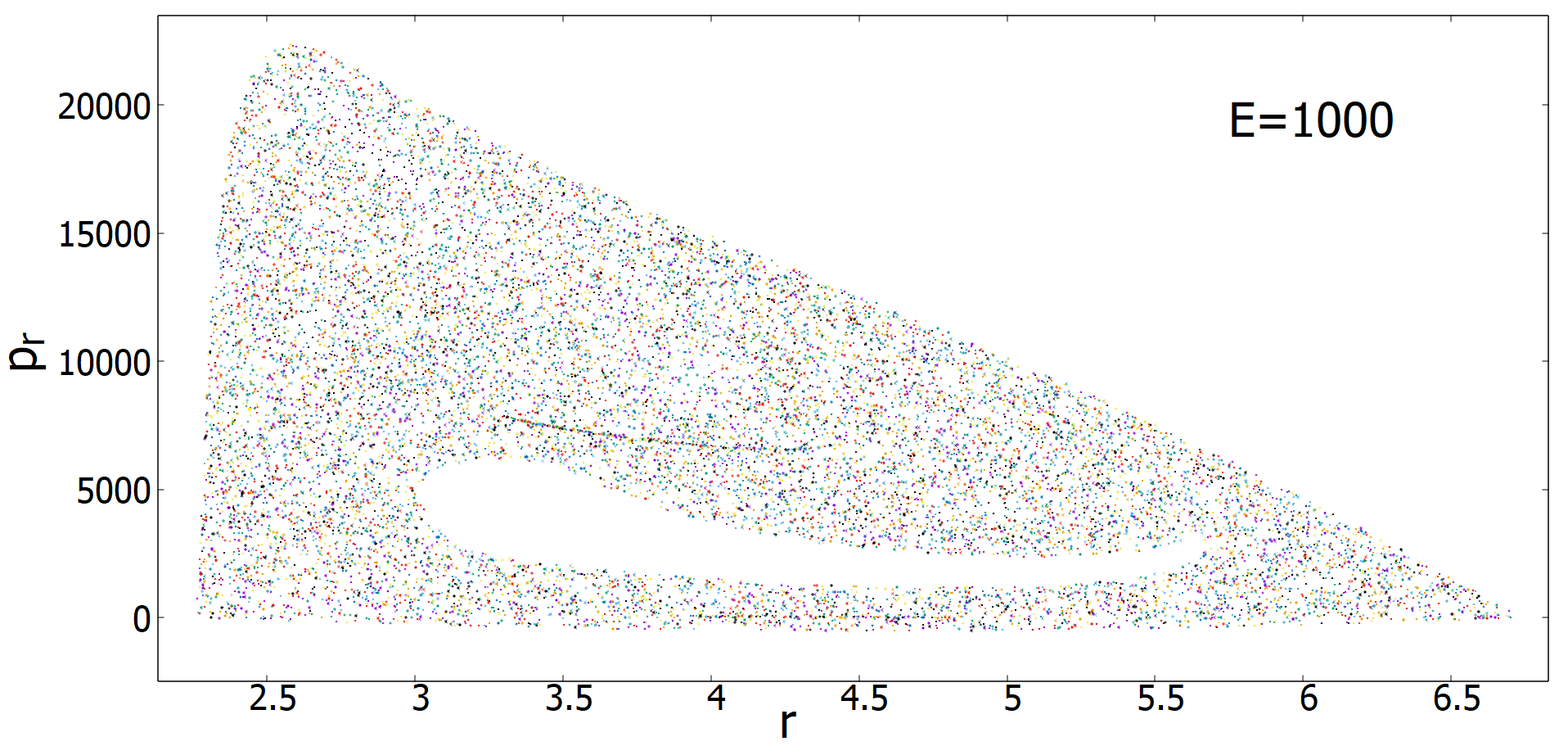}\label{6f}}
        \end{array}$
        \end{center}
        \begin{minipage}{\textwidth}
        \caption{The Poincar$\Acute{e}$ sections in the $(r-p_r)$ plane with $\theta=0$ and $p_{\theta}>0$ for different energies with fixed dimensional parameter $a=0.166$ for the SSS charged black hole.}\label{f6}   
        \hrulefill
        \end{minipage}
        \begin{center} 
        $\begin{array}{ccc}
	\subfigure[] 
        {\includegraphics[width=0.7\linewidth,height=0.5\linewidth]{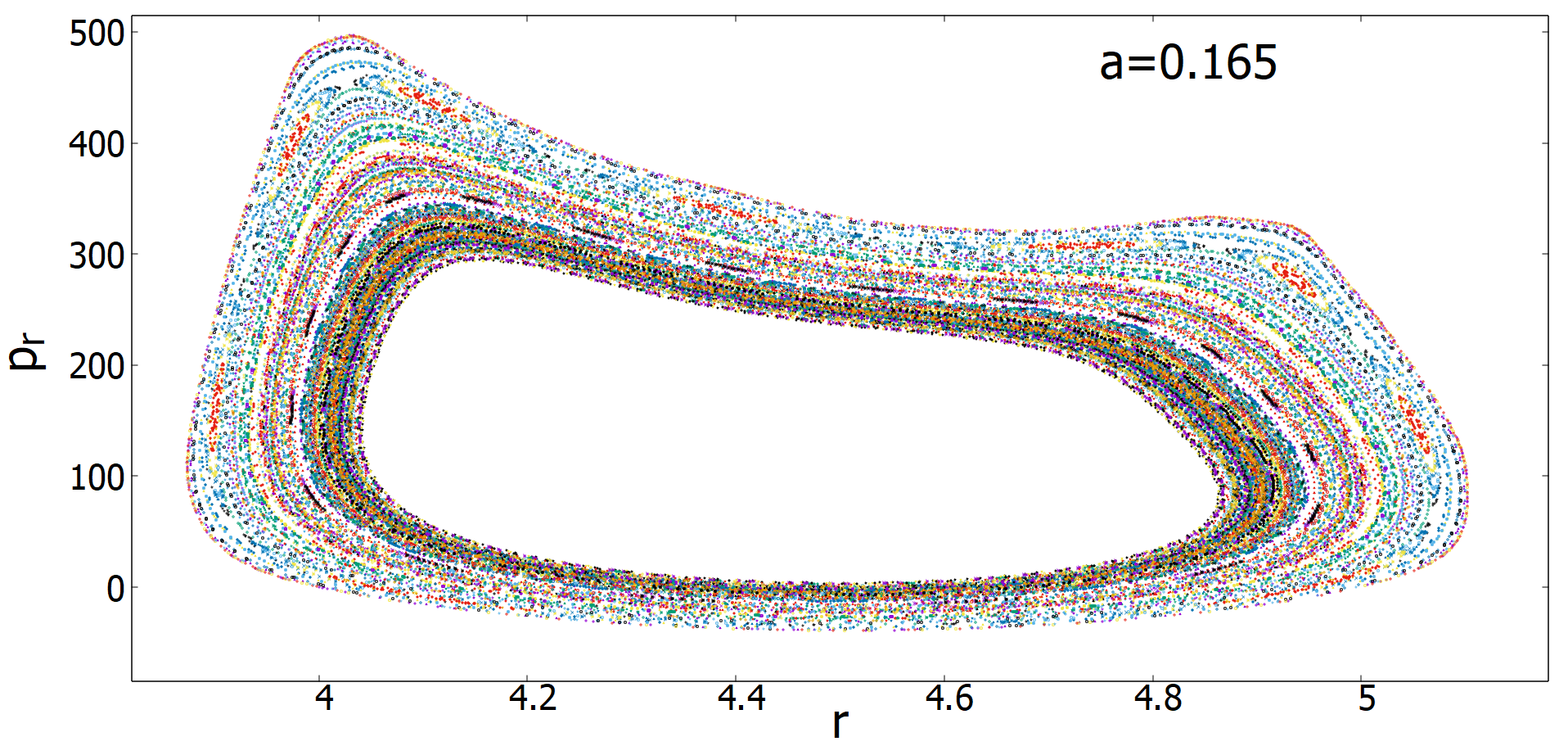}\label{7a}}
        \subfigure[] 
        {\includegraphics[width=0.7\linewidth,height=0.5\linewidth]{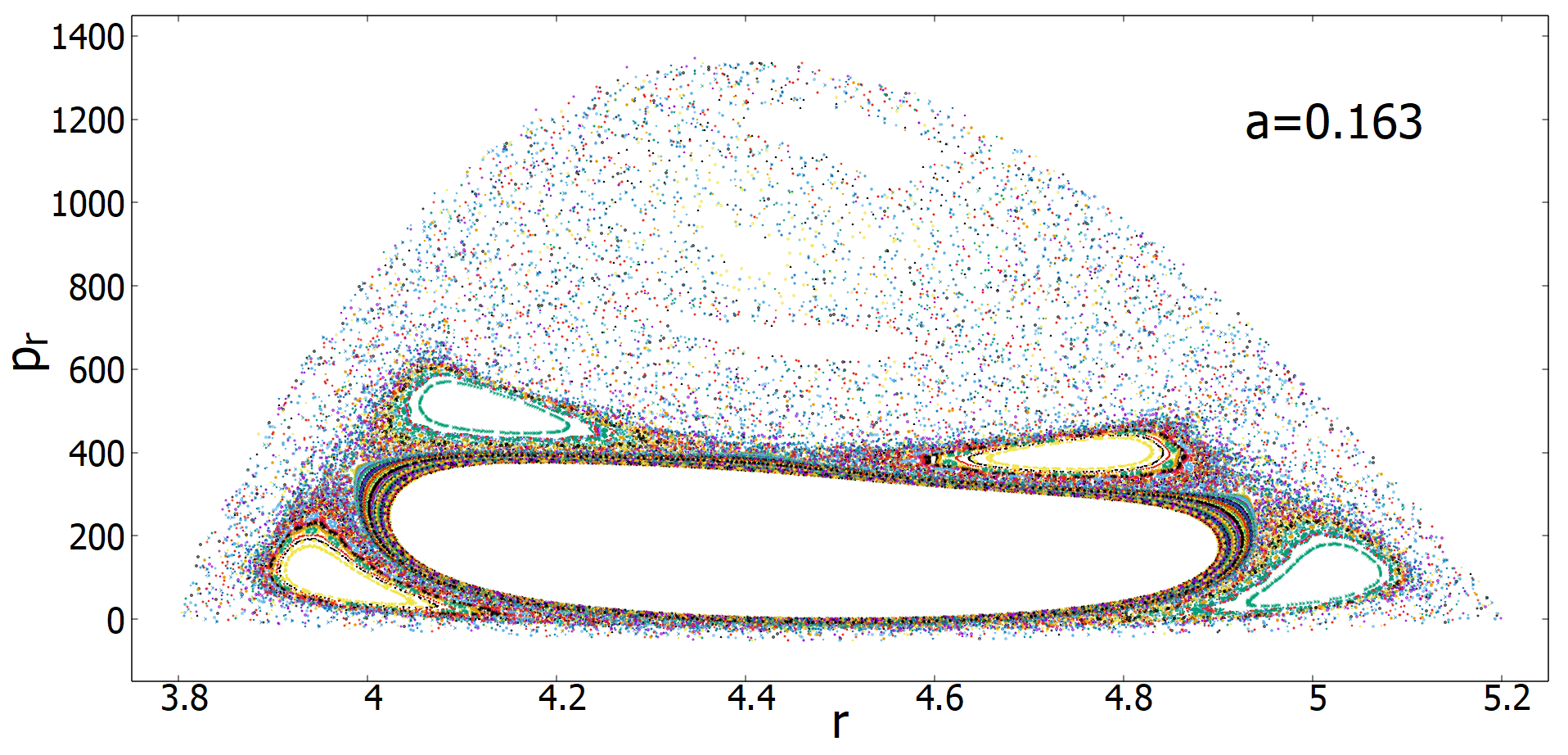}\label{7b}}
        \subfigure[]{\includegraphics[width=0.7\linewidth,height=0.5\linewidth]{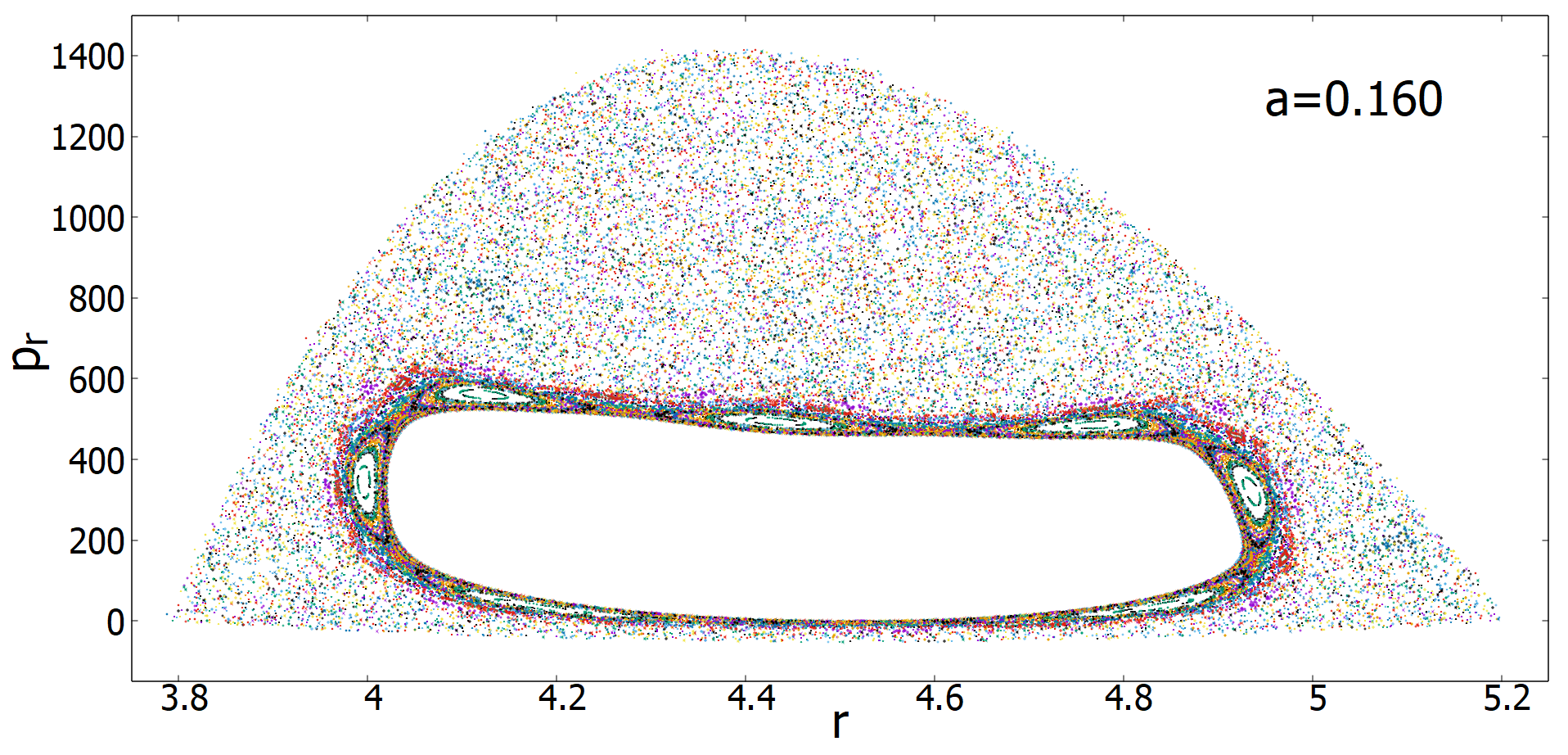}\label{7c}}\\
        \subfigure[]{\includegraphics[width=0.7\linewidth,height=0.5\linewidth]{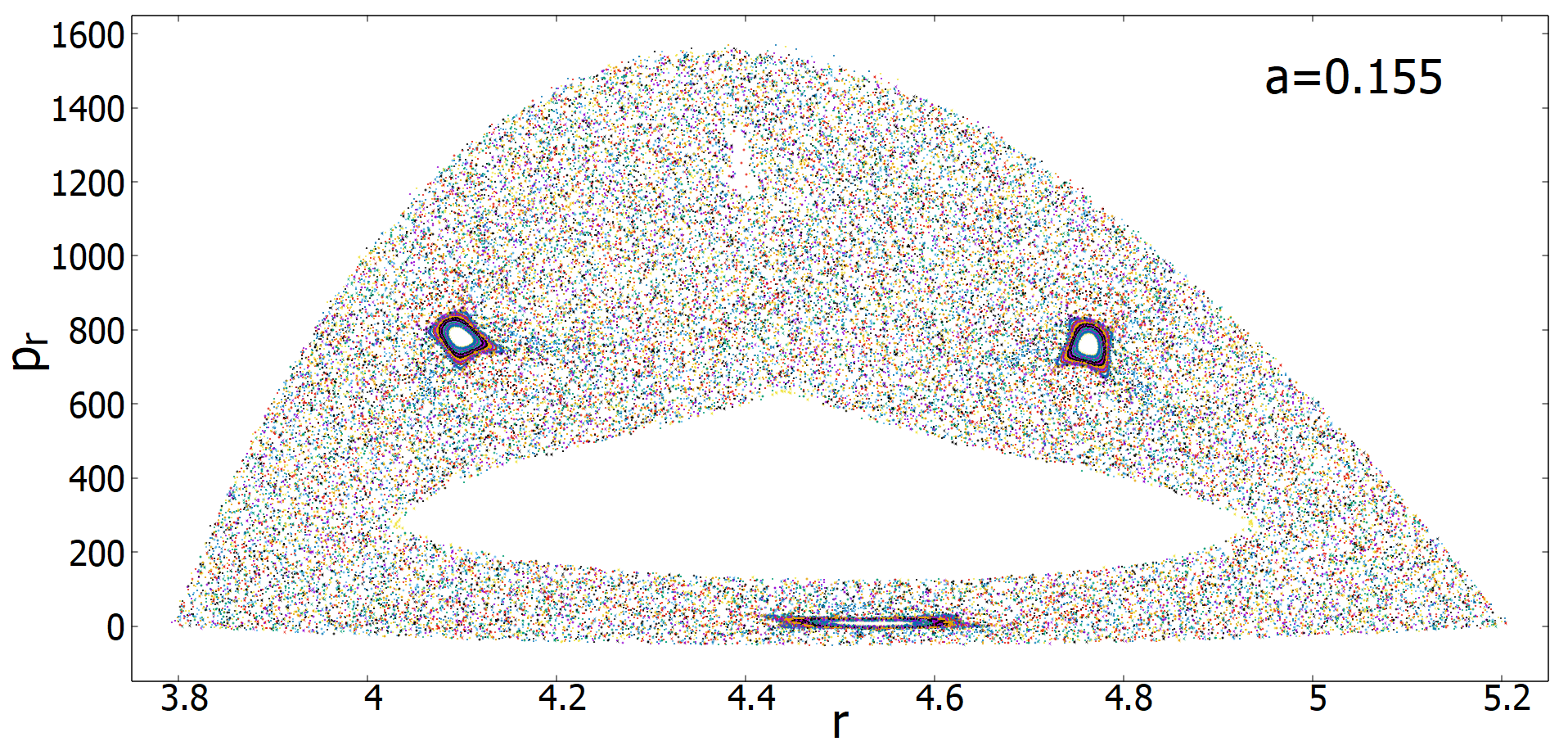}\label{7d}}
        \subfigure[]{\includegraphics[width=0.7\linewidth,height=0.5\linewidth]{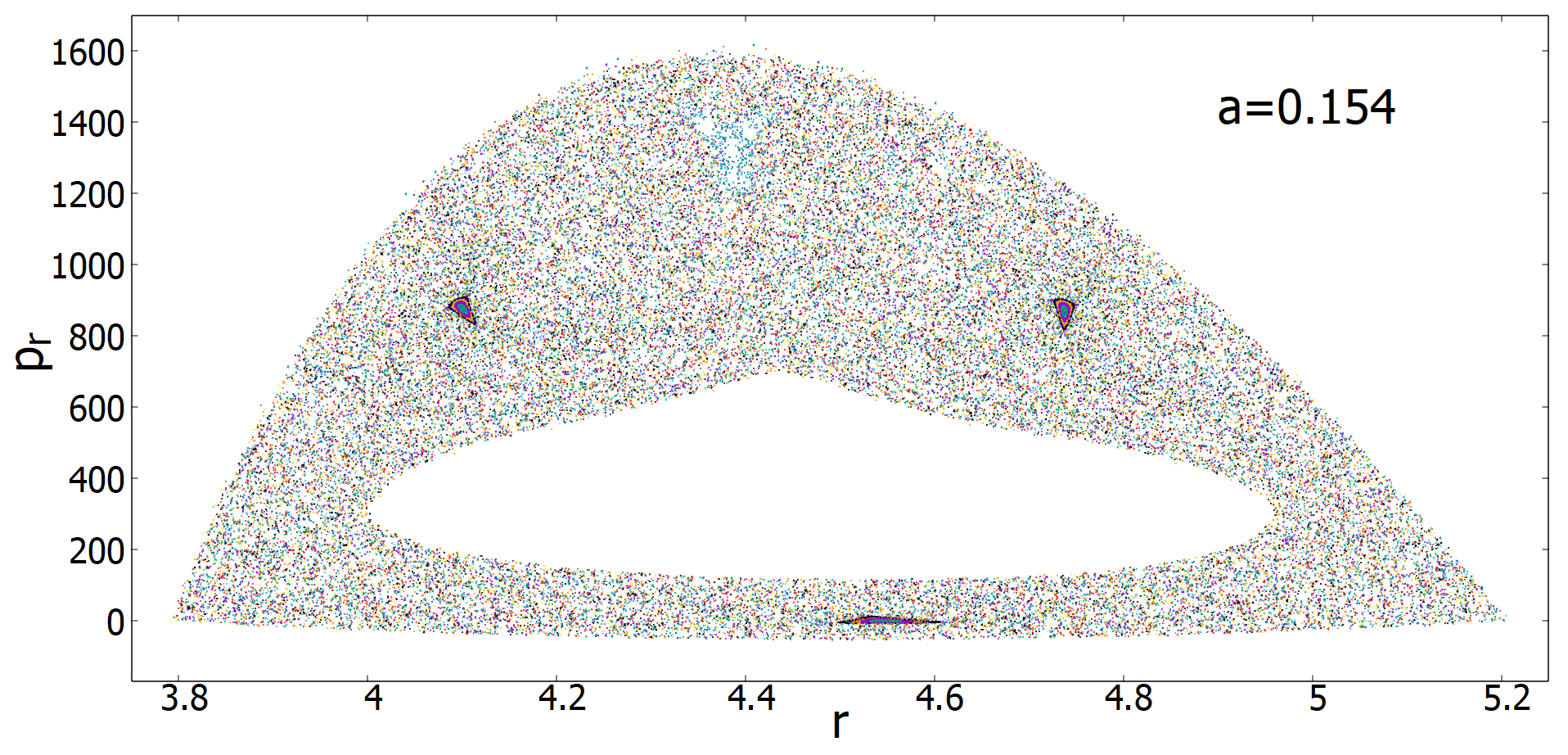}\label{7e}}
	\end{array}$
	\end{center}
        \begin{minipage}{\textwidth}
        \caption{The Poincar$\Acute{e}$ sections in the $(r-p_r)$ plane with $\theta=0$ and $p_{\theta}>0$ along with different dimensional parameter $a$ for fixed energy $E=100$ for the SSS charged black hole.}\label{f7}     
        \end{minipage}
        \end{figure}
    \end{widetext}

    In Sec. \ref{s2b-a}, we previously established the upper limit for the modified gravity parameter $a$ as $\frac{1}{6}$. Keeping this in mind, we construct two sets of Poincar$\Acute{e}$ sections. For the first set (Fig.\ref{f6}), we fix the modified gravity parameter to the maximum possible value  $a=\frac{1}{6}$ and tune the energy to the values $E=30,70,250,400,700$ and $E=1000$. Figure \ref{6a} clearly depicts that for low energy ($E=30$), the Poincar$\Acute{e}$ sections exhibit regular KAM tori, indicating that the corresponding orbit remains predominantly confined near the centre of the harmonic potential, set at $r_c=4.5$. Similar tori patterns are observed in Fig.\ref{6b} and Fig.\ref{6c} for $E=70$ and $250$, respectively. However due to the conservation of the Hamiltonian, as the total energy of the system increases, the momentum of the system also increases. Consequently, the trajectory of the particle tends to approach the black hole event horizon located at $r_{+}=2.13$ for the fixed dimensional parameter $a=0.166$. This trend becomes evident in the Poincar$\Acute{e}$ sections for $E=400$ and $E=700$ [Fig.\ref{6d} and Fig.\ref{6e}], where the KAM tori begin to distort and appear pinched. Further increasing the energy to $E=1000$ [Fig.\ref{6f}] leads to the complete breakdown of the regular tori, resulting in a random distribution of points across the phase plane. This observation indicates a shift of the corresponding orbits towards the event horizon as the energy increases, along with the disruption of the regular tori. 
    
    For the second set of figures (Fig.\ref{f7}), we fix the energy $E=100$ and tune $a$ to the values $a=0.165$, $0.163$, $0.160$, $0.155$, and $0.154$.  Similar to the effect of the modified gravity parameter $\beta$ in model I, decreasing $a$ results in an increase in the horizon radius, eventually leading to closer proximity of the particle trajectory with respect to the event horizon. Consequently, the system begins to interact with the horizon, inducing chaotic behavior.  This characteristic feature of the Poincar$\Acute{e}$ sections again supports the chaotic nature of the particle trajectory near the charged black hole's horizon, similar to our conclusions in model I.

        \subsubsection{Model II: Neutral black hole background}\label{s4b-3}
    In this section, we will explore the horizon effect on chaos in the context of model II SSS neutral black hole solution.

    For the neutral black hole solution described by Eq.\eqref{25}, the dimensional parameter $a$ does not have an upper bound, as the event horizon radius is determined by $r_H=\frac{2}{3a}$, where $a>0$. Larger values of $a$ correspond to smaller sizes of the black hole. Just like the previous investigations, we again solve the associated dynamical equations of motion \Big[refer to Eq.\eqref{3.12}, Eq.\eqref{3.13}, Eq.\eqref{3.14}, and Eq.\eqref{3.15}\Big], employing the Runge-Kutta fourth-order scheme. The initial conditions for $r$ is randomly chosen in the range $3.7<r<4.6$, and all the rest of the initial conditions are selected following a similar approach as discussed for the model II SSS charged black hole with $K_r=400$, $K_{\theta}=75$, $\theta_c=0$, and $r_c=4.2$.
    
\begin{widetext}
        \begin{figure}[H]
	\centering
	\begin{center} 
	$\begin{array}{ccc}
	\subfigure[] 
        {\includegraphics[width=0.7\linewidth,height=0.5\linewidth]{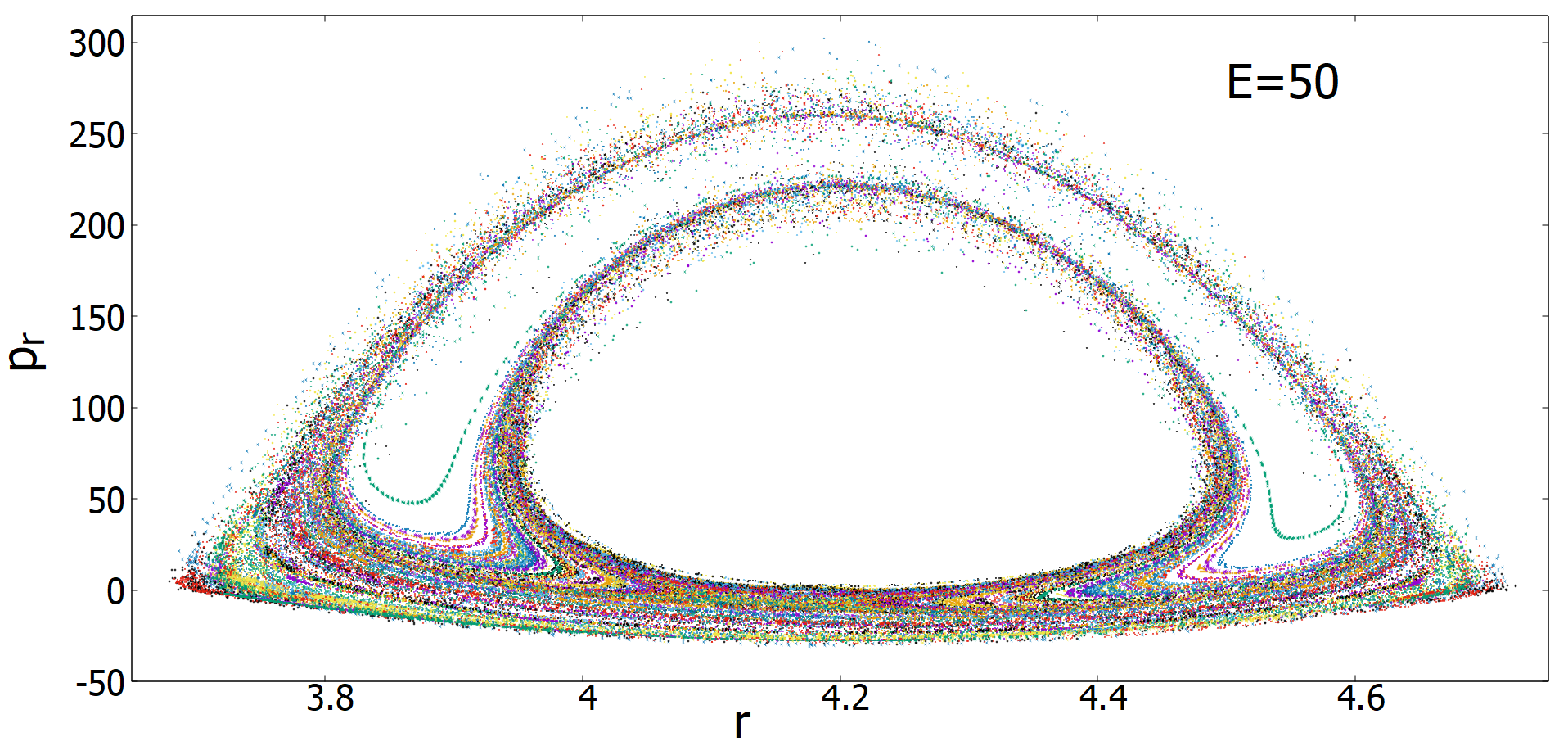}\label{8a}}
        \subfigure[]{\includegraphics[width=0.7\linewidth,height=0.5\linewidth]{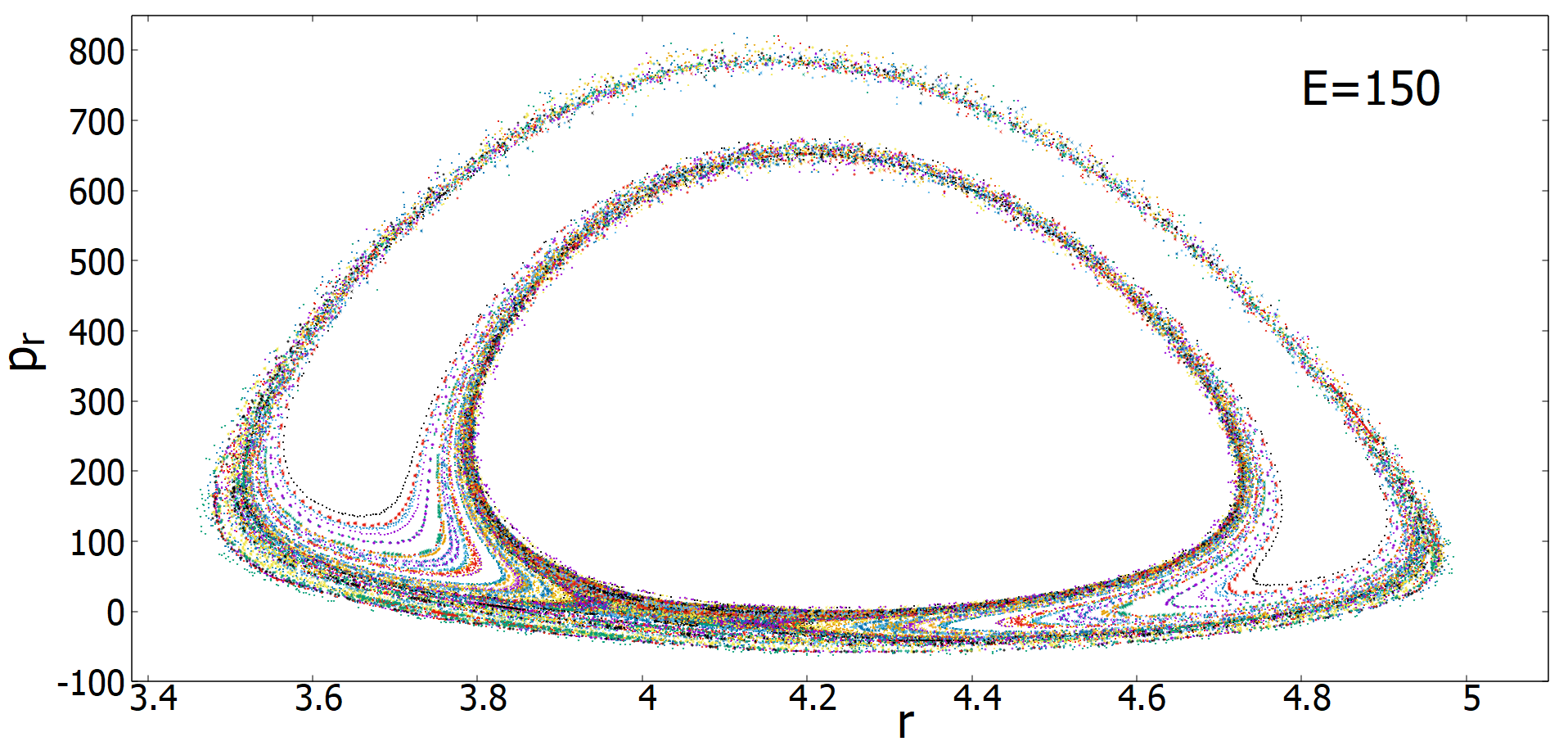}\label{8b}}
	\subfigure[] 
        {\includegraphics[width=0.7\linewidth,height=0.5\linewidth]{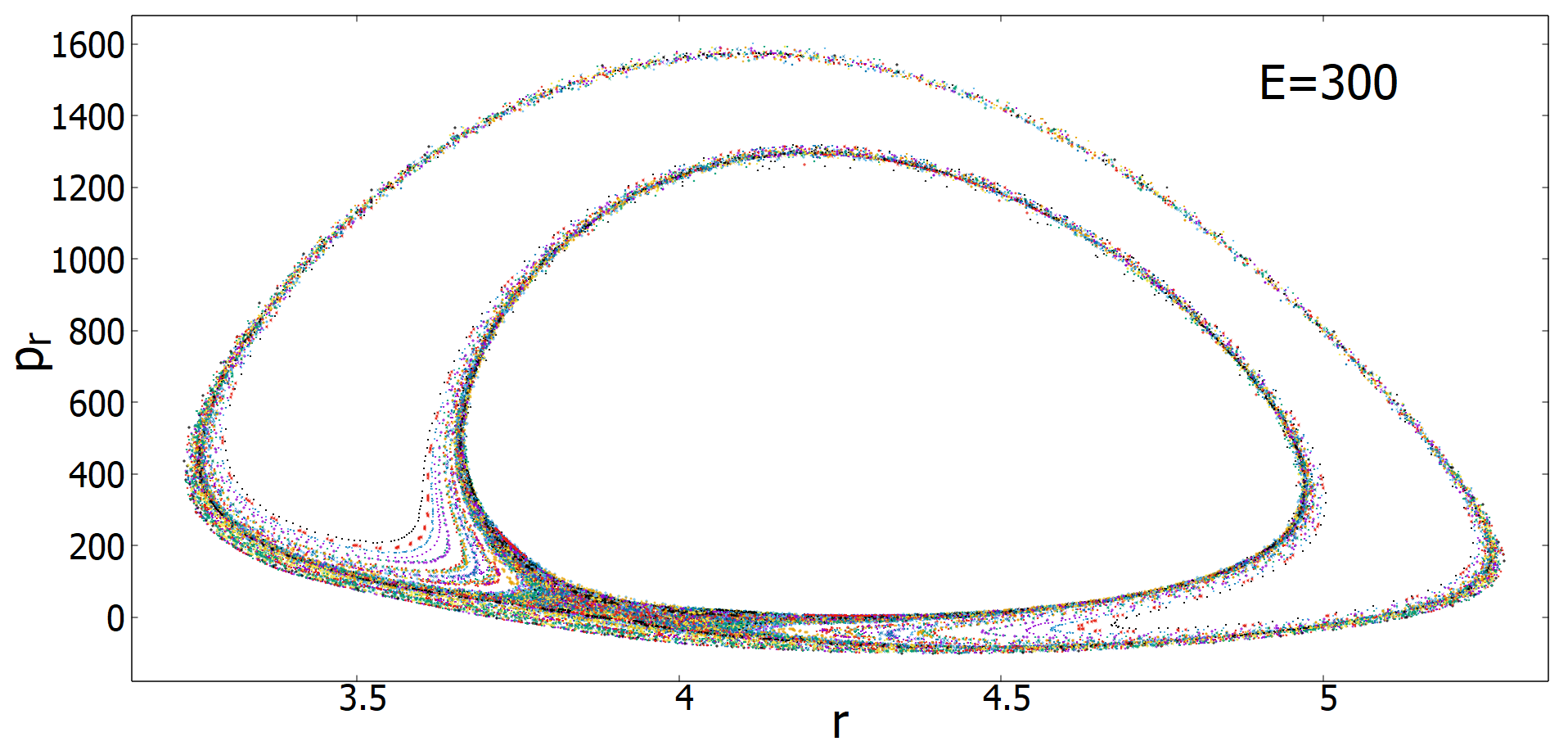}\label{8c}}\\
	\subfigure[] 
        {\includegraphics[width=0.7\linewidth,height=0.5\linewidth]{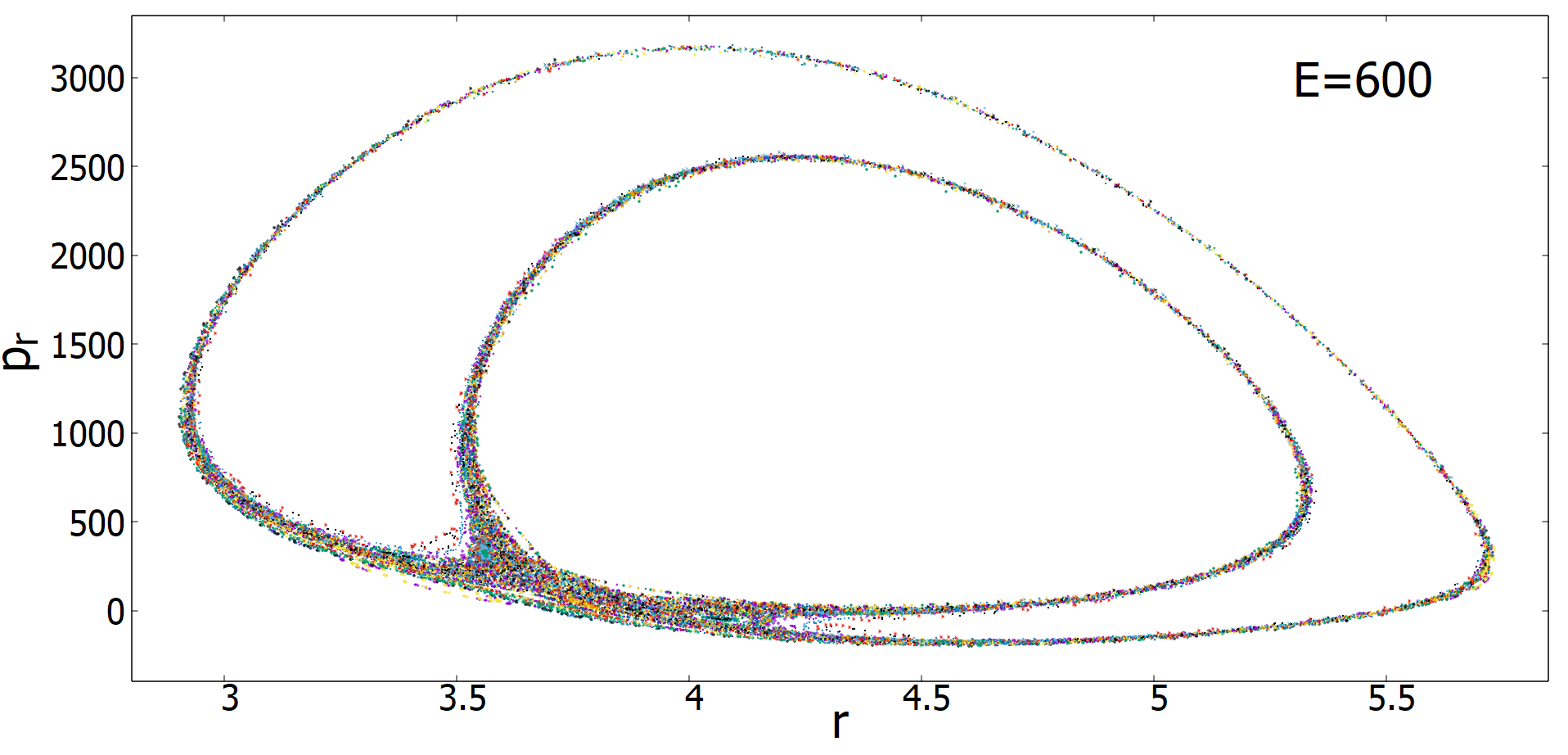}\label{8d}}
	\subfigure[] 
        {\includegraphics[width=0.7\linewidth,height=0.5\linewidth]{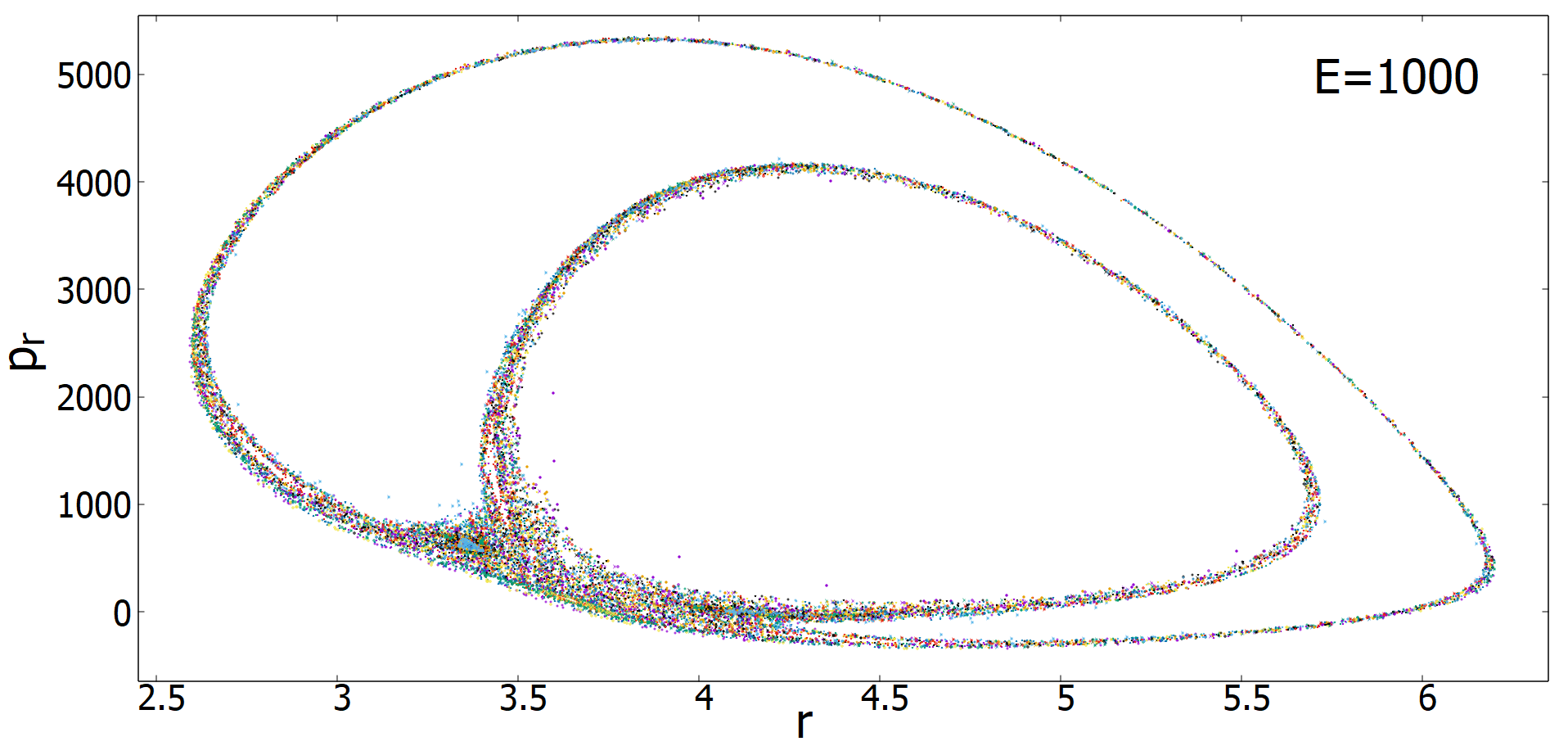}\label{8e}}
        \subfigure[] 
        {\includegraphics[width=0.7\linewidth,height=0.5\linewidth]{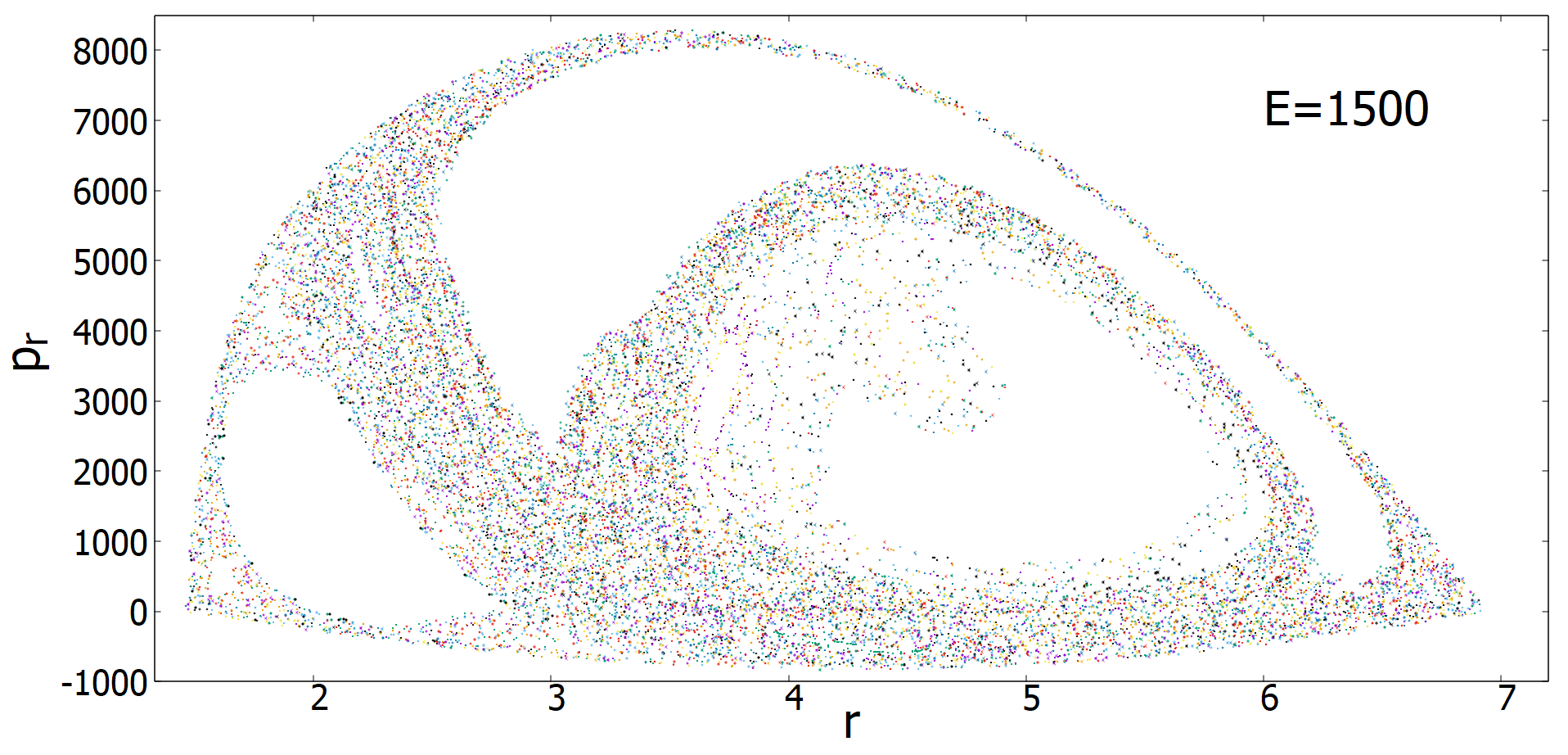}\label{8f}}
	\end{array}$
        \end{center}
        \begin{minipage}{\textwidth}
        \caption{The Poincar$\Acute{e}$ sections in the $(r-p_r)$ phase plane with $\theta=0$ and $p_{\theta}>0$ for different energies with fixed dimensional parameter $a=0.5$ for the SSS neutral black hole.}\label{f8}
        \hrulefill
        \end{minipage}
        \begin{center}
        $\begin{array}{ccc}
	\subfigure[] 
        {\includegraphics[width=0.7\linewidth,height=0.5\linewidth]{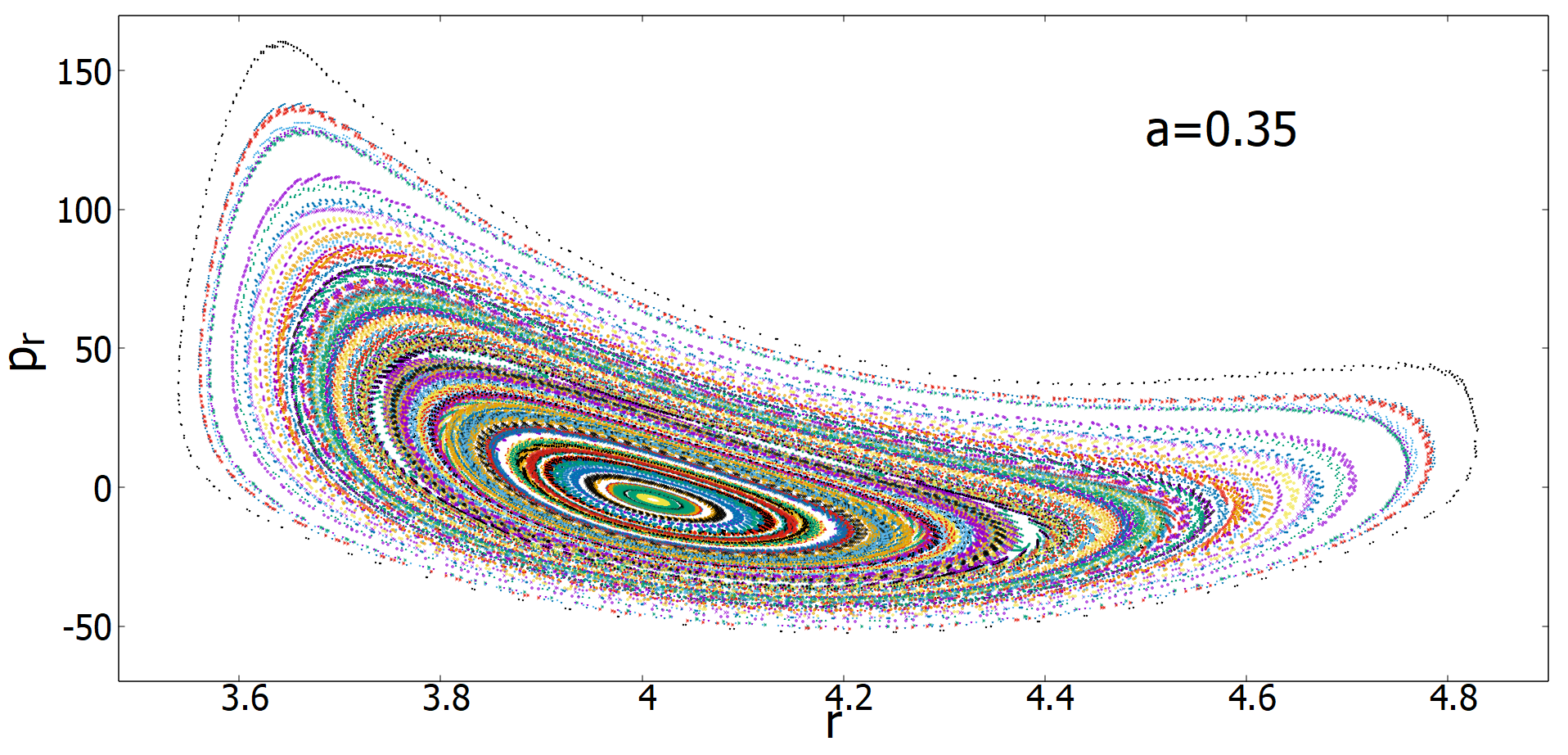}\label{9a}}
        \subfigure[]{\includegraphics[width=0.7\linewidth,height=0.5\linewidth]{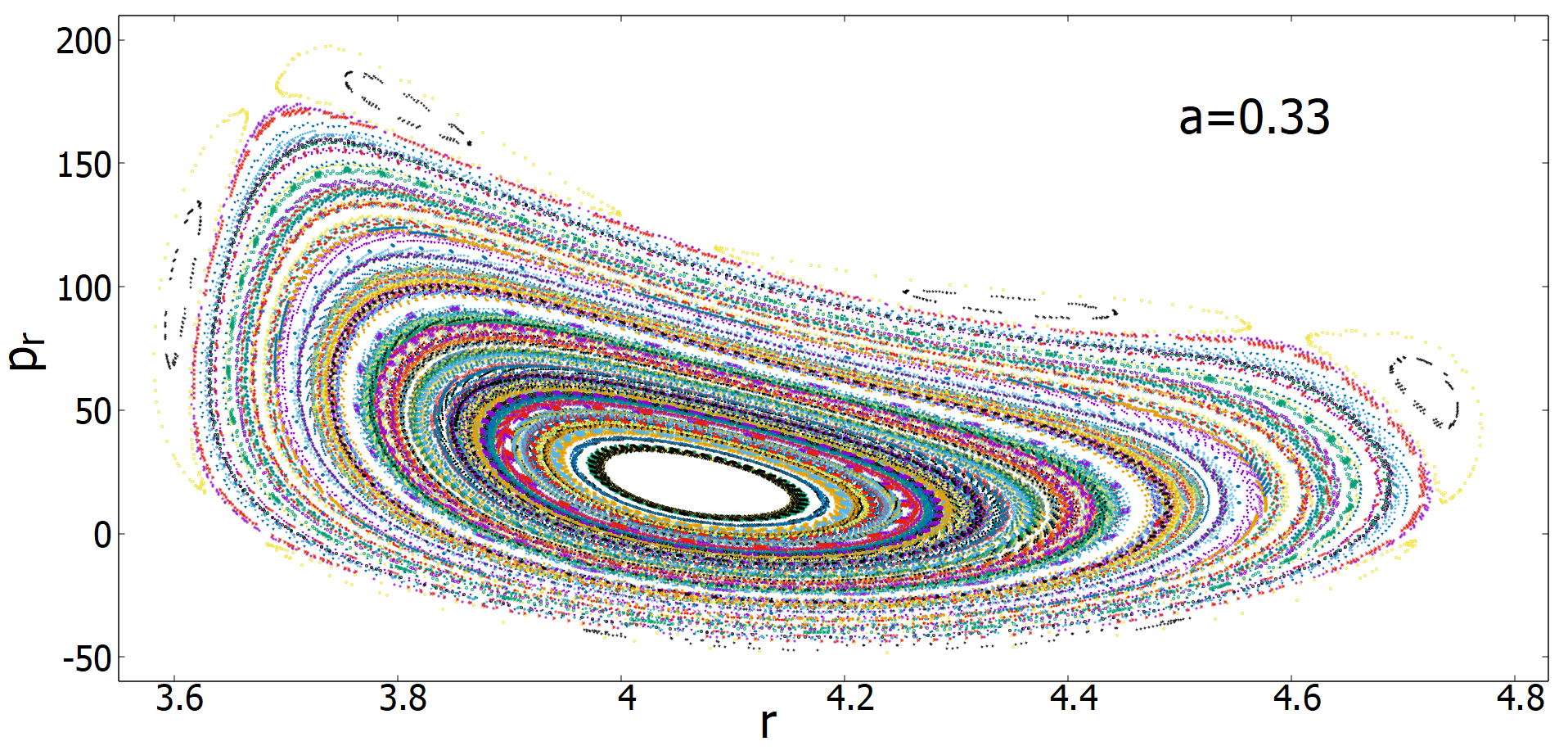}\label{9b}}
	\subfigure[] 
        {\includegraphics[width=0.7\linewidth,height=0.5\linewidth]{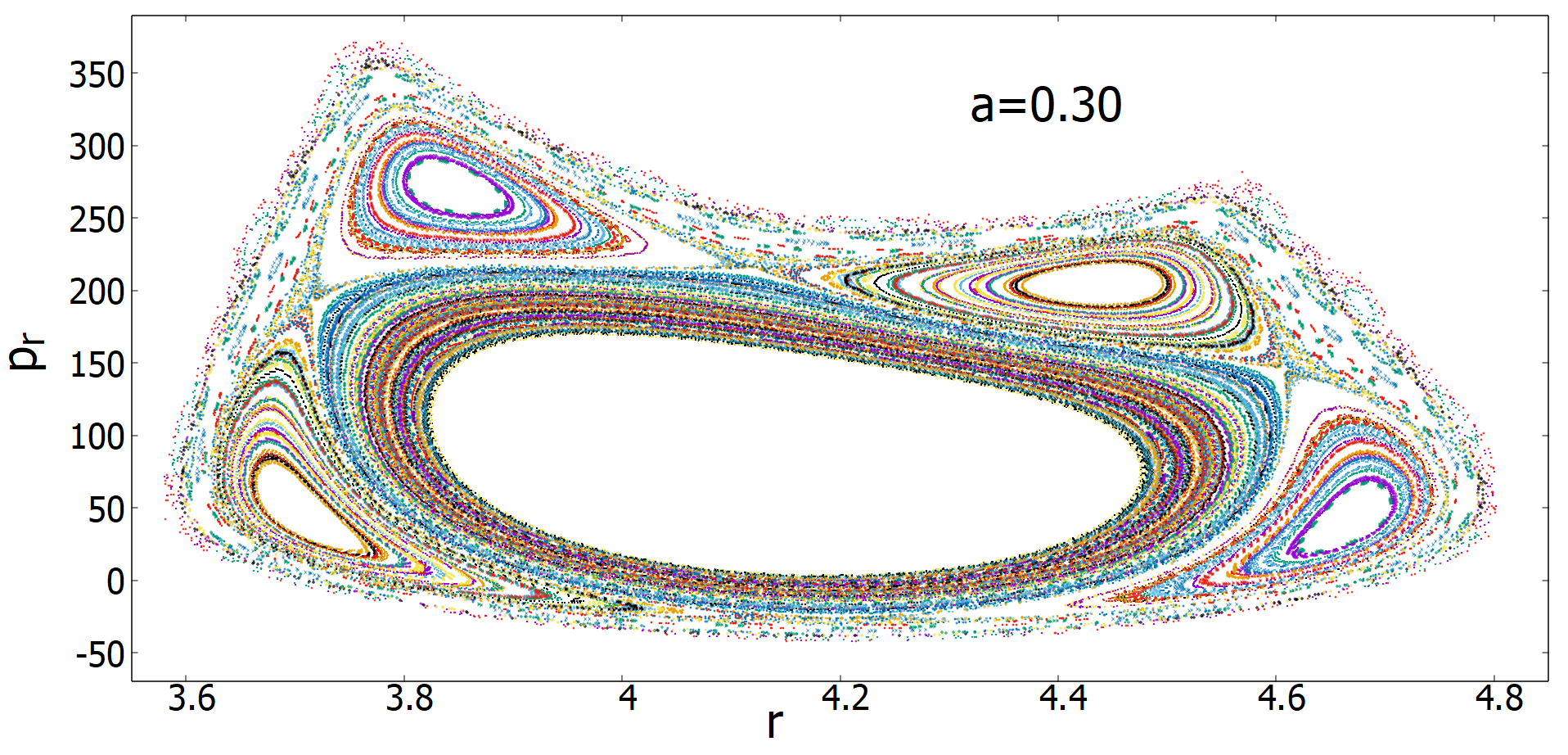}\label{9c}}\\
        \subfigure[]{\includegraphics[width=0.7\linewidth,height=0.5\linewidth]{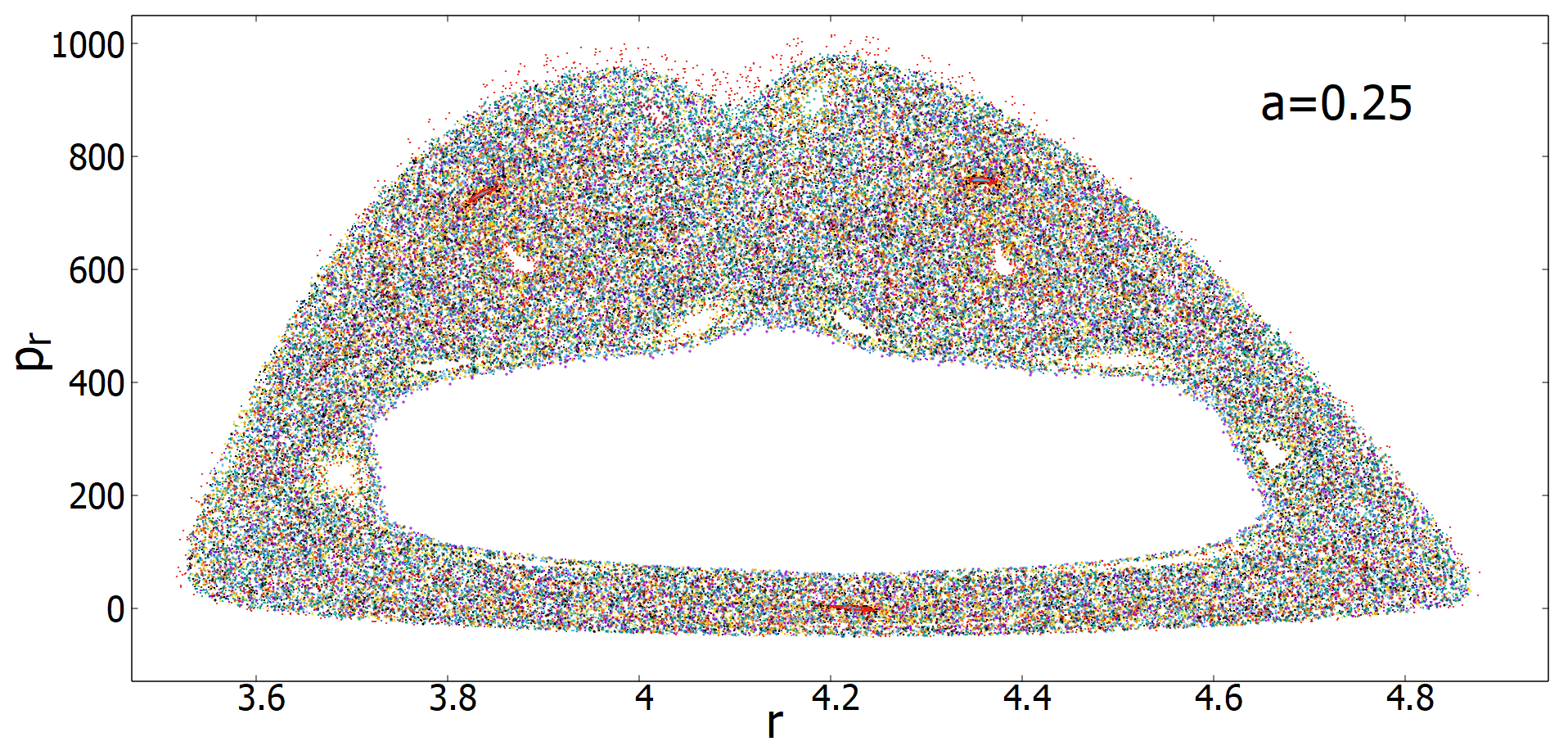}\label{9d}}
        \subfigure[]{\includegraphics[width=0.7\linewidth,height=0.5\linewidth]{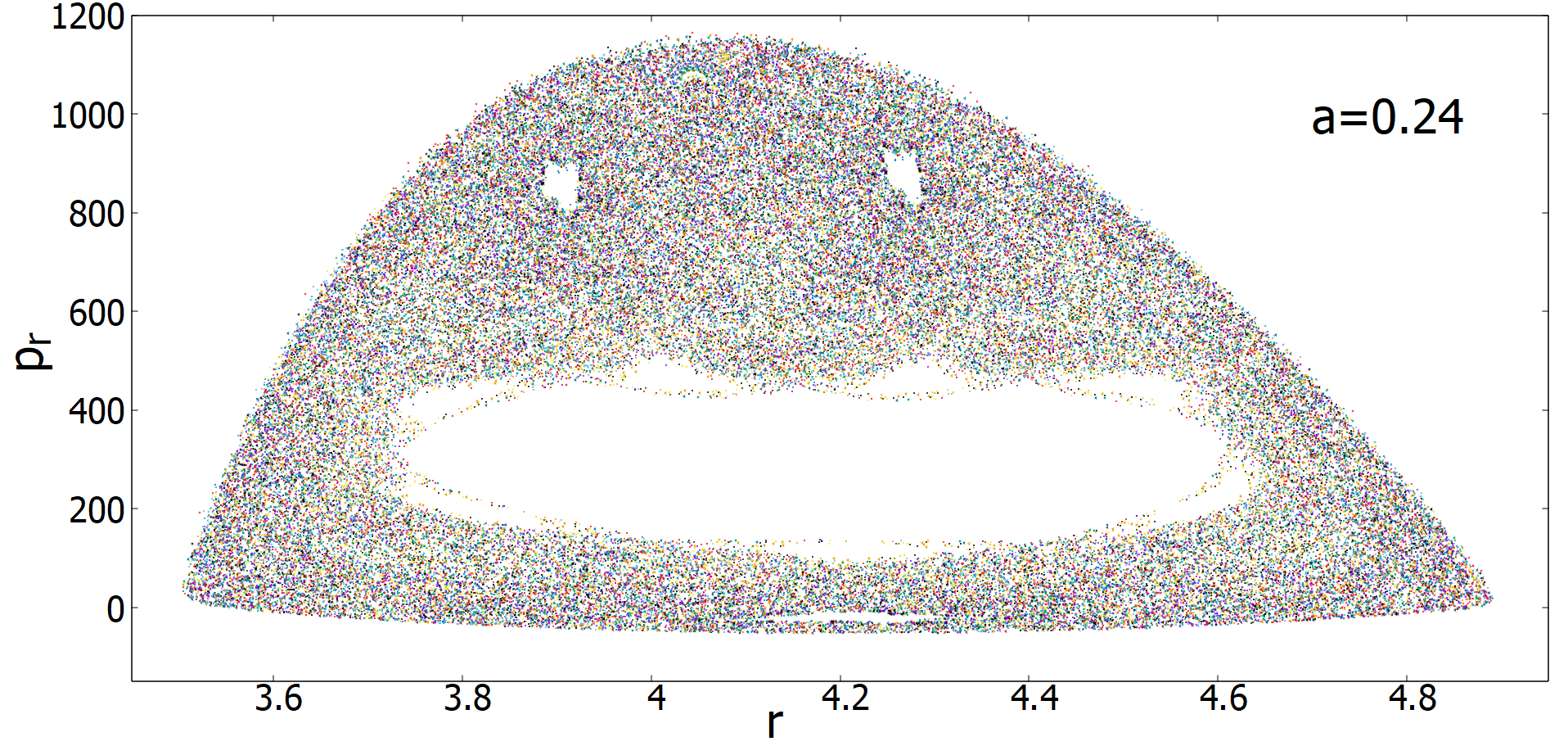}\label{9e}}
        \end{array}$
        \end{center}
        \begin{minipage}{\textwidth}
        \caption{The Poincar$\Acute{e}$ sections in the $(r-p_r)$ phase plane with $\theta=0$ and $p_{\theta}>0$ for different dimensional parameters $a$ with fixed energy $E=100$ for the SSS neutral black hole.}\label{f9}
        \end{minipage}
        \end{figure}
    \end{widetext}

    In Fig.\ref{f8}, we present Poincar$\Acute{e}$ sections corresponding to various energies while for $a=0.5$. For Fig.\ref{8a}, we notice the emergence of regular closed curves at lower energy levels, specifically $E=50$. However, as the energy rises to $E=150$, these closed curves undergo compression alongside the emergence of scattered points, leading to the disappearance of certain regions within the plot.

    As we increase $E$, the regular orbits begin to destabilize. At very high energy ($E=1500$), depicted in Fig.\ref{8f}, the radial trajectory of the particle approaches the event horizon. This results in the breakdown of closed orbits, clearly illustrated by the scattered points filling the phase plane. The dynamic motion of the system exhibits chaotic behavior, evident in the increasing radial momentum components with rising energy, akin to the previously discussed scenario. It is worth noting that in the case of a charged SSS black hole, chaotic behavior manifests at a relatively lower energy ($E=1000$) compared to the neutral black hole solution. This observation is a manifestation of the additional term $\frac{1}{3ar^2}$ that arises in the charged black hole solution [Eq.\eqref{21}]. 
    
    Next, we investigate the trajectories' by varying the dimensional parameter $a$ while keeping the energy fixed at $E=100$. In Fig.\ref{f9}, we depict the Poincar$\Acute{e}$ sections on the $(r-p_r)$ plane for various values of $a$, specifically $a=0.35,0.33,0.30,0.25$, and $0.24$. We observe consistent trends:  for higher $a$ values ($a=0.35,0.33$), a confined region of unbroken tori emerges within the phase plane. However, as $a$ decreases, the horizon interacts with the particle trajectory, leading to the onset of chaos.

        \subsection{Analysis of Lyapunov exponents}\label{s4c}
    We now proceed to investigate the Lyapunov exponents (LE) to quantify the chaos observed in the Poincar$\Acute{e}$ sections studied in the previous section. The LE of a dynamical system is defined as the quantity that characterizes the rate of separation of two infinitesimally close trajectories \cite{Sandri}. In this work we will investigate two kinds of LE for model I and model II. The first quantity is the total Lyapunov exponent $(\lambda_{T})$, which is defined as the rate of divergence between two trajectories in the whole phase space ($r,p_{r},\theta,p_\theta$).  The second quantity is denoted by $(\lambda_r)$, which is defined as the rate of divergence only along the radial $(r)$ direction between two trajectories in the phase space.

    Let us note that in this classical setting, there is an upper bound on chaos given by the surface gravity ($\kappa$) of
    
    \newpage

    \newpage

    \noindent
    the black hole \cite{hashimoto}, which arises from universal aspects of particle dynamics close to the event horizon. Over the recent years, several works have reported a violation of this bound either within Einstein gravity  \cite{Gwak,Zhao,Lei,Kan} or in $f(R),~f(T)$ modified theories of gravity \cite{Andrea1,Addazi}. This motivates us to quantify the chaos via a numerical computation of the LEs for model I and model II and compare it with the aforementioned bound. This is particularly interesting in the strong gravity regime in the vicinity of the black hole's horizon where we have started to probe the validity of Einstein gravity over the recent years. An early or delayed onset of chaos in particle dynamics close to the horizon or deviations from the LE predicted from Einstein gravity may act as an indirect probe of departure from Einstein gravity in the strong field regime. With this motivation in mind, we now proceed towards a numerical estimate of the LEs in model I and model II.

        \subsubsection{Model I}\label{s4c-1}
\begin{widetext}
        \begin{figure}[H]
	\centering
	\begin{center} 
	$\begin{array}{cc}
	\subfigure[]
        {\includegraphics[width=1.0\linewidth,height=0.6\linewidth]{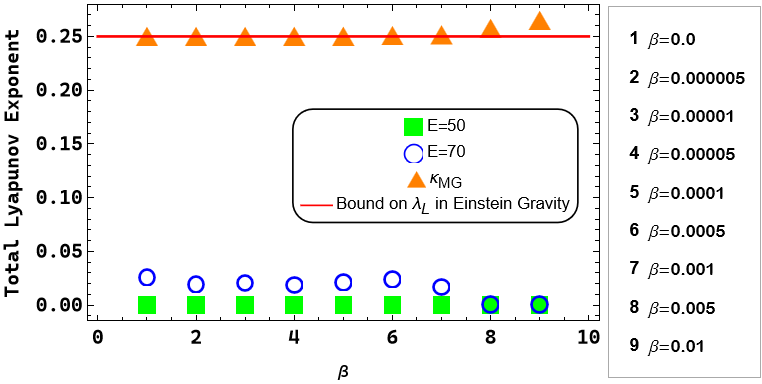}\label{10a}}
        \qquad
        \subfigure[]{\includegraphics[width=1.0\linewidth,height=0.6\linewidth]{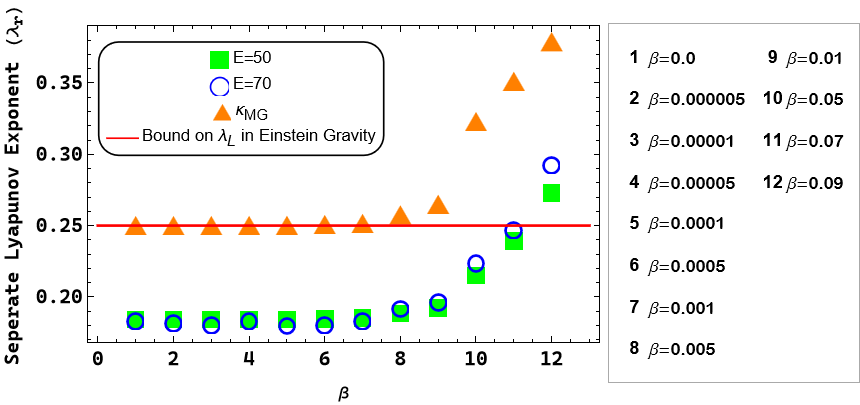}\label{10b}}
        \end{array}$
        \end{center}
        \begin{minipage}{\textwidth}
        \caption{Here, Fig.\ref{10a} represents the variation of total Lyapunov exponent $(\lambda_T)$ with modification parameter $\beta$ for two different energies, $E=50$ and $E=70$ in model I. Figure \ref{10b} represents the variation of separate Lyapunov exponent along radial direction $(\lambda_r)$ with modification parameter $\beta$ for two different energies, $E=50$ and $E=70$ in model I. For both the figures, we set the points of $\beta$ axis as given in the right panel of each figure.}\label{f10}
        \hrulefill
	\end{minipage}
        \begin{center}
        $\begin{array}{cc}
        \subfigure[] 
        {\includegraphics[width=1.0\linewidth,height=0.6\linewidth]{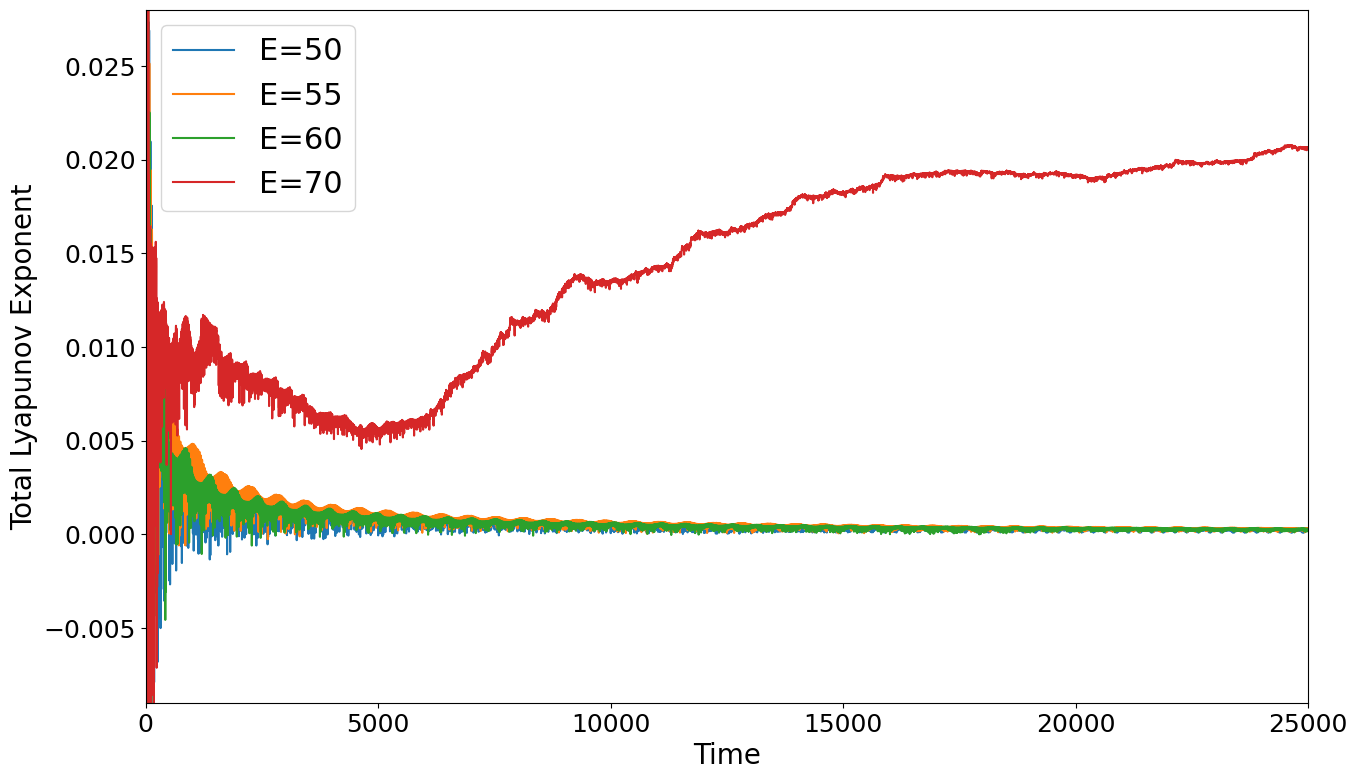}\label{11a}}
        \qquad
	\subfigure[] 
        {\includegraphics[width=1.0\linewidth,height=0.6\linewidth]{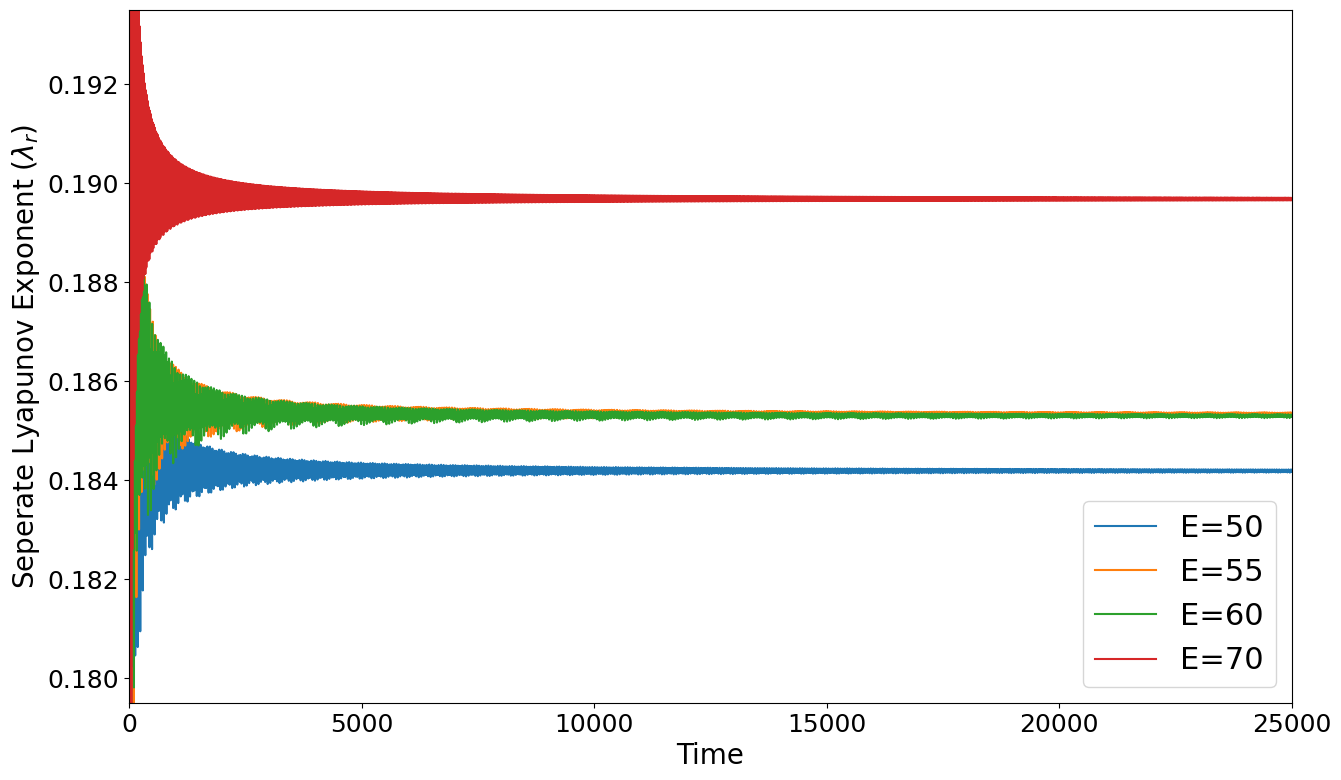}\label{11b}}
        \end{array}$
        \end{center}
        \begin{minipage}{\textwidth}
        \caption{Figure\ref{11a} represents the total Lyapunov exponent $(\lambda_T)$ for the SSS black hole of model I for $\beta=10^{-5}$ at different energy values. The exponent settles at the maximum positive value $\sim 0.0206613$ for $E=70$. Figure \ref{11b} represents the separate Lyapunov exponent $(\lambda_r)$, which is along radial direction for the SSS black hole of model I with $\beta=10^{-5}$ for different energies. The exponent settles at the maximum positive value $\sim 0.189656$ for $E=70$.
        }\label{f11}
        \hrulefill
	\end{minipage}
        \begin{center}
        $\begin{array}{cc}
        \subfigure[] 
        {\includegraphics[width=1.0\linewidth,height=0.6\linewidth]{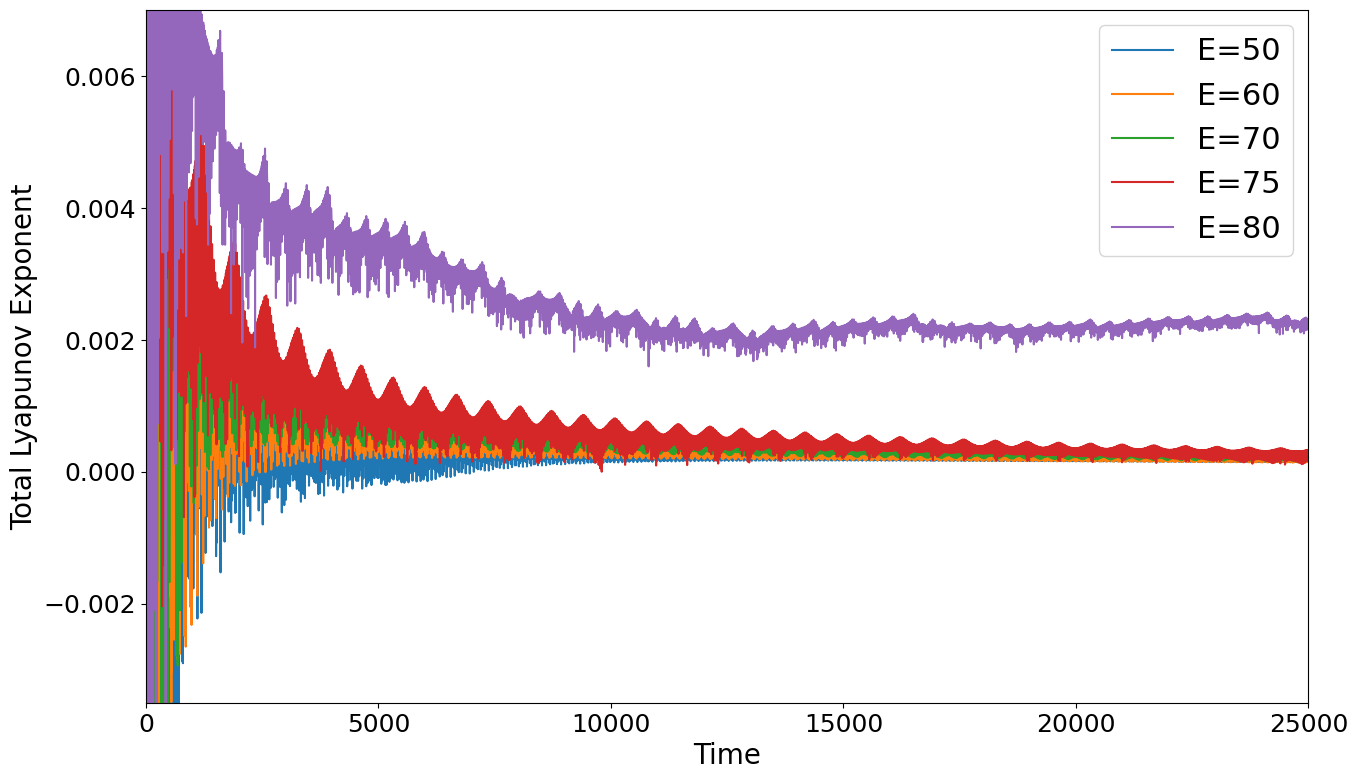}\label{12a}}
        \qquad
        \subfigure[] 
        {\includegraphics[width=1.0\linewidth,height=0.6\linewidth]{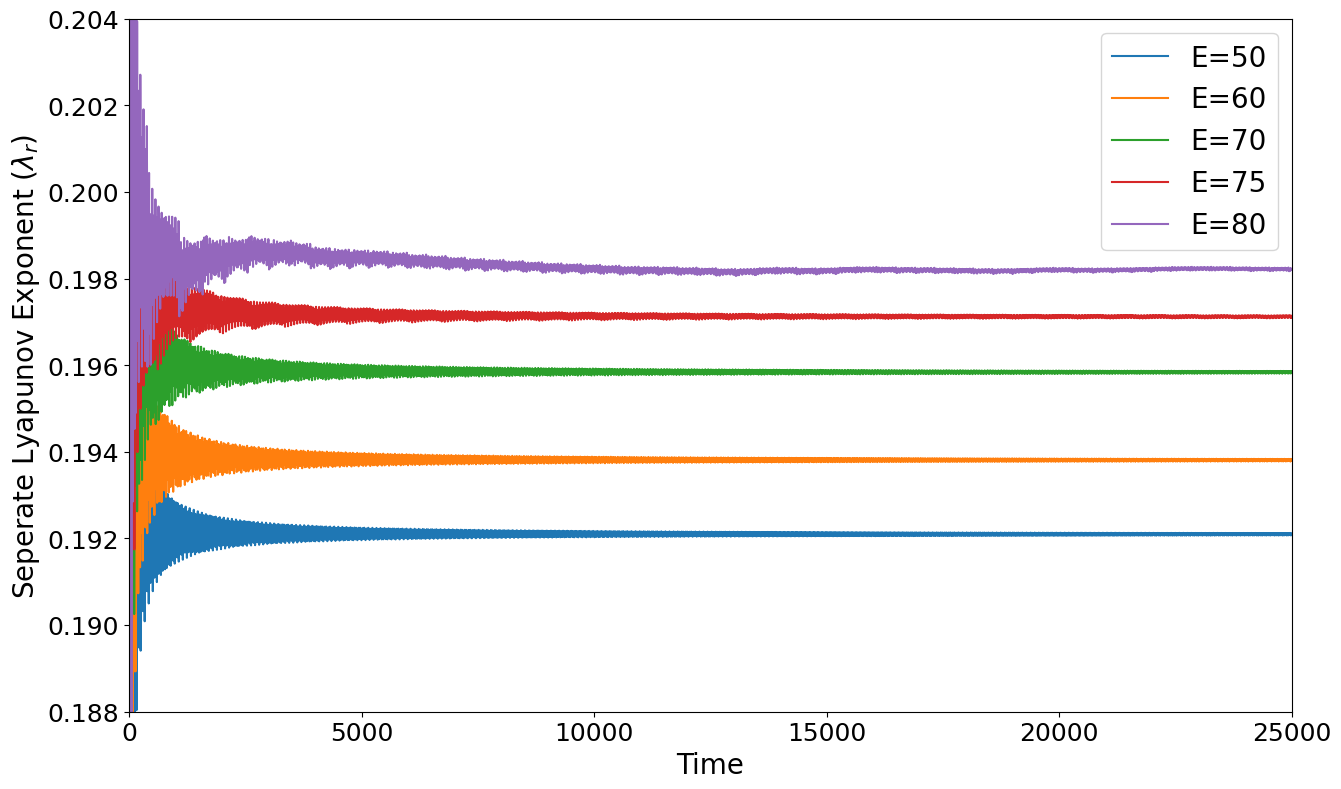}\label{12b}}
	\end{array}$
        \end{center}
        \begin{minipage}{\textwidth}
        \caption{Figure \ref{12a} represents the total Lyapunov exponent $(\lambda_T)$ for the SSS black hole of model I for $\beta=10^{-2}$ at different energy values. The exponent settles at the maximum positive value $\sim 0.002219 $ for $E=80$. Figure \ref{12b} represents the separate Lyapunov exponent $(\lambda_r)$ for the SSS black hole of model I with $\beta=10^{-2}$ for different energies. The exponent settles at the maximum positive value $\sim 0.198206$ for $E=80$.}\label{f12}
        \end{minipage}
        \end{figure}
    \end{widetext}

    We first analyze the total LE  $(\lambda_{T})$ in Fig.\ref{10a}, where we have plotted the total LE vs the modified gravity parameter,  $\beta$ for two different values of the energy, i.e., for $E=50$ and $E=70$. We observe that for both energy values, the  LE value is much lower than the chaos bound, i.e., the value of the surface gravity computed for the black hole solution in modified gravity model I ($\kappa_{MG}$). 
    The functional form of $\kappa_{MG}$ with respect to $\beta$ is obtained for model I [Eq.\eqref{8}] as follows:
        \begin{eqnarray}
            \kappa_{MG}&&=\frac{1}{2}\Big(\frac{\partial B(r)}{\partial r}\Big)\Big\rvert_{r=r_H}\nonumber\\
            &&=\frac{1}{2}\left[\beta+\frac{8M{\beta}^2}{{\big(-1+\sqrt{1+8M\beta}\big)}^2}\right].\label{39}
        \end{eqnarray}
    By series expansion of the above equation [Eq.\eqref{39}] can be written as
        \begin{equation}
            \kappa_{MG}=\frac{1}{4M}+\frac{3\beta}{2}+\mathcal{O}(\beta^2).\label{40}
        \end{equation}
    In the $\beta \rightarrow 0$ limit, Eq.\eqref{40} reduces to the surface gravity for the Schwarzschild black hole, henceforth denoted by
        \begin{equation}
            \kappa_{EG} =\frac{1}{4M},\label{40a}
        \end{equation}
    i.e., the surface gravity in the limit of unmodified Einstein gravity. It is interesting to note that the term linear in $\beta$ in the above expansion is independent of the mass of the black hole. In both Fig.\ref{10a} and Fig.\ref{10b}, we compare the numerically computed LEs with both $\kappa_{MG}$ and $\kappa_{EG}$. In Fig.\ref{10b}, we compare the individual LE in the radial direction ($\lambda_{r}$) with respect to these bounds for the same set of energies as Fig.\ref{10a}. We observe that although the total LE $(\lambda_{T})$ respects both the bounds in Fig.\ref{10a}, the individual LE $(\lambda_{r})$ only respects $\kappa_{MG}$ and not $\kappa_{EG}$ for some high values of $\beta$.
    
    Next, in Fig.\ref{11a}, we plot $\lambda_T$ and $\lambda_r$ vs time for different values of the total energy $E$ but for a fixed $\beta=10^{-5}$. As we can see from the figure, the saturated value of LE $(\lambda_T)$ increases with increasing energy, signifying larger chaos at higher energies. A similar feature is also visible in the case of $\lambda_r$, where at energy $E=70$, we get the highest peak value of $\lambda_r$ [see Fig.\ref{11b}]. However, for both cases, the saturated LEs respect both the chaos bound $\kappa_{EG}$ and $\kappa_{MG}$. Next, we increase the value of the modification parameter  $\beta=10^{-2}$ to see the effect of modified gravity (see Fig.\ref{f12}). We again observe that  $\lambda_T$ and $\lambda_r$ respect both the chaos bounds $\kappa_{EG}$ and $\kappa_{MG}$. However, like the low $\beta$ scenario in Fig.\ref{f11}, the highest energy corresponds to the highest value of the LE in Fig.\ref{f12}.

        \subsubsection{Model II: Charged black hole}\label{s4c-2}
    For the charged black hole in model II [Eq.\eqref{21}], the expression for surface gravity is given by
        \begin{eqnarray}
            \kappa_{cBH}=\dfrac{3a\left(1-6a+\sqrt{1-6a}\right)}{2{\left(1+\sqrt{1-6a}\right)}^3}\label{41}
        \end{eqnarray}
        
\begin{widetext}
        \begin{figure}[H]
	\centering
	\begin{center} 
	$\begin{array}{cc}
	\subfigure[]
        {\includegraphics[width=1.0\linewidth,height=0.6\linewidth]{14a.png}\label{13a}}
        \qquad
        \subfigure[]{\includegraphics[width=1.0\linewidth,height=0.6\linewidth]{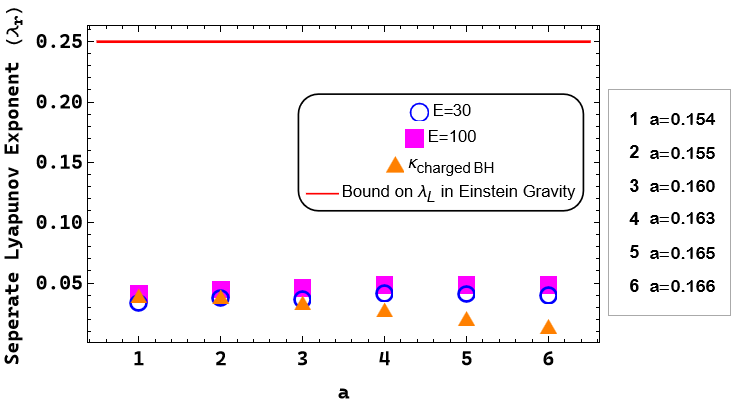}\label{13b}}
        \end{array}$
        \end{center}
        \begin{minipage}{\textwidth}
        \caption{Here, Fig.\ref{13a} represents the variation of total Lyapunov exponent $(\lambda_T)$ with dimensional parameter $a$ for two different energies, $E=30$ and $E=70$ for charged black hole solution in model II. Figure \ref{13b} represents the variation of separate Lyapunov exponent $(\lambda_r)$ with parameter $a$ for two different energies, $E=30$ and $E=70$ for charged case in model II. For both the figures, we set points of ``a" axis as given in the right panel of each figure.}\label{f13}
        \hrulefill
	\end{minipage}
        \begin{center}
        $\begin{array}{cc}
        \subfigure[] 
        {\includegraphics[width=1.0\linewidth,height=0.6\linewidth]{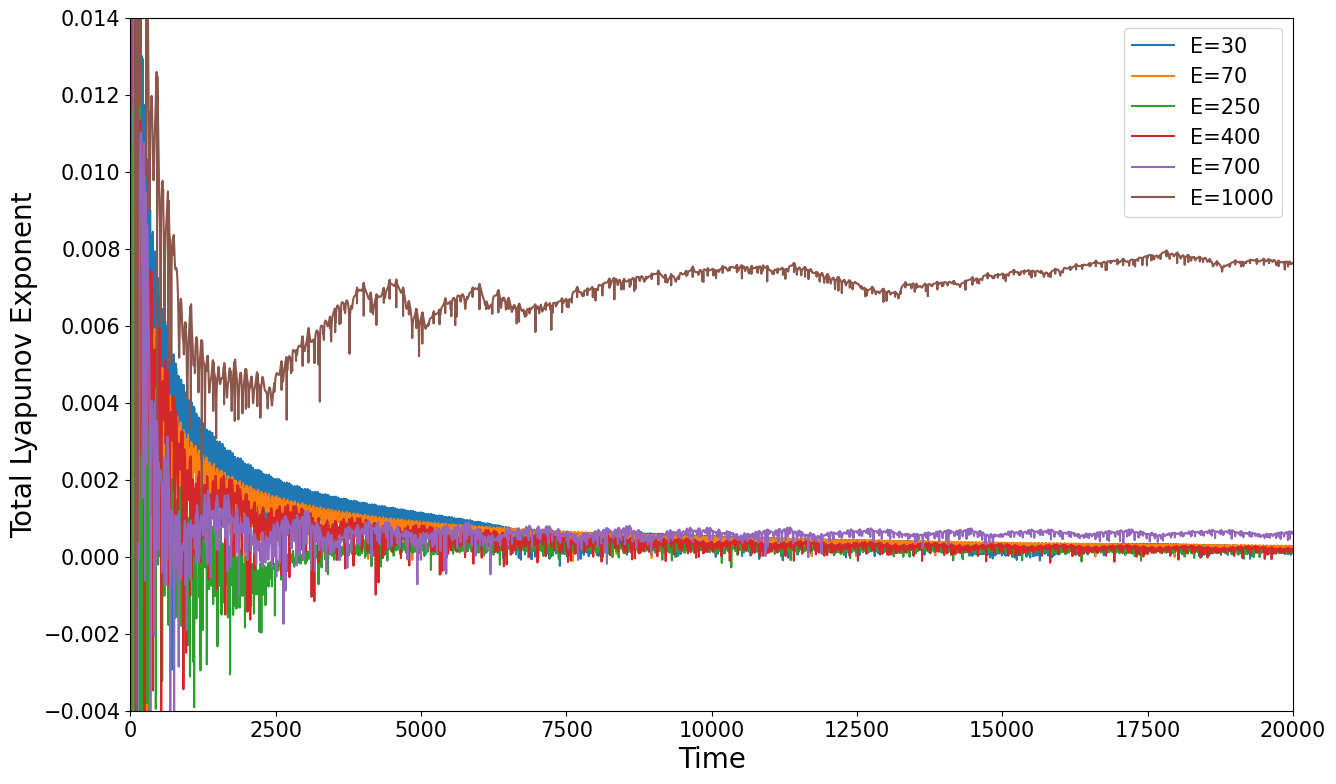}\label{14a}}
        \qquad
	\subfigure[] 
        {\includegraphics[width=1.0\linewidth,height=0.6\linewidth]{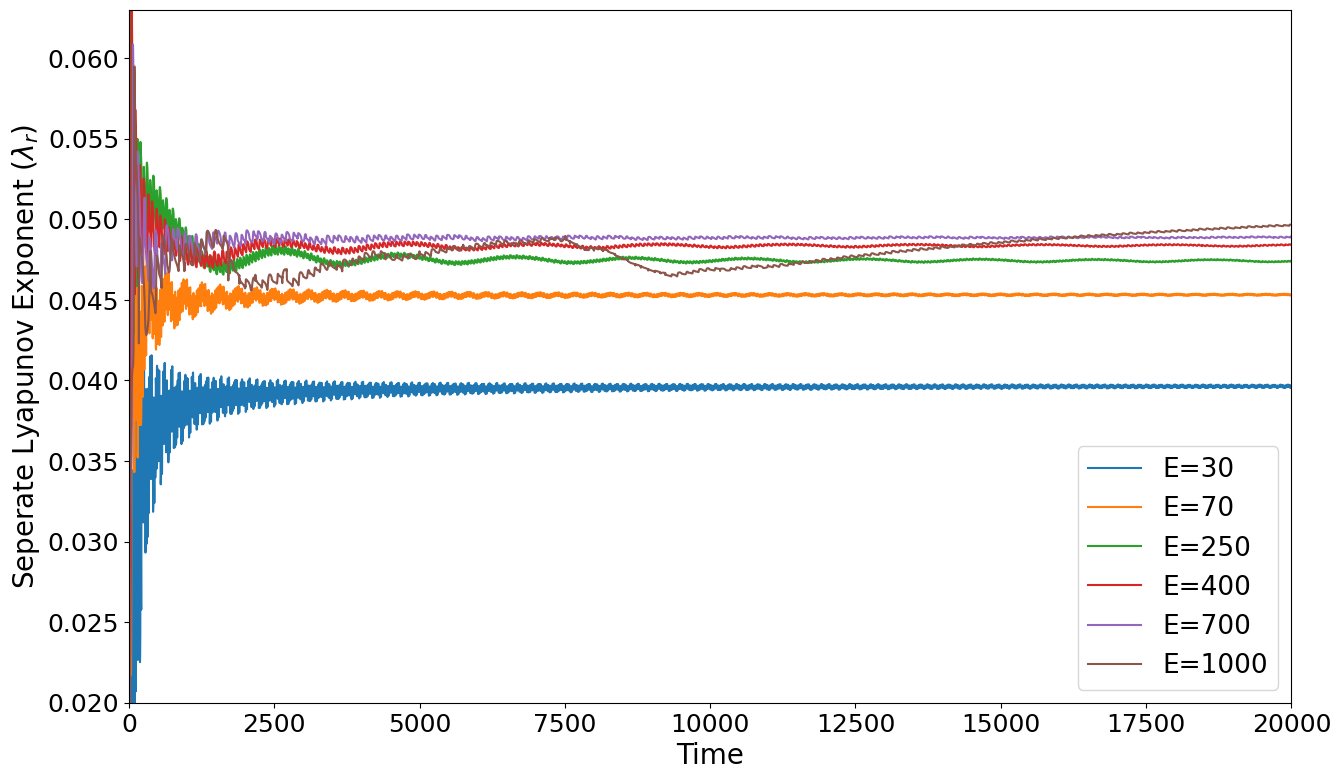}\label{14b}}
        \end{array}$
        \end{center}
        \begin{minipage}{\textwidth}
        \caption{Figure \ref{14a} represents the total Lyapunov exponent $(\lambda_T)$ for the charged black hole of model II for fixed $a=0.166$ at different energy values. The exponent settles at the maximum positive value $\sim 0.007632$ for $E=1000$. Figure \ref{14b} represents the separate Lyapunov exponent $(\lambda_r)$ for the charged black hole of model II with fixed $a=0.166$ for different energies. The exponent settles at the maximum positive value $\sim 0.0496643$ for $E=1000$.
        }\label{f14}
        \hrulefill
	\end{minipage}
        \begin{center}
        $\begin{array}{cc}
        \subfigure[] 
        {\includegraphics[width=1.0\linewidth,height=0.6\linewidth]{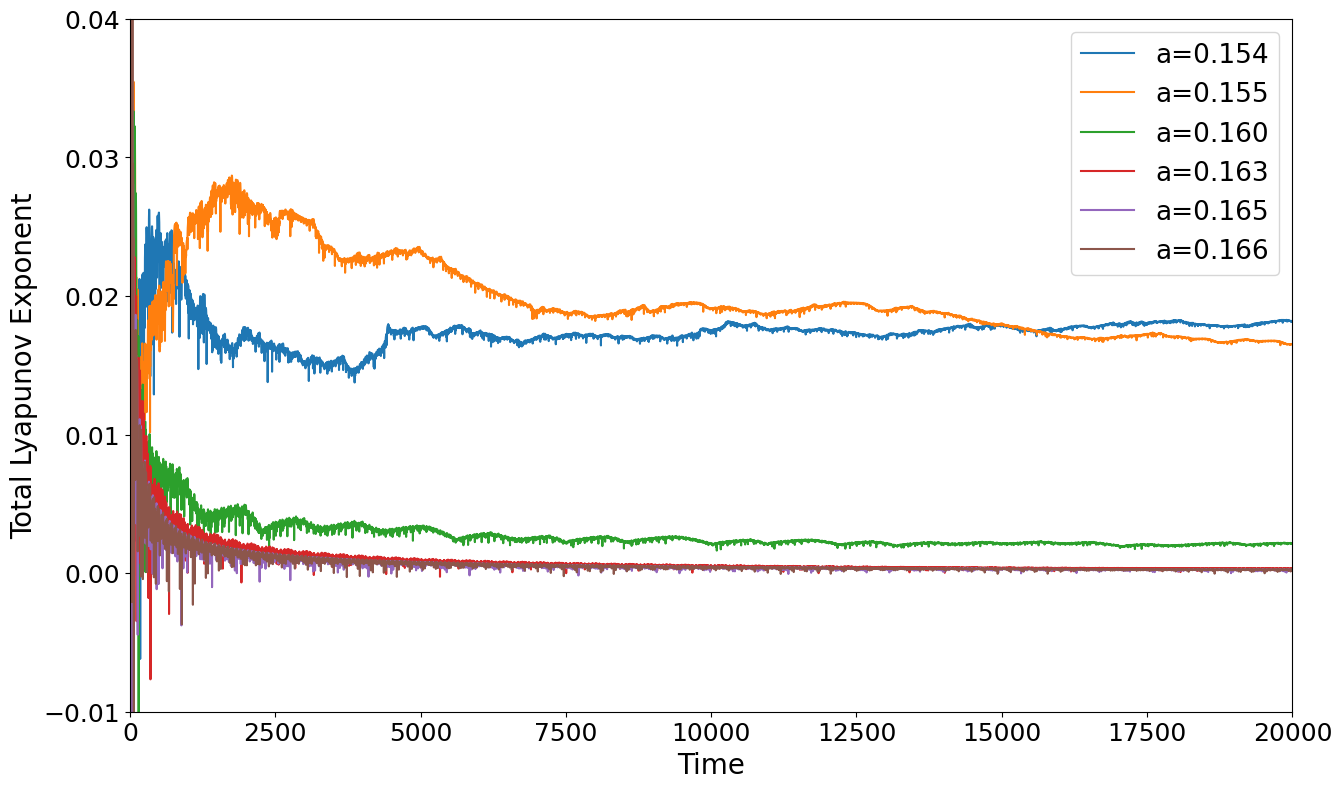}\label{15a}}
        \qquad
        \subfigure[] 
        {\includegraphics[width=1.0\linewidth,height=0.6\linewidth]{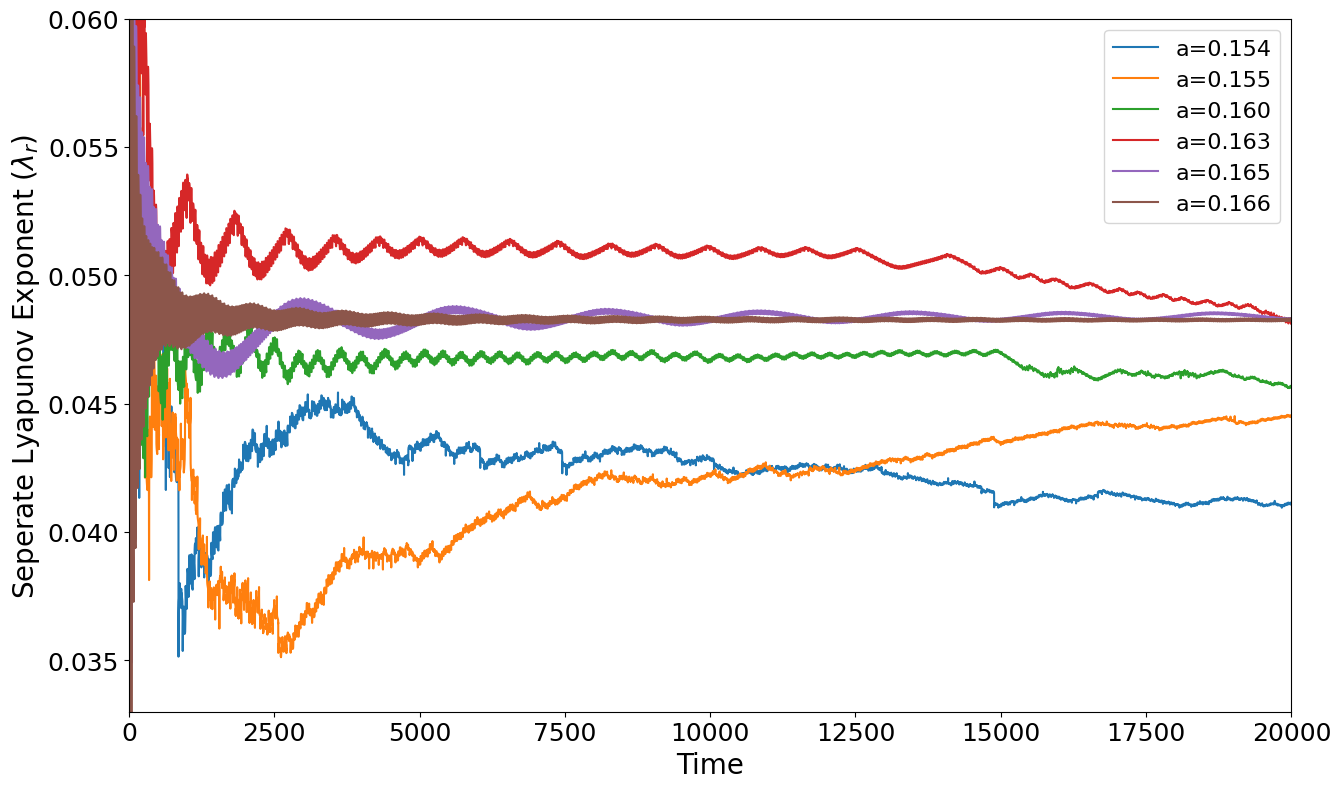}\label{15b}}
	\end{array}$
        \end{center}
        \begin{minipage}{\textwidth}
        \caption{Figure \ref{15a} represents the total Lyapunov exponent $(\lambda_T)$ for the charged black hole of model II for several $a$ at fixed energy $E=100$. The exponent settles at the maximum positive value $\sim 0.018206$ for the lowest $a=0.154$. Figure \ref{15b} represents the separate Lyapunov exponent $(\lambda_r)$ for the charged black hole of model II with fixed energy value, $E=100$ for different $a$. The exponent settles at the maximum positive value $\sim 0.048231$ for the highest $a=0.166$.}\label{f15}
        \end{minipage}
        \end{figure}
    \end{widetext}

    We now perform a similar analysis as model I. In Fig.\ref{f13}, we plot the Lyapunov exponent $(\lambda_T)$ vs the modified gravity parameter ($a$) for two values of energy ($E=30$ and $70$). Figure \ref{13a} is for $\lambda_{T}$ and Fig. \ref{13b} is for $\lambda_{r}$, respectively. We see in Fig.\ref{13a} that  $\lambda_{T}$ for $E=30$ is close to $\kappa_{cBH}$, but well below it for the case of $E=100$ \Big(let us note that since $a<\frac{1}{6}$, $\kappa_{cBH}$ is always positive\Big).  However, both the energy curves respect the chaos bound $\kappa_{EG}$ from Schwarzschild  BH in Einstein gravity. In Fig.\ref{13b}, we also see that for higher values of $a$, the individual Lyapunov $(\lambda_{r})$ exceeds $\kappa_{cBH}$ for both energy values ($E=30$ and $100$). However, they do respect the bound from Einstein's gravity  $\kappa_{EG}$.

    In Fig.\ref{14a}, we plot $\lambda_{T}$ vs time for the charged black hole of model II for the maximum value of $a=1/6$ for different values of energy. We observe that $\lambda_{T}$ saturates to a positive value $\sim 0.007632$ at highest energies $E=1000$. Similarly, in Fig.\ref{14b}, we observe that $\lambda_{r}$ also attains maximum positive value at $E=1000$.

    On the other hand, in Fig.\ref{f15}, we plot both $\lambda_{T}$ and $\lambda_{r}$ for fixed value of energy $E=100$ for various values of the parameter $a$. In Fig.\ref{15a}, we see that a decrease in ``$a$" corresponds to an increase in $\lambda_{T}$. This is consistent with the fact that small values of ``$a$" corresponds to an increase in horizon size, and hence, the particle resides closer to the horizon, leading to increased chaos and high $\lambda_{T}$. However, in Fig.\ref{15b}, we see that the individual LEs do not share this property.

        \subsubsection{Model II: Neutral black hole}\label{s4c-3}
    The surface gravity for the neutral black  hole solution (henceforth, denoted by $ \kappa_{nBH} $) in model II [Eq.\eqref{25}] is given by
        \begin{eqnarray}
            \kappa_{nBH}=\frac{3a}{8}\label{42}
        \end{eqnarray}
    
\begin{widetext}
        \begin{figure}[H]
	\centering
	\begin{center} 
	$\begin{array}{cc}
	\subfigure[]
        {\includegraphics[width=1.0\linewidth,height=0.6\linewidth]{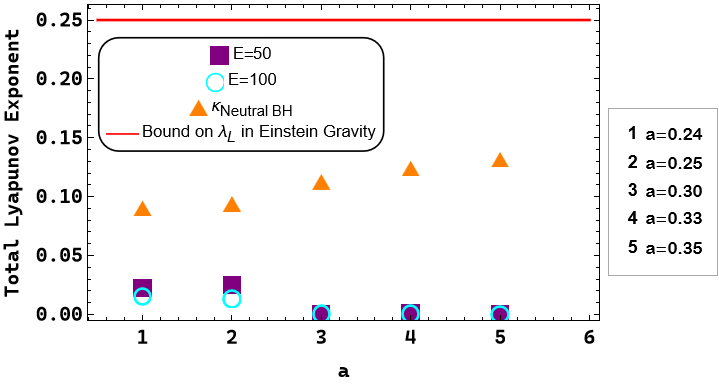}\label{16a}}
        \qquad
        \subfigure[]{\includegraphics[width=1.0\linewidth,height=0.6\linewidth]{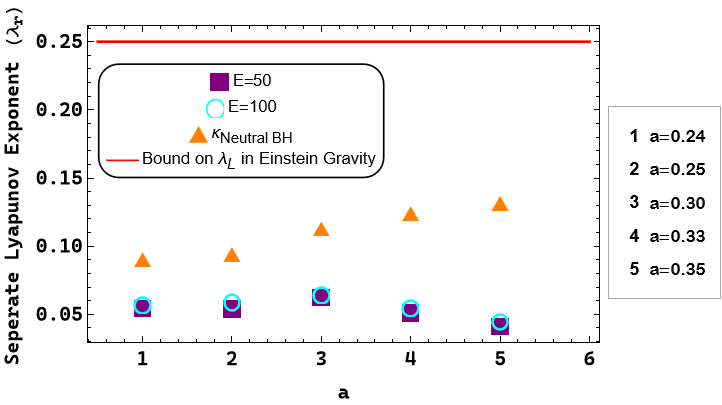}\label{16b}}
        \end{array}$
        \end{center}
        \begin{minipage}{\textwidth}
        \caption{Here, Fig.\ref{16a} represents the variation of total Lyapunov exponent $(\lambda_T)$ with dimensional parameter $a$ for two different energies, $E=50$ and $E=100$ for neutral black hole solution in our model II. Figure \ref{16b} represents the variation of separate Lyapunov exponent $(\lambda_r)$ with parameter $a$ for two different energies, $E=50$ and $E=100$ for the same neutral black hole case in model II. For both the figures, we set the points of ``a" axis as given in the right panel of each figure.}\label{f16}
        \hrulefill
	\end{minipage}
        \begin{center}
        $\begin{array}{cc}
        \subfigure[] 
        {\includegraphics[width=1.0\linewidth,height=0.6\linewidth]{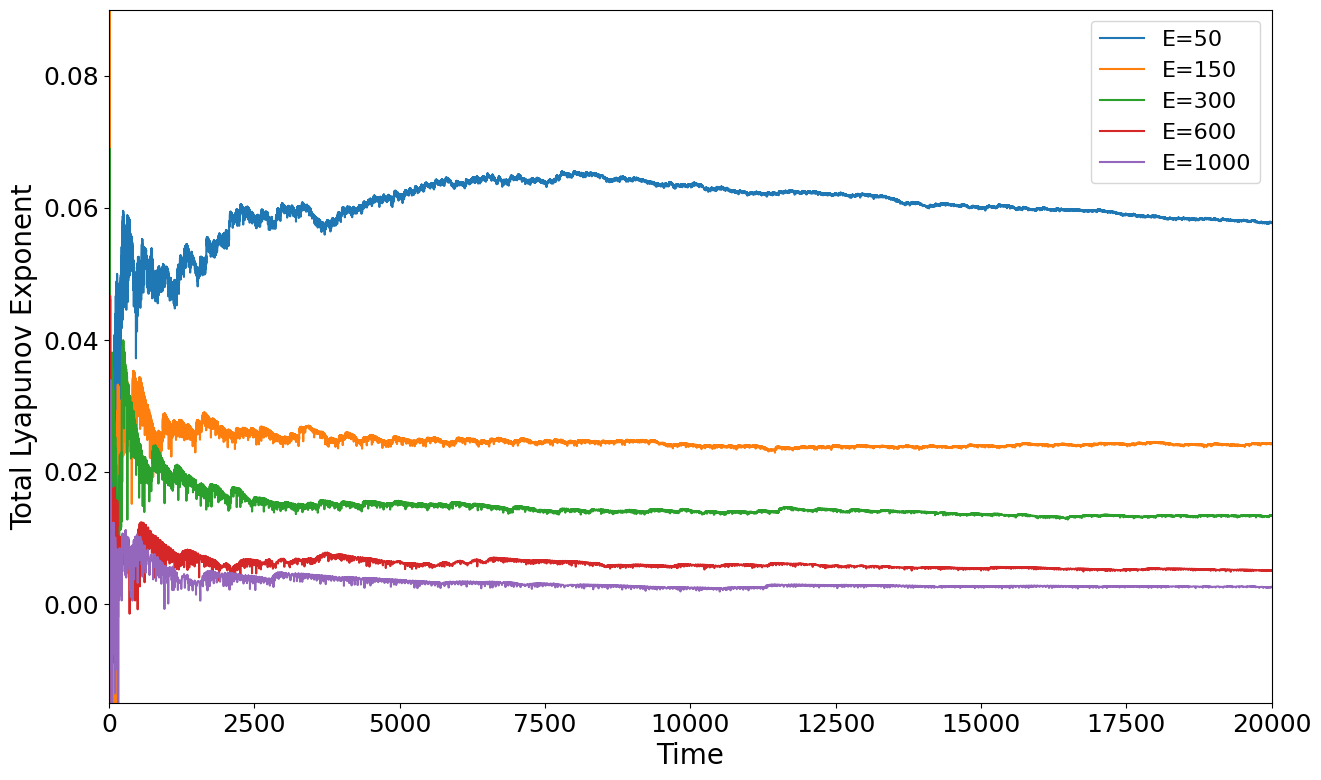}\label{17a}}
        \qquad
	\subfigure[] 
        {\includegraphics[width=1.0\linewidth,height=0.6\linewidth]{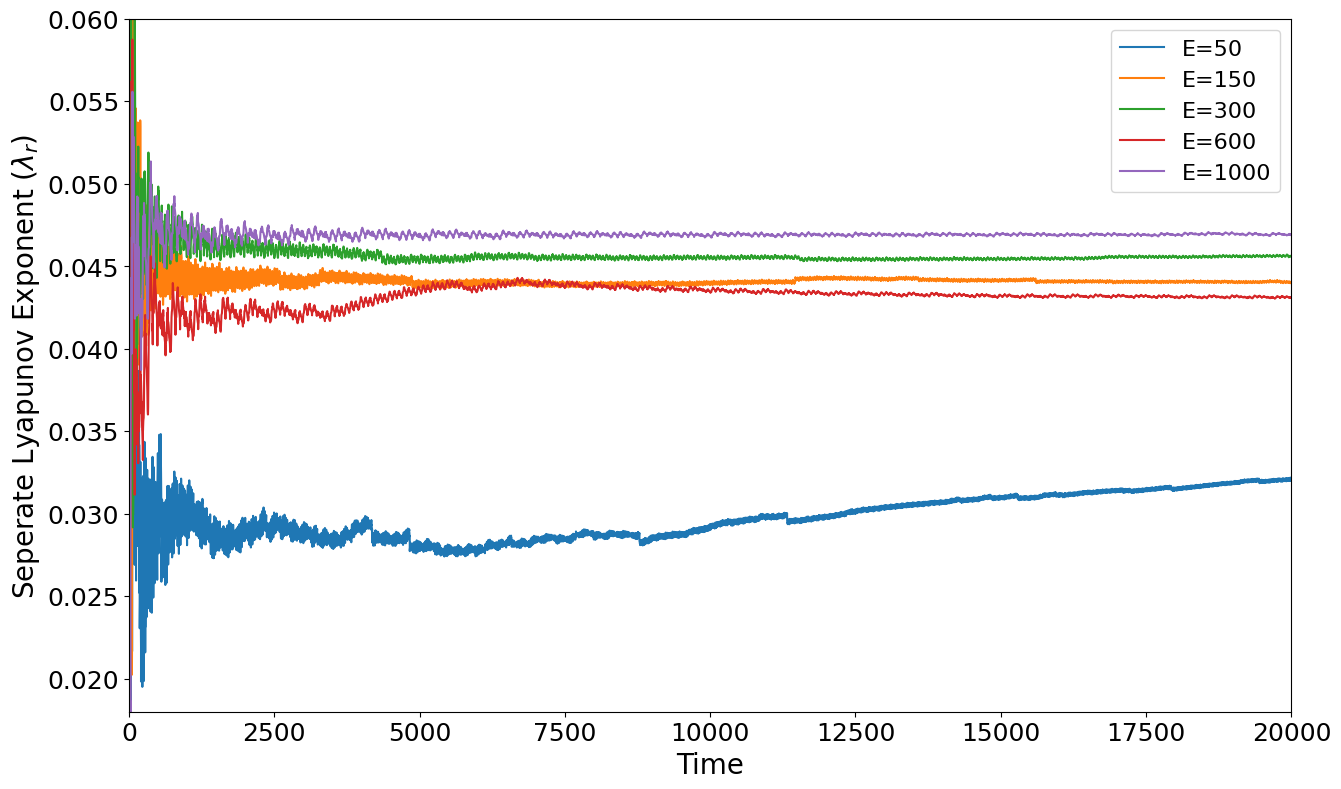}\label{17b}}
        \end{array}$
        \end{center}
        \begin{minipage}{\textwidth}
        \caption{Figure \ref{17a} represents the total Lyapunov exponent $(\lambda_T)$ for the neutral black hole of model II for fixed $a=0.5$ at different energy values. The exponent settles at the maximum positive value $\sim 0.057819$ for $E=50$. Figure \ref{17b} represents the separate Lyapunov exponent $(\lambda_r)$ for the neutral black hole of model II with fixed $a=0.5$ for different energies. The exponent settles at the maximum positive value $\sim 0.046896$ for $E=1000$.
        }\label{f17}
        \hrulefill
	\end{minipage}
        \begin{center}
        $\begin{array}{cc}
        \subfigure[] 
        {\includegraphics[width=1.0\linewidth,height=0.6\linewidth]{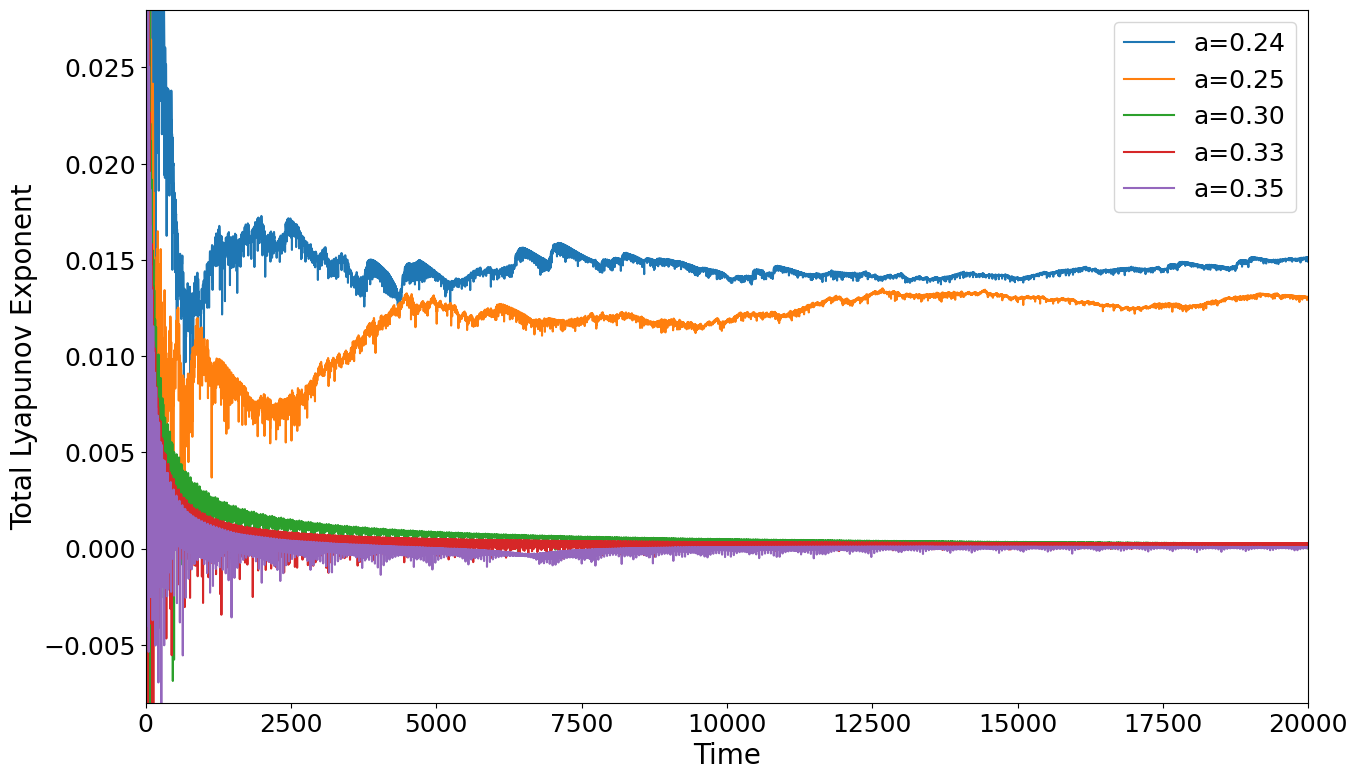}\label{18a}}
        \qquad
        \subfigure[] 
        {\includegraphics[width=1.0\linewidth,height=0.6\linewidth]{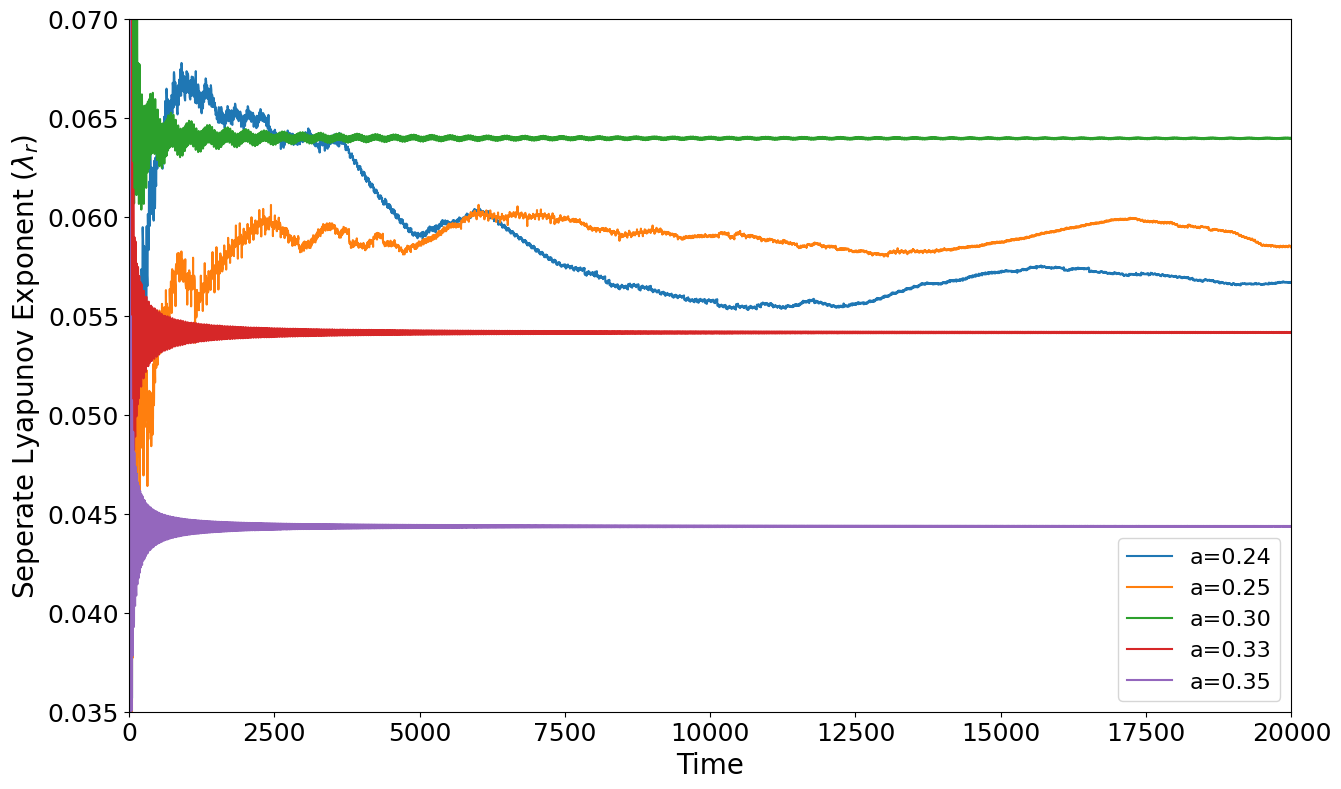}\label{18b}}
	\end{array}$
        \end{center}
        \begin{minipage}{\textwidth}
        \caption{Figure \ref{18a} represents the total Lyapunov exponent $(\lambda_T)$ for the neutral black hole of model II for several $a$ at fixed energy $E=100$. The exponent settles at the maximum positive value $\sim 0.015106$ for the low $a=0.24$. Figure \ref{18b} represents the separate Lyapunov exponent $(\lambda_r)$ for the neutral black hole of model II with fixed energy value, $E=100$ for different $a$. The exponent settles at the maximum positive value $\sim 0.063981$ for $a=0.30$.}\label{f18}
        \end{minipage}
        \end{figure}
    \end{widetext}

    In Fig.\ref{f16}, we plot the variation of $\lambda_{T}$ and $\lambda_{r}$ vs $a$ for two different energies, $E=50$ and $E=100$ for the neutral black hole solution. In both cases, we observe that they violate neither the chaos bound in the modified gravity model $\kappa_{nBH}$ nor the bound in Einstein gravity $\kappa_{EG}$ \Big[see both Fig.\ref{16a} and Fig.\ref{16b}\Big].

    In Fig.\ref{17a}, we plot $\lambda_{T}$ vs time for the neutral black hole of model II for fixed $a = 0.5$ at different energy values. The saturated LE attains maximum positive value $\sim 0.057819$ for $E=50$, whereas in Fig.\ref{17b} represents the $\lambda_{r}$, attains the maximum positive value $\sim 0.046896$ for $E=1000$.

    In Fig.\ref{18a}, we plot $\lambda_{T}$ vs time for the neutral black hole for different values of $a$ at a fixed energy $E=100$. The exponent settles at the maximum positive value for the lowest value of $a=0.24$, which demonstrates enhancement of chaos close to the horizon (i.e., with the decrement of $a$). However, in Fig.\ref{18b}, the individual $\lambda_{r}$ does not follow this trend.
    
\begin{widetext}
    \begin{table*}[htbp]
    \centering
    \rowcolors{2}{pink!30}{myolive!18}
    \begin{tabular}{p{0.35\linewidth} p{0.51\linewidth}}
        \toprule
        \rowcolor{mymaroon!30}
        \textbf{~~~~~~~~~~~~~~Einstein's GR} & \textbf{~~~~~~~~~~~~~~~~~~~~~~~~~~~~Modified GR} \\
        \midrule
        \begin{tabular}[t]{@{}>{\centering\arraybackslash}p{5.5cm}@{}}
            The onset of chaos is observed at $E\sim55$. \\
            Here $\lambda_{T}=0.00028487$ for $t=25000.$
        \end{tabular}    
        &
        \begin{tabular}[t]{@{}>{\centering\arraybackslash}p{8.95cm}@{}}
        \textit{\textbf{Model I with $\beta=10^{-5}$:}}\\Onset of chaos at $E\sim 55$. Here, $\lambda_{T}=0.00029556$ for $t=25000.$ \\
        \hline
        \textit{\textbf{Model I with $\beta=10^{-2}$:}}\\Onset of chaos at $E\sim75$. Here, $\lambda_{T}=0.00032495$ for $t=25000.$
        \end{tabular}    
        \\
        \begin{tabular}[t]{@{}>{\centering\arraybackslash}p{5.5cm}@{}}
            Same as above.
        \end{tabular}    
        &
       \begin{tabular}[t]{@{}>{\centering\arraybackslash}p{8.95cm}@{}}
        \textit{\textbf{Model II (Charged BH with fixed a=0.166):}}\\Onset of chaos at $E\sim 400$. Here, $\lambda_{T}=0.00015088$ for $t=20000.$ \\
        \hline
        \textit{\textbf{Model II (Charged BH with fixed E=100):}}\\Onset of chaos at $a \sim0.163 $. Here, $\lambda_{T}=0.00031707$ for $t=20000.$
        \end{tabular}     
        \\
        \begin{tabular}[t]{@{}>{\centering\arraybackslash}p{5.5cm}@{}}
            Same as above.
        \end{tabular}    
        &
       \begin{tabular}[t]{@{}>{\centering\arraybackslash}p{8.95cm}@{}}
        \textit{\textbf{Model II (Neutral BH with fixed a=0.5):}}\\Onset of chaos at $E\sim 150$. Here, $\lambda_{T}=0.02429723$ for $t=20000$.  \\
        \hline
        \textit{\textbf{Model II (Neutral BH with fixed E=100):}}\\Onset of chaos at $a \sim 0.30$. Here, $\lambda_{T}=0.00025345$ for $t=20000.$
        \end{tabular} 
        \\
        \bottomrule 
        \hline
    \end{tabular}
    \caption{Comparison for the onset of chaos (i.e., the first appearance of broken tori in the associated Poincar$\Acute{e}$ section) in Schwarzschild background in Einstein gravity with modified theories of gravity, model I and model II.}
    \label{T2}
    \end{table*}
\end{widetext}

        \section{Discussions and conclusion}\label{sec5}
    We now summarize the main results of the last few sections, along with potential avenues to explore in future. Firstly, in Table \ref{T2} we present a comparative study of chaos in modified gravity model I and model II with the corresponding results for the Schwarzschild background in Einstein gravity, for the same set of initial conditions. In this context, let us note that the Schwarzschild background arises in the $\beta \rightarrow 0$ limit of model I. On the left-hand column, we report the energy value and the corresponding Lyapunov exponent at which we first observe the onset of chaos (i.e., the first appearance of broken tori in the associated Poincar$\Acute{e}$ section). In the right-hand column, we report the corresponding values in modified GR model I and model II  (charged and neutral black hole solutions). The table illustrates the impact of the horizon in modified gravity scenarios, namely the horizon radius shrinks upon increasing the modified gravity parameters $\beta$ and ``$a$'' in model I and model II, respectively, resulting in the onset of chaos at higher values of energy so that the particle trajectories ``feel" the presence of the horizon; see also Fig.\ref{T1}. The Poincar$\Acute{e}$ sections constructed in Sec. \ref{s4b} are in agreement with the Kolmogorov-Arnold-Moser (KAM) theory, which asserts that nonlinear perturbations in an integrable system introduce chaos. Far from the horizon, one gets regular tori, which progressively disintegrate into a scattered array of points as the system's energy escalates or, equivalently, as the particle trajectories come closer to the horizon. It is also clear that upon reducing the strength of the harmonic potentials denoted by $K_r$ and $K_{\theta}$ along the $r$ and $\theta$ directions, respectively, the influence of the horizon becomes more prominent, leading to chaotic dynamics at lower energy values.

    Additionally, it is crucial to emphasize that the

    \newpage

    \newpage

    \newpage

    \noindent
    harmonic oscillators in the radial and angular directions are incorporated by hand to prevent the particle from falling into the black hole horizon. Our methodology closely mirrors the framework outlined in Sec. IV ``Numerical Analyses" and the corresponding Eq.(38) of Hashimoto {\it{et al}}. \cite{hashimoto}. Two comments are worth mentioning in this context: First, for model I, our numerical investigations indicate that the total Lyapunov exponent increases with increasing energy and attains the largest value at the highest possible energies, for all values of $\beta$ allowed by current observations \cite{Saffa,Anderson}. Secondly, in the context of model II, the Lyapunov exponent decreases with an increase in modified gravity parameter ``$a$" for both the charged and neutral black hole solutions, consistent with physical expectations. Let us note that as the particle approaches the horizon with higher values of the conserved energy, its motion experiences a redshift relative to an asymptotic observer, akin to the phenomenon described in \cite{hashimoto}, where the metric component $g_{rr}$ diverges very near the event horizon. Thus, it is reasonable to anticipate the presence of a chaotic nature near the horizon. We have also studied the total Lyapunov exponent as well as the separate Lyapunov exponent (along radial direction) for both models and compared them with the MSS bound on chaos. In contrast to some other $f(R)$ models in the existing literature \cite{Andrea1}, we find that the models considered in this work do respect the MSS bound. The present work needs to be expanded in several directions in future: for example, one can now perform a similar analysis with massive particles instead of massless particles studied in this work. We also wish to investigate the effect of adding other external potentials in future, in addition to harmonic oscillators studied in this work. Moreover, another plan is to work on different spacetimes, such as Kerr or Kerr-Newmann solutions in $f(R)$ gravity. Also, the onset of chaos and complete breakdown of KAM tori needs to be studied in a plethora of $f(R)$ gravity models \cite{Andrea1} as well as other modified gravity theories \cite{Addazi}. An early or delayed onset of chaos or the observation of LE is different from that obtained in Einstein GR may be used as an additional diagnostic to identify departure from Einstein action in the strong gravity regime close to the black hole horizon. Moreover, investigation of instabilities in circular photon trajectories has been shown to influence black hole quasinormal modes in the ringdown phase of black hole mergers \cite{Andrea1,Addazi} and associated gravitational waves. It will be interesting to explore these instabilities in model I and model II of our study. Additionally, one can examine the power spectral density (PSD) \cite{rindler} in these models since the PSD is known to be related to the Lyapunov exponent \cite{Sigeti}. Further, the PSD serves as a useful measurable quantity, thereby enhancing the observational aspect of our results. Interestingly, Tian {\it{et al}}. \cite{Fis} has recently proposed an experimental probe for the MSS bound in ion traps using an analogue-gravity setup, which motivates experimental investigations based on the results described in this work. We plan to answer these intriguing questions in subsequent work.

        \section*{Acknowledgment}
    We are very thankful to Subir Ghosh, Koji Hashimoto, Uwe R. Fischer and Leopoldo A. Pando Zayas for discussions, comments, and suggestions on our manuscript. The authors would like to acknowledge the anonymous Referee for suggesting investigations in  Schwarzschild coordinates. S. Das would like to acknowledge all the members of Physics and Applied Mathematics Unit, Indian Statistical Institute, Kolkata, India for the hospitality during his longterm academic visit, where a part of this work has been done and he also thanks BITS-Pilani, Hyderabad Campus, India for the institute fellowship. Additionally, S. Das would like to thanks the support of USTC Fellowship Level A---CAS-ANSO Scholarship 2024 for PhD candidates. S. Dalui thanks the Department of Physics, Shanghai University for providing postdoctoral funds during the period of this work. R.S thanks DST Inspire Faculty (Grant No:IFA19-PH231), Grants No. BITS - NFSG, and No. BITS - OPERA research grants.\\\\

    \textbf{Data Availability :} The analytical computations that support the findings of this study are available within the article. The numerical code is currently not publicly available but will be made available by the authors upon reasonable request.\\\\

    \appendix

        \section{Visualization of chaos in Schwarzschild coordinates for Model I}
    Chaos in particle motion near the event horizon of a black hole is not merely a coordinate artifact; it arises due to the properties of the horizon itself. The chaotic behavior can also be observed directly in the usual Schwarzschild coordinate system, without the need to transform to the Painlevé-Gullstrand coordinate system. To explore this, one can begin with the SSS metric [Eq.\eqref{3.1}] and derive the energy of a null probe particle from the covariant dispersion relation [Eq.\eqref{3.4}], which takes the form,
    \begin{equation}
		E=\sqrt{A(r)\Big[A(r)p_r^2+\dfrac{p_{\theta}^2}{r^2}\Big]},\label{a1}
	\end{equation}
    where, we have assumed the motion of the particle is in the poloidal plane with $p_{\phi}=0$.\\
    It is noteworthy to mention that in Eq.\eqref{a1}, with fixed $E$ and $p_{\theta}$, there must be an instability in the phase space of the motion of a massless particle. The phase portrait generally looks like a rectangular hyperbola type, resembles a $xp$-type Hamiltonian, which provides instability into the motion of the particle dynamics (a possible inherent property of the black hole horizon), as long as near horizon dynamics of the massless particle is concerned \cite{sd2,bery}. On the other hand, one may find a homoclinic orbit in the phase space of the motion of a relativistic massive particle with fixed energy and angular momentum \cite{i1,i4,Soy,KAM} and hence, when a particle is in a homoclinic orbit, any small perturbation can cause the particle to deviate significantly from its trajectory, leading to chaotic or divergent motion. Therefore, it refers that instability is enhanced by the effect of the event horizon, making the orbits particularly sensitive to even minor disturbances, ultimately resulting in the particle falling into the black hole or escaping to infinity. Therefore, to prevent the particle from falling into the black hole or escaping to infinity, we introduce two harmonic potentials along the radial ($r$) and cross-radial ($\theta$) directions. This allows us to better visualize the chaotic dynamics near the horizon, providing a more controlled framework for studying the particle's behavior in such an unstable region \cite{hashimoto}.\\
    Therefore, with the perturbations of harmonic potentials, one may have the Hamiltonian of probe particle as follows:
    \begin{eqnarray}
		&&E=\sqrt{A(r)\Big[A(r)p_r^2+\dfrac{p_{\theta}^2}{r^2}\Big]}+\frac{1}{2}K_r(r-r_c)^2\nonumber\\
		&&~~~~~~~~~~~~+\frac{1}{2}K_\theta~r^2_{H}(\theta-\theta_c)^2.\label{a2}
    \end{eqnarray}
    Correspondingly, the dynamical equations of motion have the following form:
	\begin{eqnarray}
		\dot{r}&=&\dfrac{\partial E}{\partial p_r}=\frac{A^2(r)p_r}{\sqrt{A^2(r)p_r^2+\dfrac{A(r)p_{\theta}^2}{r^2}}},\label{a3}
		\\
		\dot{p_r}&=&-\frac{\partial E}{\partial r}=-\dfrac{A(r)A'(r)p_r^2}{\sqrt{A^2(r)p_r^2+\dfrac{A(r)p_{\theta}^2}{r^2}}}-K_r(r-r_c)\nonumber\\
            &&-\dfrac{A'(r)p_{\theta}^2}{2r^2\sqrt{A^2(r)p_r^2+\dfrac{A(r)p_{\theta}^2}{r^2}}}+\dfrac{A(r)p_{\theta}^2}{r^3\sqrt{A^2(r)p_r^2+\dfrac{A(r)p_{\theta}^2}{r^2}}},\nonumber\\
            \label{a4}
		\\
		\dot{\theta}&=&\frac{\partial E}{\partial p_{\theta}}=\dfrac{A(r)p_{\theta}}{r^2\sqrt{A^2(r)p_r^2+\dfrac{A(r)p_{\theta}^2}{r^2}}},\label{a5}
		\\
		\dot{p_{\theta}}&=&-\frac{\partial E}{\partial\theta}=-K_{\theta}~r^2_H(\theta-\theta_c).\label{a6}
	\end{eqnarray}
    We solve the dynamical equations of motion \Big[Eq.(\eqref{a3}), Eq.(\eqref{a4}), Eq.(\eqref{a5}), and Eq.(\eqref{a6}\Big] for model I [Eq.(\eqref{8})] using the fourth order Runge-Kutta method with fixed step size $h=0.01$, $K_r=100$, $K_{\theta}=25$, $r_c=4.3$, and $\theta_c=0$, same as the previous section Sec. \ref{s4b-1}.

    It is important to note that a relativistic massive particle exhibits an unstable circular orbit around a Schwarzschild black hole. Each one of these generates a homoclinic trajectory in phase space for time $t\rightarrow\pm\infty$, approaches the unstable orbit \cite{i1,KAM}. For a relativistic massless particle, there also exist an instability. This provides a clear illustration of unstable behavior of particle motion in the vicinity of a black hole's event horizon. Consequently, when subjected to periodic perturbations in the total energy of the probe particle, the orbit tends to become chaotic. This approach provides a clear framework for visualizing the chaotic dynamics near the event horizon of a black hole.

    \newpage
    \begin{widetext}
        \begin{figure}[H]
	\centering
	\begin{center} 
	$\begin{array}{cc}
	\subfigure[] 
        {\includegraphics[width=1.0\linewidth,height=0.5\linewidth]{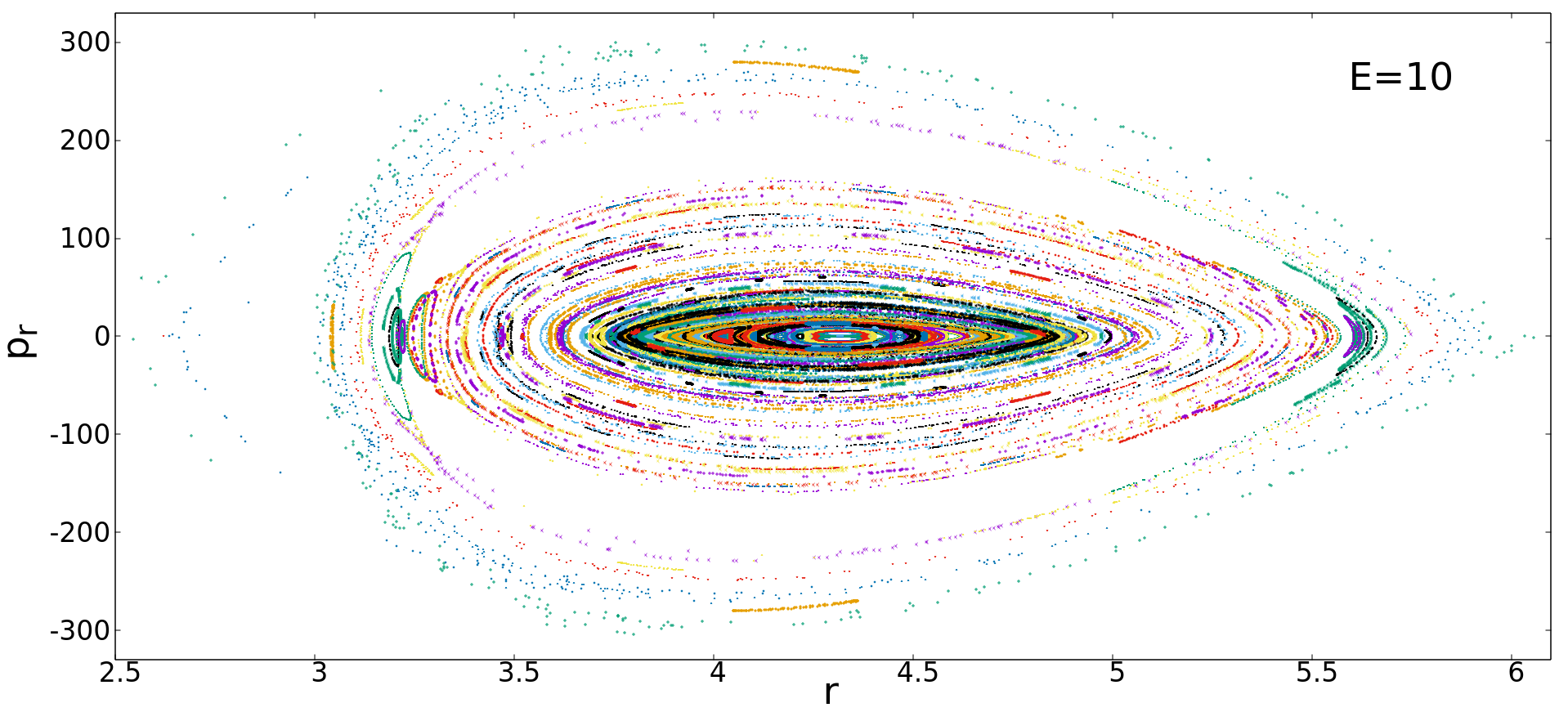}\label{ap1}}
        \subfigure[]
        {\includegraphics[width=1.0\linewidth,height=0.5\linewidth]{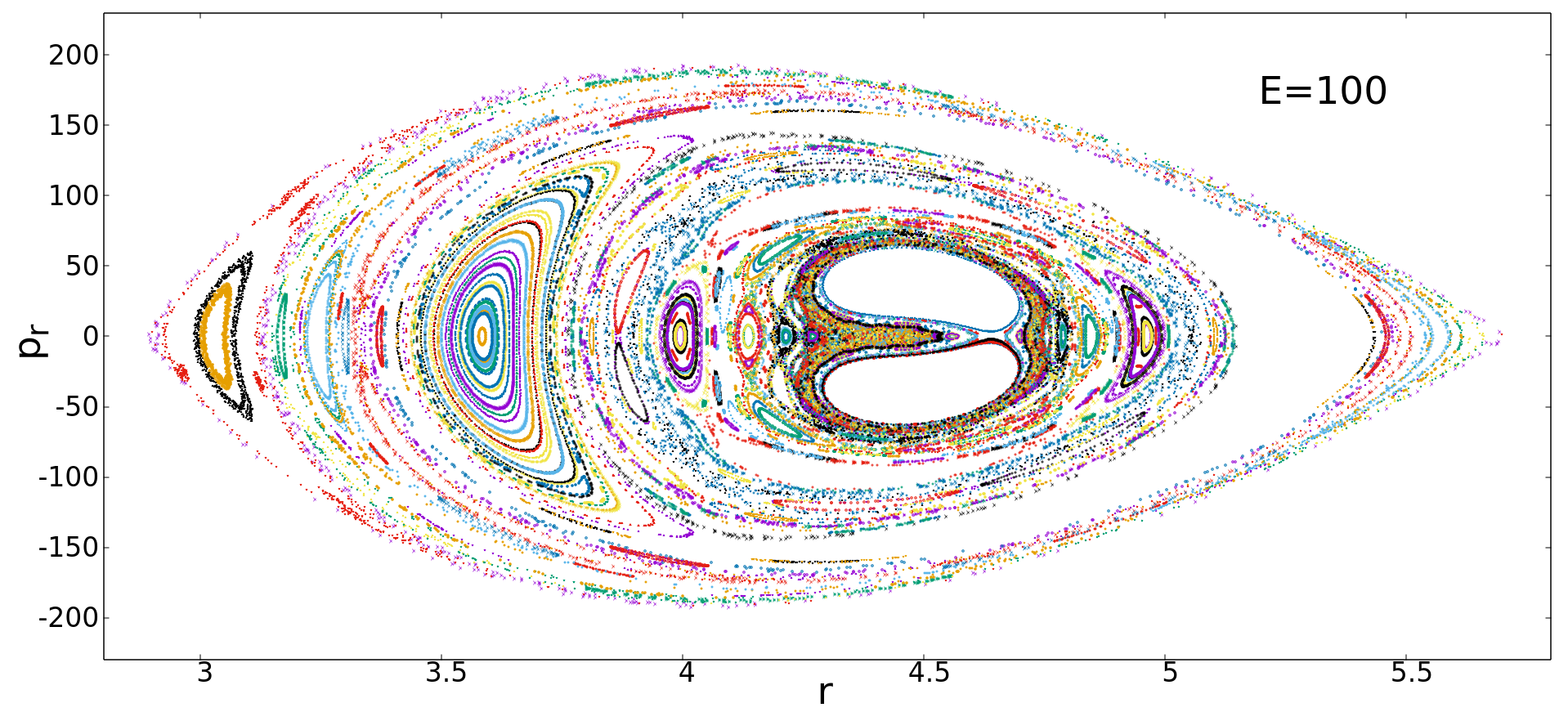}\label{ap2}}\\
	\subfigure[] 
        {\includegraphics[width=1.0\linewidth,height=0.5\linewidth]{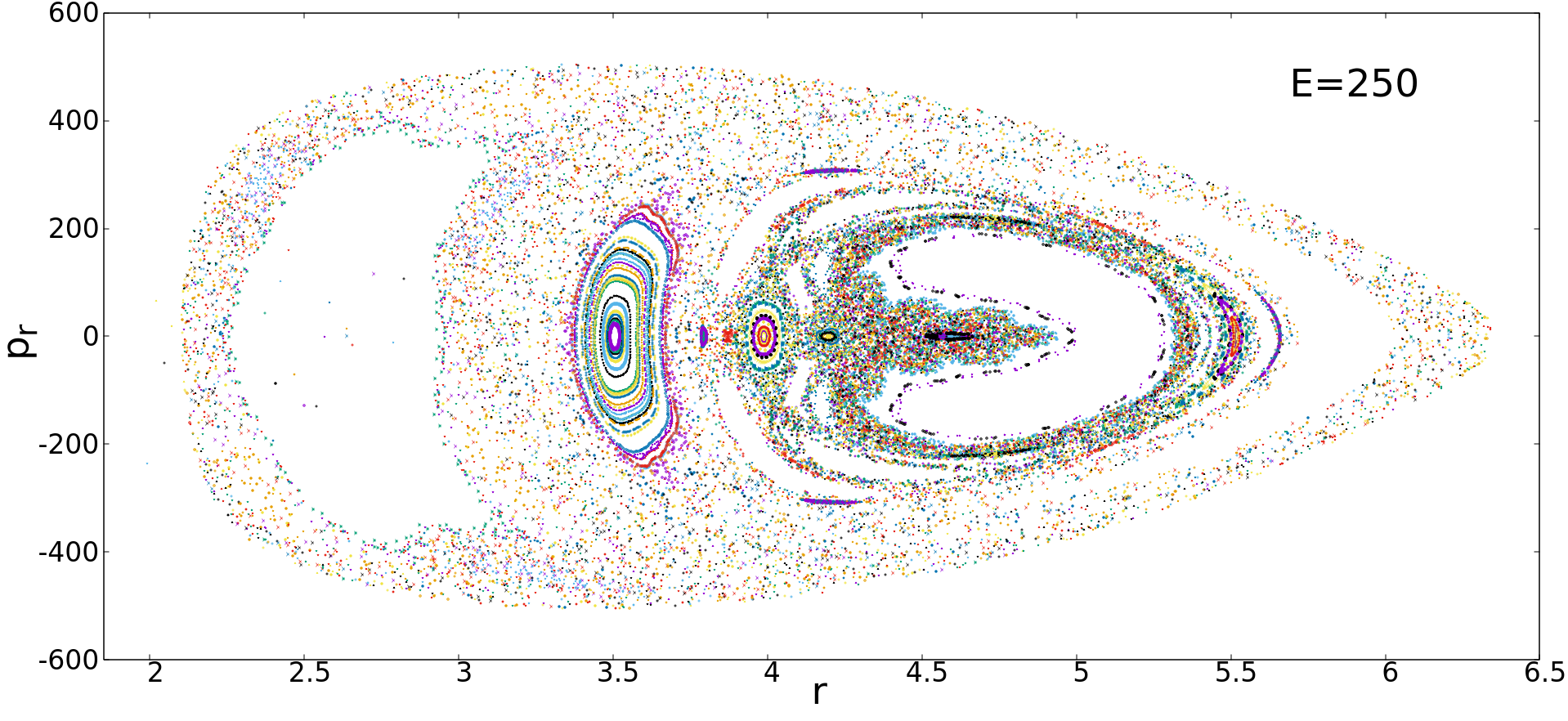}\label{ap3}}
	\subfigure[] 
        {\includegraphics[width=1.0\linewidth,height=0.5\linewidth]{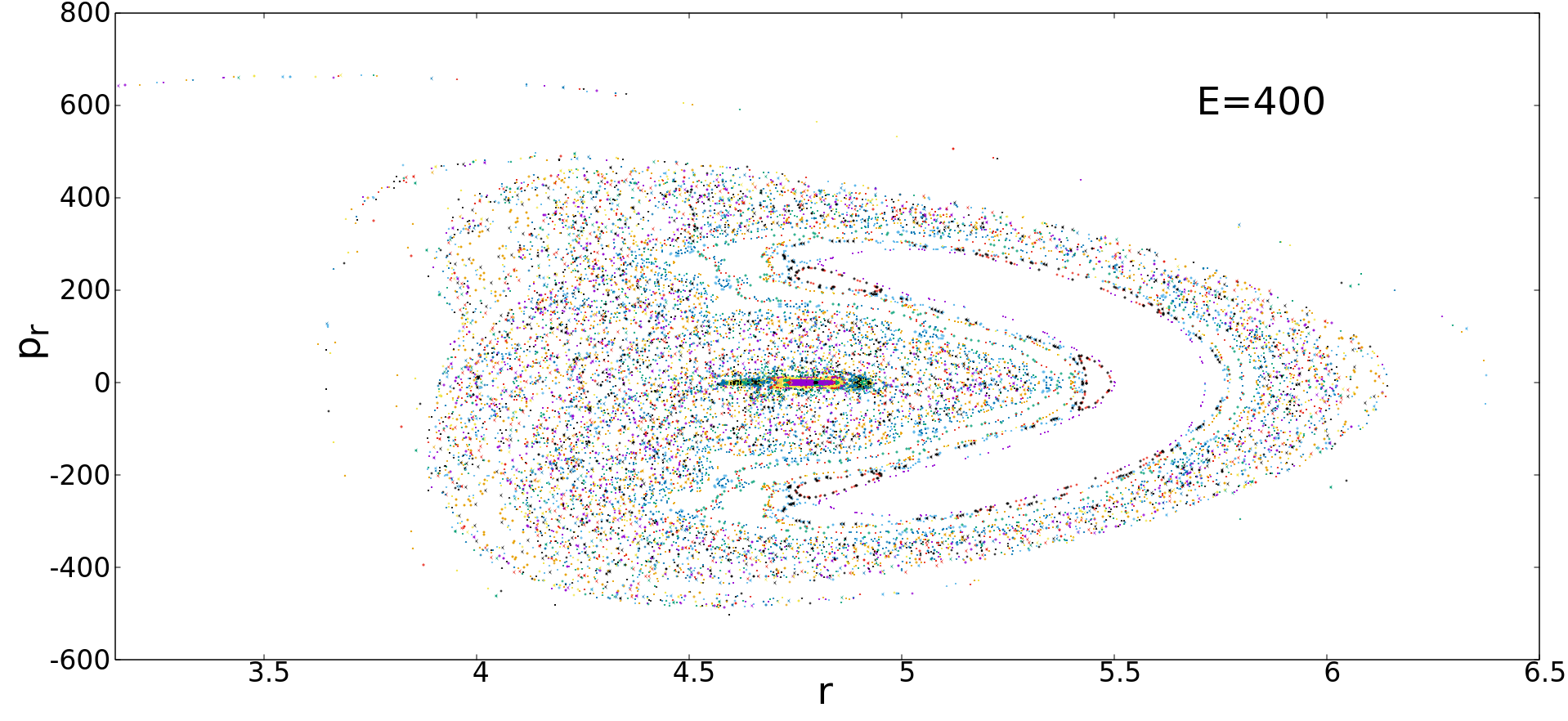}\label{ap4}}
	\end{array}$
        \end{center}
        \begin{minipage}{\textwidth}
        \caption{The Poincar$\Acute{e}$ sections in the $(r-p_r)$ phase plane in Schwarzschild coordinates with $\theta=0$ and $p_{\theta}>0$ for different energies with $\beta=10^{-5}$. For $E=100$, there is onset of chaos, and for energy ($E=400$), the KAM Tori break and the entire region gets filled with the scattered points indicating the presence of chaos.}\label{af1}
        \hrulefill
        \end{minipage}
        \begin{center}
        $\begin{array}{cc}
	\subfigure[] 
        {\includegraphics[width=1.0\linewidth,height=0.5\linewidth]{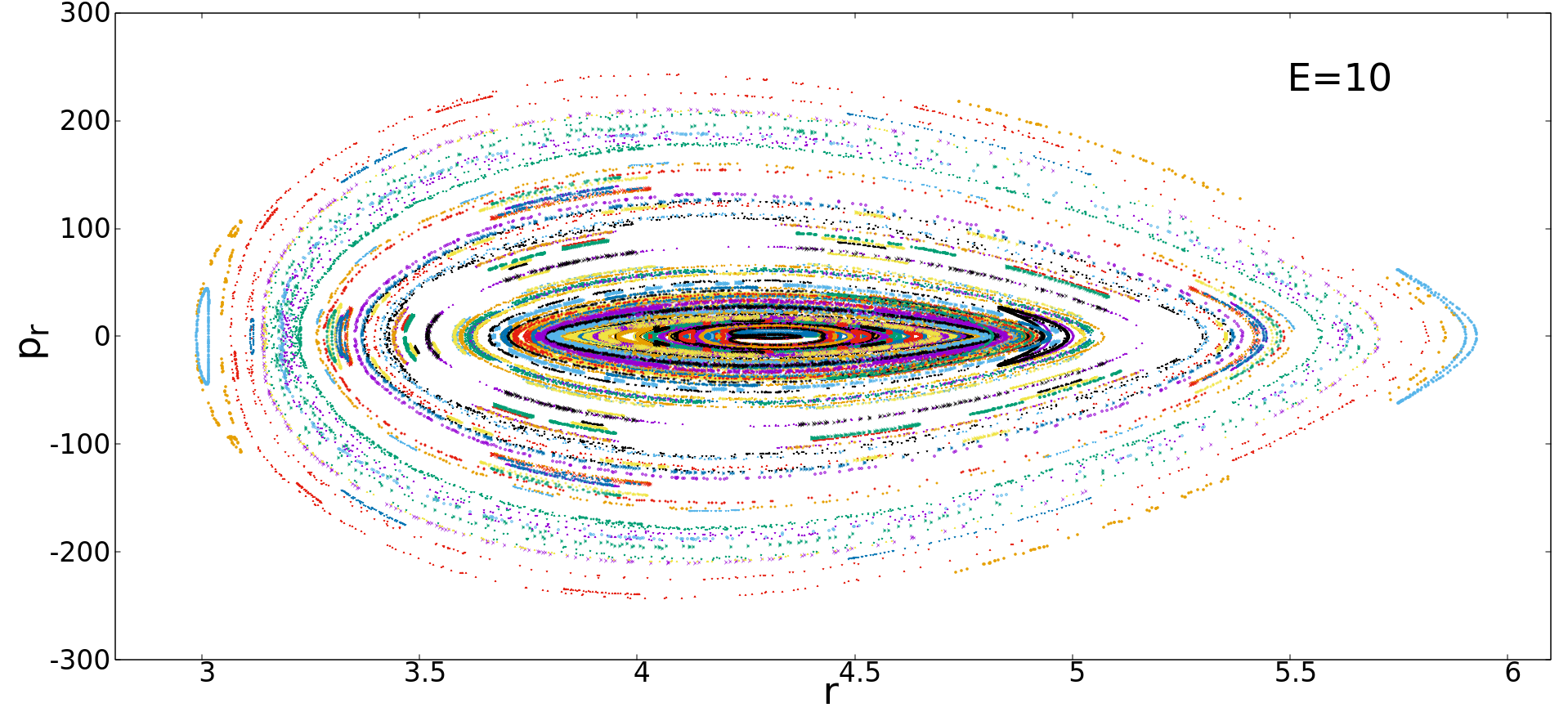}\label{a5}}
        \subfigure[]
        {\includegraphics[width=1.0\linewidth,height=0.5\linewidth]{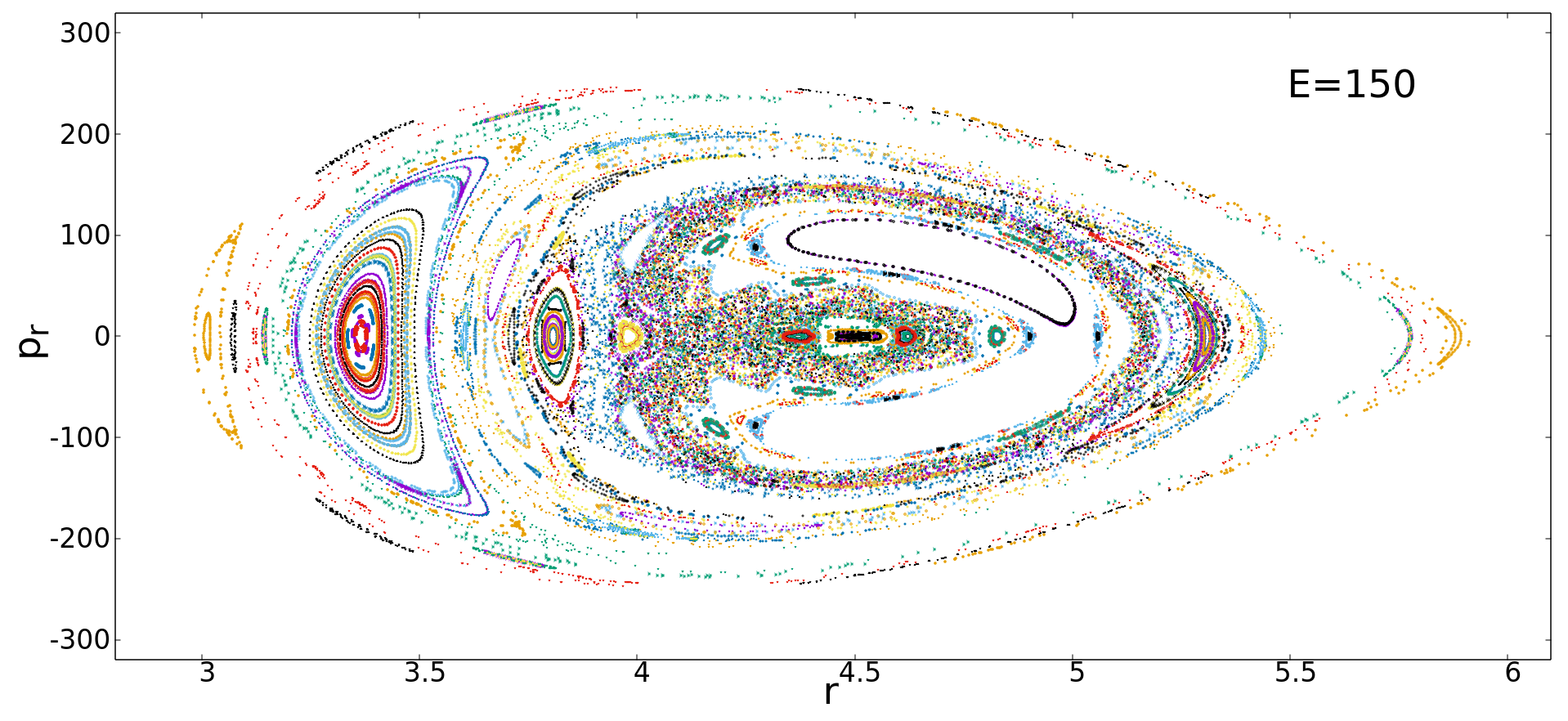}\label{a6}}\\
	\subfigure[] 
        {\includegraphics[width=1.0\linewidth,height=0.5\linewidth]{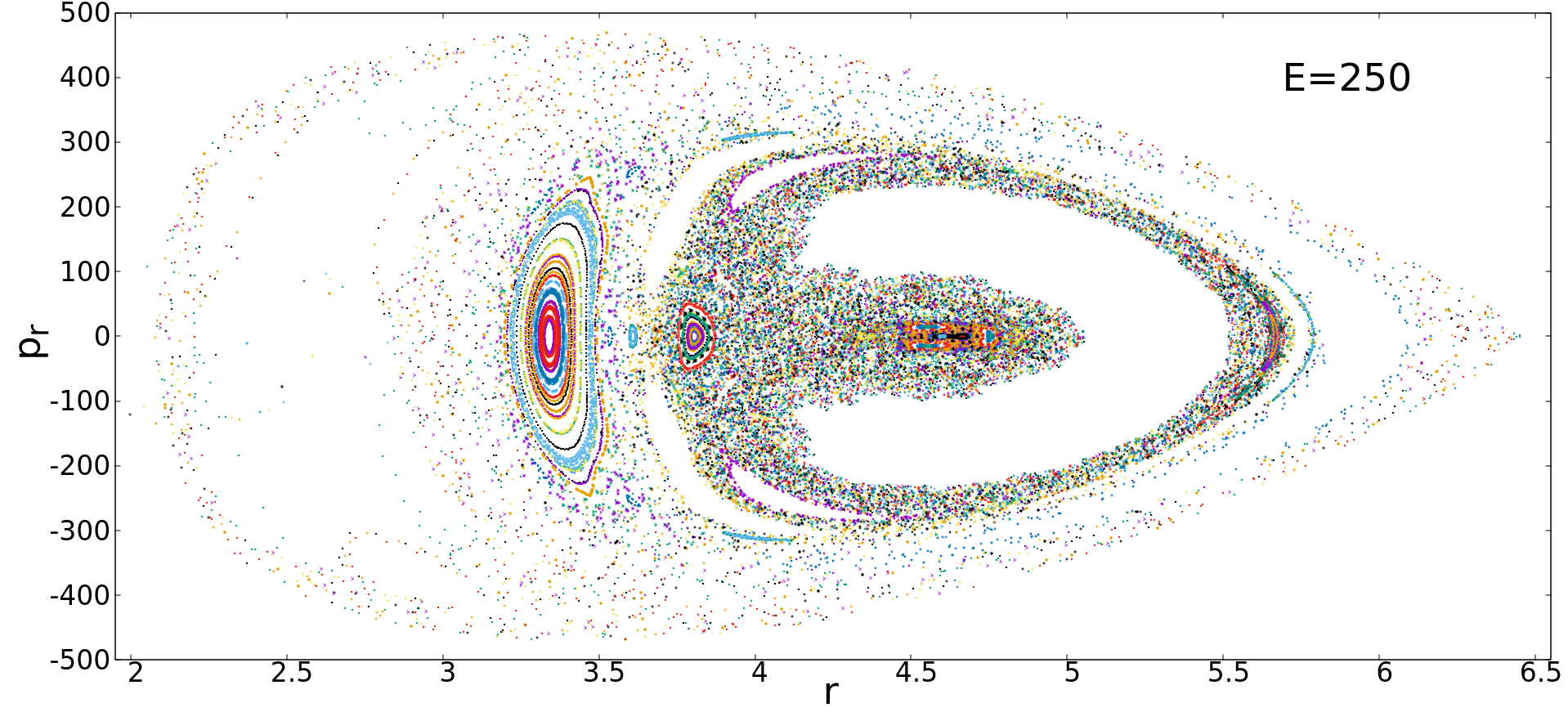}\label{a7}}
        \subfigure[]
        {\includegraphics[width=1.0\linewidth,height=0.5\linewidth]{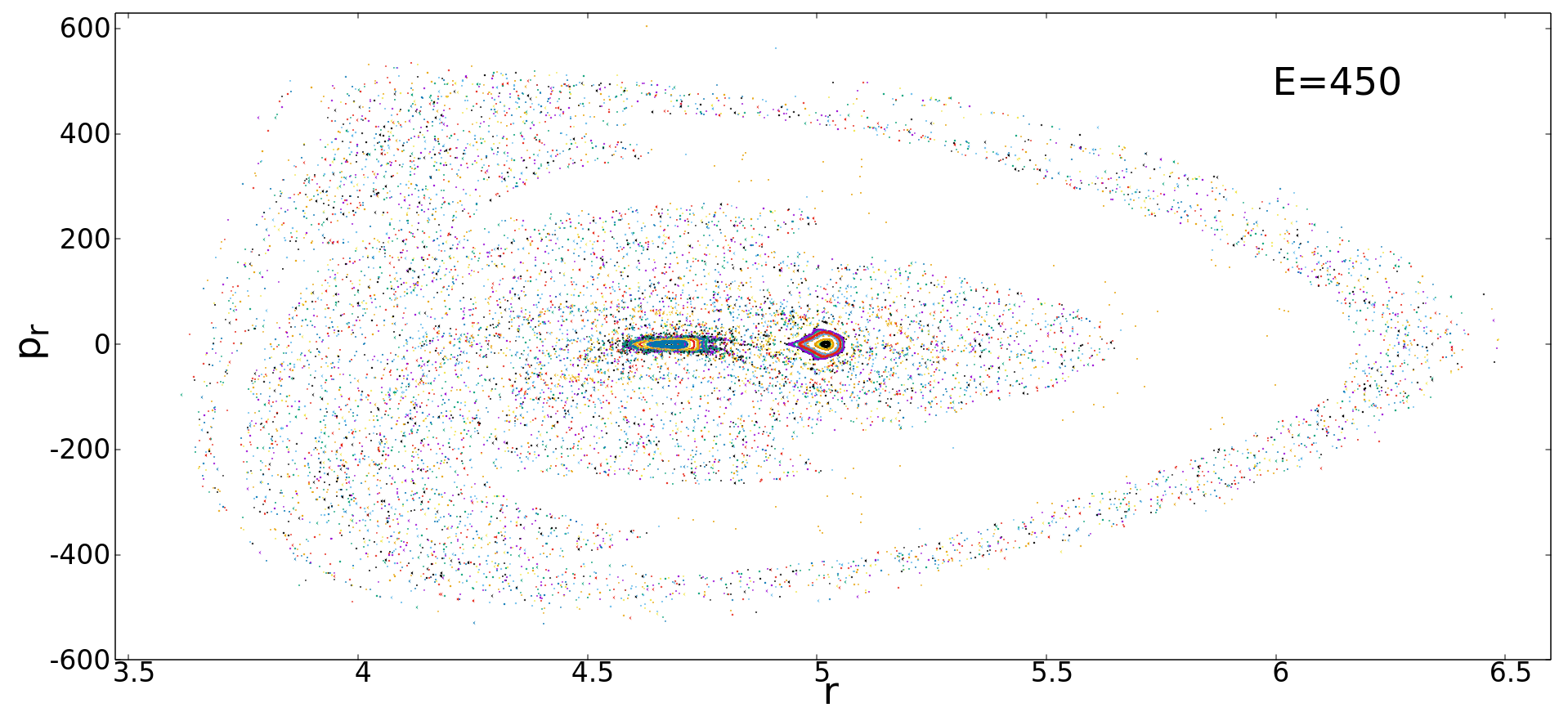}\label{a8}}
        \end{array}$
        \end{center}
        \begin{minipage}{\textwidth}
        \caption{The Poincar$\Acute{e}$ sections in the $(r-p_r)$ phase plane in Schwarzschild coordinates with $\theta=0$ and $p_{\theta}>0$ for different energies with $\beta=10^{-2}$. For $E=150$, there is onset of chaos, and for large energy ($E=450$) comparative to Fig.\ref{af1}, the KAM Tori break and the entire region gets filled with the scattered points indicating the presence of chaos.}\label{af2}
        \end{minipage}
        \end{figure}
    \end{widetext}

\end{document}